%% file: bare_conf_NDSS2025.tex
\documentclass[conference]{IEEEtran}
\ifCLASSINFOpdf
\else
\fi
\input{package}

\begin{document}
%
\title{\toolname: A Dual Prevention Approach against Singing Voice Conversion based Illegal Song Covers}

\author{\IEEEauthorblockN{Guangke Chen\IEEEauthorrefmark{1}, 
Yedi Zhang\IEEEauthorrefmark{2},
Fu Song\textsuperscript{(\Letter)}\IEEEauthorrefmark{3},\IEEEauthorrefmark{4}
Ting Wang\IEEEauthorrefmark{5}, 
Xiaoning Du\IEEEauthorrefmark{6} and 
Yang Liu\IEEEauthorrefmark{7}}
\IEEEauthorblockA{
\IEEEauthorrefmark{1}Pengcheng Laboratory 
\IEEEauthorrefmark{2}National University of Singapore
\IEEEauthorrefmark{3}Key Laboratory of System Software (Chinese Academy of Sciences) \\ and State Key
Laboratory of Computer Science, 
Institute of Software, Chinese Academy of Science}
\IEEEauthorblockA{\IEEEauthorrefmark{4}
 Nanjing Institute of Software Technology
 \IEEEauthorrefmark{5}Stony Brook University \IEEEauthorrefmark{6}Monash University \IEEEauthorrefmark{7}Nanyang Technological University}
  \IEEEauthorblockA{chengk@pcl.ac.cn\qquad yd.zhang@nus.edu.sg\qquad\ \ songfu@ios.ac.cn} \IEEEauthorblockA{inbox.ting@gmail.com\quad xiaoning.du@monash.edu\quad yangliu@ntu.edu.sg}
 }


\IEEEoverridecommandlockouts
\makeatletter\def\@IEEEpubidpullup{6.5\baselineskip}\makeatother
\IEEEpubid{\parbox{\columnwidth}{
Network and Distributed System Security (NDSS) Symposium 2025\\
24-28 February 2025, San Diego, CA, USA\\
ISBN 979-8-9894372-8-3\\
https://dx.doi.org/10.14722/ndss.2025.240747\\
www.ndss-symposium.org
}
\hspace{\columnsep}\makebox[\columnwidth]{}}

\maketitle

\input{abstract}


%

\input{introduction}

\input{background-related_work}
\input{overview}
\input{methodology}

\input{evaluation}

\input{discussion}

\input{conclusion}

\section*{Acknowledgments}
We thank our anonymous discussion lead and the reviewers for their constructive feedback and suggestions. 

This research is partially supported by Joint Funds of the National Natural Science Foundation of China (Grant No. U22A2036), National Natural Science Foundation of China (No. 62102202), CAS Project for Young Scientists in Basic Research (Grant No. YSBR-040), ISCAS New Cultivation Project (Grant No. ISCAS-PYFX-202201), and ISCAS Basic Research (Grant No. ISCAS-JCZD-202302). 

It is also supported by the National Research Foundation, Singapore, and DSO National Laboratories under the AI Singapore Programme (AISG Award No: AISG2-GC-2023-008). 
It is also supported by the National Research Foundation, Singapore, and the Cyber Security Agency under its National Cybersecurity R\&D Programme (NCRP25-P04-TAICeN). 
Any opinions, findings and conclusions or recommendations expressed in this material are those of the author(s) and do not reflect the views of National Research Foundation, Singapore and Cyber Security Agency of Singapore.

\bibliographystyle{IEEEtran}
\bibliography{ref_used}

\input{appendix}

\end{document}

%% file: package.tex
\usepackage{amsmath}
\usepackage{amsthm,amsfonts}
\usepackage{pifont}
\usepackage{multirow}
\DeclareMathOperator*{\argmin}{argmin}

\newcommand{\cmark}{\ding{51}}
\newcommand{\xmark}{\ding{55}}

\usepackage{enumitem}

\makeatletter
  \newcommand\figcaption{\def\@captype{figure}\caption}
  \newcommand\tabcaption{\def\@captype{table}\caption}
\makeatother

\newcommand{\toolname}{{SongBsAb}\xspace}

\usepackage{verbatim}
\usepackage{hyperref}
\usepackage{balance}

\usepackage{colortbl,xcolor}

\usepackage{flexisym}
\usepackage{wrapfig,lipsum}

\usepackage{cleveref}
\crefname{section}{\S}{\S}
\Crefname{section}{\S}{\S}

\usepackage{subcaption}

\usepackage[ruled,linesnumbered,vlined,noend]{algorithm2e}
\makeatletter
\newcommand{\removelatexerror}{\let\@latex@error\@gobble}
\makeatother
\usepackage[noend]{algpseudocode}

\usepackage{makecell}

\usepackage[flushleft]{threeparttable}

\definecolor{goldenrod}{rgb}{0.85, 0.65, 0.13}

\SetKw{Continue}{continue}

\usepackage{diagbox}

\usepackage{graphicx}

\usepackage{booktabs}

\usepackage{tcolorbox}
\tcbuselibrary{breakable}

\usepackage[normalem]{ulem}

\usepackage{marvosym}

\renewcommand{\footnoterule}{
    \kern -3pt 
    \hrule width 0.49\textwidth height 0.6pt 
    \kern 2pt 
}

\usepackage{fancyhdr}

%% file: abstract.tex
\begin{abstract}
    Singing voice conversion (SVC) automates song covers by converting a source singing voice from a source singer into a new singing voice with the same lyrics and melody as the source, but sounds like being covered by the target singer of some given target singing voices. However, it raises serious concerns about copyright and civil right infringements. 
    We propose \toolname\footnote{BsAb stems from ``{\bf B}i{\bf s}pecific {\bf A}nti{\bf b}ody'' which has two different antibodies (\toolname's prevention of a song being used as both the target and source songs by identity and lyric disruptions), respectively neutralizing two different types of antigen (SVC's infringement of civil rights and copyrights).},  
    the first proactive approach to tackle 
    SVC-based illegal song covers.
    \toolname adds perturbations to singing voices before releasing them, so that when they are used, the process of SVC will be interfered, leading to \emph{unexpected} singing voices. 
    {Perturbations are carefully crafted to
    (1) provide a dual prevention, i.e., preventing the singing voice from being used as the source and target singing voice in SVC,
    by proposing a gender-transformation loss and a high/low hierarchy multi-target loss, respectively;
    and (2) be harmless,  i.e., no side-effect on the enjoyment of protected songs,  
    by refining a psychoacoustic model-based loss with the backing track as an additional masker, 
     a unique accompanying element for singing voices compared to ordinary speech voices.}
    We also adopt a frame-level interaction reduction-based loss and encoder ensemble  
    to enhance the transferability of \toolname to unknown SVC models. 
    We demonstrate the prevention effectiveness, harmlessness, and robustness of \toolname on {{five} {diverse} and 
    promising
    SVC models, using both English and Chinese datasets}, and both objective and human study-based subjective metrics. 
    Our work fosters an emerging research direction for mitigating illegal automated song covers.
\end{abstract}

%% file: introduction.tex
\section{Introduction}\label{sec:intro}
The advent of generative AI has revolutionized the realm of AI-generated art, including AI-generated song covers based on singing voice conversion (SVC)~\cite{SVC-challenge-2023}. 
Unlike human-based song covers, SVC empowers individuals without exceptional singing and vocal imitation abilities to 
create song covers. Consequently, the internet has seen a surge in SVC-covered singing voices and songs. One of the most notable examples is ``AI Sun Yanzi'', a virtual singer that imitates the singing voice of the famous Mandopop female singer Stefanie Sun (Chinese name Yanzi Sun) and 
has covered over 1,000 songs {from other singers}, far more than the total number of songs {by Stefanie} in her past 23-year career. The most popular {cover} has {garnered} millions of views and thousands of shares on Bilibili, China’s largest user-generated video streaming site~\cite{ai_singer_bilibili,bilibili}. Another cover is the 
 song ``Heart on My Sleeve'', which imitates the singing voices of the singers Drake and The Weeknd. It has garnered over 15 million views on TikTok in just two days, and was submitted for 
 a Grammy Award consideration~\cite{drake_grammy}. 

However, it raises serious concerns about copyright and civil right infringements~\cite{ai_singer_bilibili,Global_Times,NoAIFraudAct} (cf. \cref{sec:rights} for details), because a song is an intellectual property composed of key elements such as lyrics, melody, and the singer's rendition. 
Recently, 
the ``Elvis Act'' was signed into state law for the first time to protect against exploitative use of generative AI~\cite{Elvis_law}, and
an open letter issued by the Artist Rights Alliance and signed by more than 200 artists (e.g., Billie Eilish, Katy Perry) calls for responsible AI music practices~\cite{openLetter}.  

Thus, it is increasingly crucial for the music industry and society at large to safeguard the interests and rights of song owners and singers facing potential infringements whenever  
songs are used as source or target singing voices in SVC. One may detect SVC-covered singing voices after infringements have already been committed, but, this passive solution becomes inefficient and cumbersome with the surge of SVC-covered singing voices and songs due to its low entry barriers. In this work, we propose \toolname, the first prevention approach, to effectively tackle  
SVC-based illegal song covers. 
\toolname is a proactive, dual prevention  solution that can fundamentally prevent infringements from {happening} by adding a subtle perturbation to a singing voice. 
The song owners (defenders) 
can employ \toolname on singing voices prior to their release. When protected singing voices
are used, the process of SVC will be
interfered,
producing unexpected singing voices to the SVC users. 
The design of \toolname faces and solves 
the following technical challenges, especially compared to (ordinary) speech voices and their conversion.

\noindent
\textbf{{Challenge-1}: More Involved Rights to Protect.}
The protection requirements for songs are complex and distinct, involving more intricate rights than those of speech voices or images. 
Indeed, songs can be used as either source or target singing voices in SVC, unknown to song owners in advance, thus requiring protection of various rights (e.g., singer’s civil rights, and copyrights of lyrics and melodies; cf.~\Cref{sec:rights}). 
In contrast, there are no copyright issues regarding melody or textual content for speech voices and images. 
Therefore, prior works on speech voices~\cite{AntiFake, Attack-VC, VSMask, V-Cloak, VoiceCloak}, which only protect speaker identity, 
or on images~\cite{MIST, Glaze, UnGANable}, which only protect the identity in faces or artists' painting styles, 
cannot be ported for the protection of songs. 
To tackle this challenge,  
\toolname is designed to provide a dual prevention by adding subtle perturbations to singing voices to prevent them from being used as source/target singing voices in SVC. 
By doing so, SVC-covered singing voices (songs) neither preserve the original lyrics (lyric disruption) nor imitate the singer (identity disruption), 
thus directly protecting the copyrights of lyrics and the civil rights of the singer. 
The copyrights of melodies and copyrights to reproduce and distribute songs are indirectly protected as SVC users are discouraged to release unexpected SVC-covered singing voices and gradually abandon SVC due to its weird behavior. 
We remark that \toolname is also effective in disrupting lyrics (resp. identity) only. 
Inspired by adversarial attacks~\cite{Intriguing}, 
\toolname formulates the perturbations searching as an optimization problem with novel designated loss functions, 
including a gender-transformation loss and a high/low hierarchy multi-target loss to maximize identity disruption and lyric disruption, respectively. 

\noindent
\textbf{{Challenge-2}: Higher Quality Requirements.}
In contrast to speech voices that are primarily used for conversation, songs are music arts  for appreciation and entertainment,  
and are highly expected to meet high-quality standards~\cite{carterette1989science, titze1998principles, cook2021music}.
Thus, the prevention should be harmless for the song  (including melody, lyrics, and singing style). 
To tackle 
this challenge, 
we harness the simultaneous masking~\cite{auditorymasking} 
which entails that a faint yet audible sound (the maskee) becomes inaudible when another louder audible sound (the masker) is concurrently occurring~\cite{audio-watermark}. 
In the real world, a singing voice is typically accompanied with a backing track in the song. We treat both a singing voice and its backing track as maskers and a perturbation as maskee, and use a loss to control the magnitude of the perturbation. It refines the prior simultaneous masking-based loss~\cite{Qin_Psy, psychoacoustic_hiding_attack} that only uses the speech voice as the masker, thus significantly improving the harmlessness, as the perturbation will be inaudible as long as it is weaker than any of two maskers. 

\noindent
\textbf{{Challenge-3}: More Challenging for Transferability.}
In practice, SVC models may use distinct encoders from \toolname. Hence, the
prevention should generalize and transfer to unknown SVC models. 
Adversarial voices inherently exhibit low transferability~\cite{AWBPT20}, 
and due to the dual prevention, \toolname involves more possibly distinct encoders than prior works in the speech domain~\cite{AntiFake,Attack-VC,VSMask}. Therefore, it is more challenging for \toolname to achieve high transferability of both identity and lyric disruptions. 
To tackle this challenge, we  
adopt a frame-level interaction reduction-based (FL-IR) loss~\cite{IRA_attack} and  encoder ensemble~\cite{LCLS17,QFA2SR, AntiFake}. They improve the transferability from two different perspectives, thus are complementary, i.e., their combination further boost the transferability, making \toolname more practical and useful.

We conduct an extensive evaluation 
to demonstrate the efficacy 
of \toolname. 
We first evaluate the prevention effectiveness on {5 diverse} and promising SVC models using both English and
Chinese datasets via 5 objective metrics. 
{\toolname can reduce the identity similarity 
between SVC-covered singing voices and 
the target singer and enlarge the lyric word error rate,
together reducing the singing voice conversion success rate by over 97\%.}
It significantly outperforms two recent promising methods \cite{Attack-VC,AntiFake} that were designed for preventing ordinary speech voice conversion in terms of both prevention effectiveness and harmlessness.
The subjective human study with 3 tasks also confirms the prevention effectiveness and utility of \toolname on the enjoyment of protected 
songs. 

We then evaluate transferability on 8 distinct identity encoders and 5 distinct lyric encoders. 
{\toolname shows a strong ability to transfer to unknown SVC models, while also surpassing previous works~\cite{Attack-VC, AntiFake}.}

We finally demonstrate the robustness of \toolname in over-the-air scenario and against adaptive SVC users who completely know and aim to bypass \toolname 
by pre-processing protected singing voices via existing voice transformations and tailored optimizations, {or by fine-tuning SVC models.}

In summary, the main contribution of this work includes: 
\begin{itemize}[leftmargin=*]
    \item We present \toolname, 
    the first proactive solution to prevent right infringements caused by SVC-based illegal song covers. 
    It features a dual prevention, capable of causing both the identity disruption and lyric disruption in SVC-covered singing voices,
    for which we devise a gender-transformation loss and a high/low hierarchy multi-target loss, respectively.
     \item We propose to utilize backing tracks, a unique accompanying element with singing voices in songs compared to speech voices, as maskers to further improve harmlessness. Our simultaneous masking-based loss effectively enhances the quality of protected songs and thus the utility of \toolname.  
    \item While \toolname exhibits transferability, we further 
    utilize FL-IR loss 
    and encoder ensemble to enhance transferability for causing both the identity disruption and lyric disruption on unknown SVC models in a complementary way.
    \item Our work makes the first significant step towards coping with illegal automated song covers. We release our code and audio samples, and discuss possible future works to foster exploration in this emerging research direction.
    
\end{itemize}
For convenience, we summarize the abbreviations in \tablename~\ref{tab:abbr}. 
Our code and audio samples are available at~\cite{dual-pre-svc}.

%% file: background-related_work.tex
\section{Background \& Related Work}\label{sec:background}

\subsection{Singing Voice Conversion (SVC)}\label{sec:background_svc}
\begin{figure}[t]\centering
    \includegraphics[width=0.48\textwidth]{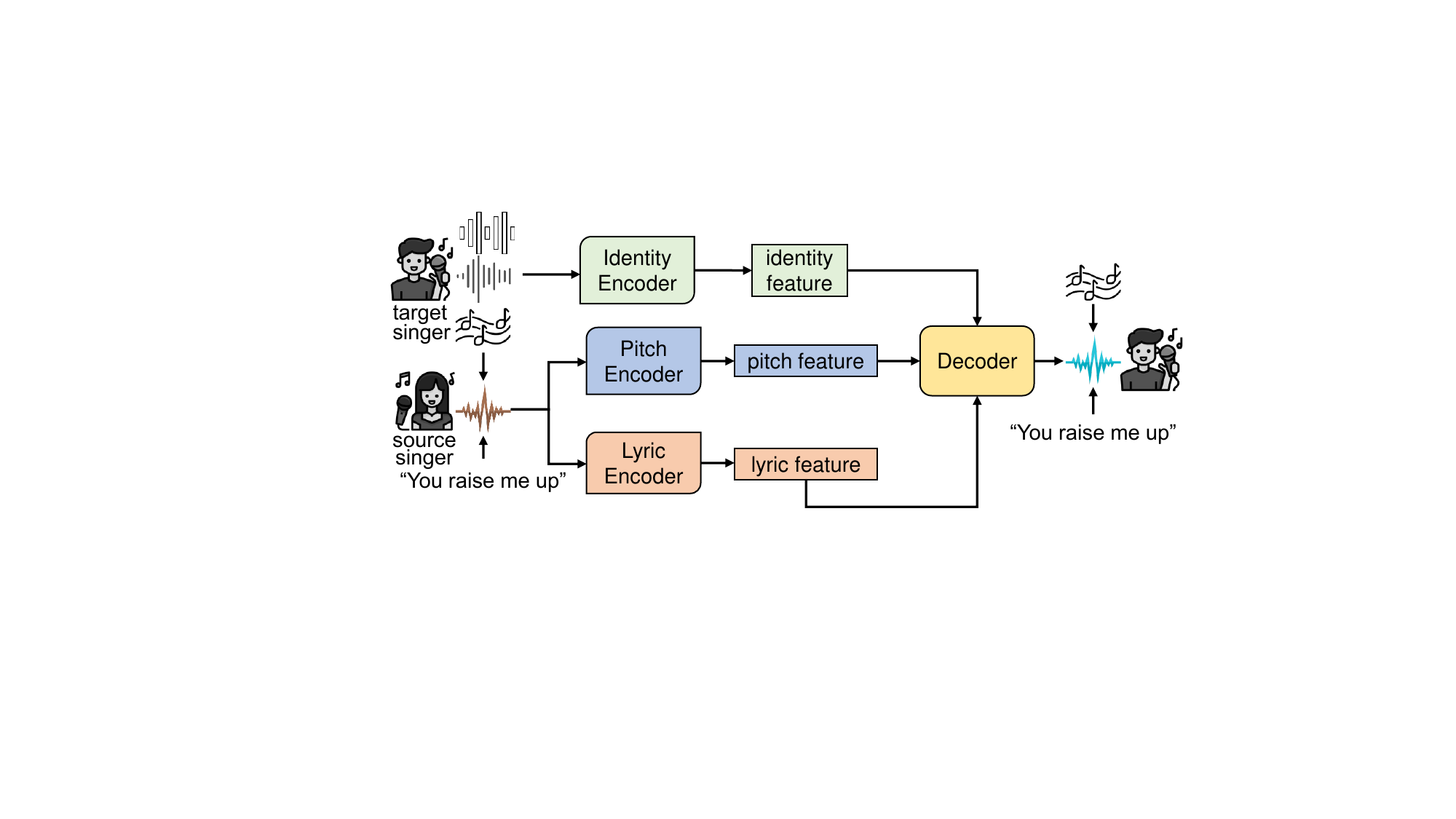}
    \caption{Mainstream Singing Voice Conversion Systems.}
    \label{fig:svc}
\end{figure}

A song consists of a singing voice and a backing track, stored in separate channels. 
Singing voice conversion transforms a song's vocal rendition from one singer to another's style and timbre 
while preserving the original lyrics and melody~\cite{SVC-challenge-2023}. 
The backing track is removed during conversion and is not part of the process.
Mainstream SVC systems use an encoder-decoder architecture~\cite{SVC-challenge-2023}, as shown in \figurename~\ref{fig:svc}. 
There are three common encoders: the identity encoder extracts the identity feature from a few target singing voices representing the target singer's singing style and voiceprint, while the pitch encoder and lyric encoder extract pitch and lyric features from the source singing voice of the source singer, characterizing the melody and lyrics. The decoder then fuses these features to produce a singing voice that resembles the target singer covering the source singing voice.

\noindent {\bf Explicit \& Implicit SVC.} 
{Since both target and source singing voices contain identity and lyric information, SVC relies on information disentanglement, achieved through either explicit or implicit methods~\cite{SVC-challenge-2023}. 
Explicit methods use pre-trained encoders with disentanglement capabilities
and the decoder is then trained with these frozen encoders. Lyric encoding is typically handled by speaker-independent models such as Whisper~\cite{Whisper} or Hubert~\cite{HuBERT}, while identity encoding uses content-independent models such as GE2E~\cite{GE2E}. 
In contrast, implicit methods adopt encoders that originally 
lacked of disentanglement capabilities and employ specialized strategies for disentanglement. 
For example, NeuCoSVC~\cite{NeuCoSVC} uses the same encoder for both voices 
and a KNN-based matching module to retain the source lyrics while shifting identity to target singers. StarGANv2~\cite{StarGANv2_VC} employs adversarial training to supervise the decoder to capture only the target's identity and source's lyrics. 
Both methods use signal-processing based (e.g., WORLD~\cite{WORLD}) or neural networks-based (e.g., Crepe~\cite{Crepe}) pitch encoders, 
and use generative models such as GANs~\cite{GANs} or diffusion models~\cite{Diffusion-Model-Survey} as decoders due to their strong generative capacity.}

\noindent {\bf Few-shot \& non-few-shot.} 
{
Few-shot trains the encoders and decoder without the target singers used during inference. 
In contrast, non-few-shot predefines target singers during training,
and to align with an unseen target singer in inference, models must be trained from scratch or fine-tuned, 
requiring more computational resources and a large number of singing voices from the target singer
to avoid overfitting~\cite{Lora-SVC,Grad-SVC,NeuCoSVC}.
For both, it is common to use a few samples from the target singer during inference 
and feed the aggregated identity features to the decoder to make outputs sound more like the target singer.}

\noindent
{{\bf Key differences between SVC and ordinary voice conversion} include (1) challenging task~\cite{SVC-challenge-2023}: singing voices differ fundamentally~\cite{sv_japan_dataset, differ_singing_speaking, science_singing_voice} and vary more in phoneme duration, pitch, expression, singing style, and speaker characteristics~\cite{svc_phoneme, OpenSinger, NUS-48E, svc_speaker}, 
requiring more proper information disentanglement~\cite{SVC-challenge-2023}; 
(2) severe rights infringements (cf.~\Cref{sec:intro}): singing voices involve more complex copyright concerns~\cite{OpenSinger}; 
(3) architecture: SVC uses specialized pitch encoders; and 
(4) inputs: SVC source voices should be professionally sung~\cite{OpenSinger}.}

\begin{table}[t]
   \centering\setlength\tabcolsep{4pt} 
     \caption{Main Abbreviations.}
    \scalebox{1.1}{
    \begin{tabular}{|c|c|c|}
    \hline
         {\bf Abbr.} & {\bf Full Form} & {\bf Meaning} \\ \hline
         SVC & singing voice conversion & N/A \\ \hline
        
         $\mathcal{I}$ & \makecell[c]{target singing voice} & \makecell[c]{input of SVC providing \\ identity information} \\ \hline
        $\mathcal{L}$ & \makecell[c]{source singing voice} & \makecell[c]{input of SVC providing \\ lyric and melody} \\ \hline
        $\tilde{\mathcal{I}}$ & \makecell[c]{protected \\ target singing voice} & protected version of $\mathcal{I}$ \\ \hline
        $\tilde{\mathcal{L}}$ & \makecell[c]{protected \\ source singing voice} &  protected version of $\mathcal{L}$ \\ \hline 
         - & target singer & the singer of $\mathcal{I}$ \\ \hline
         - & source singer & the singer of $\mathcal{L}$ \\ \hline 
            $y/\tilde{y}$ & \makecell[c]{undefended/defended \\ output singing voice} & \makecell[c]{output of SVC \\ without/with \toolname} \\ \hline 
            - & {destination singer} & the singer of $\tilde{y}$ \\ \hline 
       FL-IR loss & \makecell[c]{frame-level interaction \\
reduction-based loss} & \makecell[c]{a loss for enhancing the\\  transferability of \toolname}\\      \hline 
    \end{tabular}
    }
    \label{tab:abbr}
\end{table}

\subsection{The Rights Infringed by SVC}\label{sec:rights}
A song is an intellectual property created by multiple contributors, including the lyricist, composer, singer, and record company. The lyricist and composer write the lyrics and melody, typically transferring their copyrights to the record company while retaining authorship and sharing royalties. The record company hires singers to perform the song and becomes the owner by acquiring performance rights through copyright transfer. In rare cases, such as with online singers, the lyricist, composer, and singer are the same individual. SVC may harm the following rights and interests of song owners and singers when their songs are used as source or target singing voices.

\noindent {\bf Copyrights to reproduce and distribute songs.} 
Copyright laws protect song owners' exclusive rights to reproduce and distribute their songs, as outlined in Article 9 of China's copyright law~\cite{china_copyright_law}, §106 of Title 17 of the U.S. Code~\cite{us_copyright_law}, and Section 6 of U.K. copyright law~\cite{uk_copyright_law}. During singing voice conversion, both source and target singing voices are transferred from their original publication platforms (e.g., music platforms) to a computational platform performing SVC. Without copyright licenses, this process infringes on the rights of song owners for both the source and target singing voices. 

\noindent {\bf Copyrights to perform and display lyrics and melodies.} 
Copyright laws protect song owners' exclusive rights to display and perform their melodies and lyrics~\cite{china_copyright_law, us_copyright_law, uk_copyright_law}. They must be informed and compensated for any usage of their works. Many singers have faced charges for covering songs without permission, and SVC-based automated covers similarly threaten the rights of song owners regarding the lyrics and melodies of the source singing voices. 
Additionally, song owners often seek to ensure exclusive performances by singers to maintain their reputation and profits, a goal undermined by SVC. Finally, releasing SVC-covered songs without crediting the lyricists or composers violates their authorship rights.

\noindent {\bf Civil rights of singers over voices and reputation.} 
First, singers have civil rights over their voices (e.g., Article 1023 of the Civil Code of China~\cite{china_civil_law} {and  ELVIS Act of Tennessee U.S.A~\cite{Elvis_law}}), akin to rights over their likeness, which prohibits the production, use, or publication of their voices without permission. SVC violates this regulation by producing singing voices that sound like the target singer. Second, malicious SVC users may exploit sensitive source singing voices, such as those with political bias, discrimination, or violence, leading to reputational damage and infringement under civil codes~\cite{china_civil_law} or defamation/privacy laws~\cite{uk_defamation_privacy_law, us_defamation_law, us_privacy_law}. 
Third, exceptional vocal skills and unique performance styles are vital for singers' livelihood and careers. SVC-enabled AI singers, with lower entry barriers and costs, could replace traditional singers, potentially breaching unfair competition laws~\cite{us_unfair_competition_law, china_unfair_competition_law}. Finally, within contracts, a singer's voice and public image are tools for the song owner's profit, 
so imitating a singer's voice or degrading his/her reputation could harm the owner's revenue.

\subsection{Adversarial Examples}\label{sec:adver_example}
\noindent {\bf Adversarial examples for good.} Adversarial examples are deliberately crafted inputs 
to deceive models 
and have been widely studied~\cite{Intriguing, CW, FakeBob, Qin_Psy, psychoacoustic_hiding_attack, AS2T, yuan2018commandersong, QFA2SR, SpeakerGuard, PhoneyTalker, FenceSitter}.
They also have been utilized for beneficial applications 
(cf. \tablename~\ref{tab:related_work} in 
\Cref{sec:summary_works} for a summary).

Error-minimizing noise is applied to personal data so that models trained on them are tricked into believing there is ``nothing'' to learn~\cite{Unlearnable}. 
Such noises are improved later to make them robust
against adversarial training \cite{Robust_Unlearnable}. 
Glaze~\cite{Glaze} and MIST~\cite{MIST} added 
perturbations to artists' artworks such that text-to-image models 
fine-tuned on these artworks fail to mimic the painting styles of the protected artists. 
UnGANable~\cite{UnGANable} perturbed face images of a target user so that the face images reconstructed from the face manipulator do not contain the user's identity. 
V-cloak~\cite{V-Cloak} and VoiceCloak~\cite{VoiceCloak} added perturbations to human voices to hide speakers' identities from speaker recognition models, thus achieving voice anonymity. 
Glaze, MIST, and UnGANable target AI-generated images,  V-cloak and VoiceCloak target human-generated speech voices, while our work targets AI-generated singing voices. 

AttackVC~\cite{Attack-VC}, VSMask~\cite{VSMask}, and
AntiFake~\cite{AntiFake} are the closest works to ours,
{all of which target the voice modality and generative models}. 
They add perturbations to ordinary speech voices of 
a target speaker to make speech voice conversion or synthesis to generate voices not recognized as the target speaker by both speaker recognition models and human perception. 
Our work focuses on singing voice conversion, a more challenging task than speech voice conversion~\cite{SVC-challenge-2023}. 
\toolname differs from them in the following aspects: 
(1) They only affected the identity of crafted voices and thus cannot prevent singing voices from being used as source singing voices in SVC,
while \toolname prevents the singing voice from being used as the source 
or target singing voice (dual prevention). 
This enables broader applications, protecting not only the civil rights of voices and performing rights of singers but also the copyright of lyrics.  
Even for protecting singers only, 
experiments show that \toolname significantly outperforms 
the publicly available AttackVC and AntiFake (cf.~\Cref{sec:overall_dual}).   
(2) To decide the destination speaker for better identity disruption, 
AttackVC and VSMask randomly selected an opposite-gender speaker, 
and AntiFake utilized the Analytic Hierarchy Process (AHP) to balance computational embedding deviation and human judgment. 
They both represent a singer with a single voice embedding. 
Instead, we propose a gender-transformation loss to optimize towards a destination speaker that has the least objective identity similarity with the target singer among a pool of opposite-gender singers
and use the centroid of multiple voice embeddings to represent a singer, 
resulting in the best identity disruption (cf. \Cref{sec:exper_gt}).
(3) To improve harmlessness, AttackVC and VSMask enforced an $L_\infty$ norm-based constraint that may not correlate with human auditory perception~\cite{AWBPT20}. 
AntiFake used different gain functions for the perturbation strength in different frequency bands and maximized the signal-to-noise ratio. 
Instead, we utilize the psychoacoustics model~\cite{audio-watermark} to hide perturbations under the auditory perception threshold of humans. Notably, motivated by the fact that a singing voice is commonly accompanied by a backing track in a song, we propose using backing tracks as additional markers, which improves perturbations hiding capacities (cf.~\tablename~\ref{tab:overall_performance}). 
Backing tracks are unique elements of singing voices and have never been explored in the literature to strengthen harmlessness. 
(4) AttackVC and VSMask evaluated transferability on unknown models, while AntiFake enhanced transferability via encoder ensemble. 
Besides the encoder ensemble, we propose a frame-level interaction reduction-based (FL-IR) loss to enhance transferability further.
The rationales behind the two methods are different (cf.~\Cref{sec:tr-enloss}), 
so their combination yields the best results (cf.~\cref{sec:exper_transfer}), 
and \toolname exhibits superior transferability.

\noindent {\bf Interaction vs. transferability.} 
Adversarial examples 
crafted on one surrogate model often can transfer to other target models. 
However, the transferability may be limited especially {when} there is a large gap between the surrogate and target models~\cite{AS2T, QFA2SR}. 
Wang et al. \cite{IRA_attack} interprets the transferability from the perspective of interaction $I$ inside perturbations. The interaction between two perturbation units $i$ and $j$, denoted by $I_{ij}$, is the change of the importance of the unit $i$ after perturbing unit $j$. 
The average interaction over all pairs of perturbation units is defined as:  
\[\frac{\mathbb{E}_{i}(v(\Omega)+v(\emptyset)-v(\Omega\setminus \{i\})-v(\{i\}))}{n-1}\] 
where $v$ is a utility function measuring the importance 
of perturbation units for deceiving models, $n$ is the number of perturbation units, and $\Omega$, $\emptyset$, $\Omega\setminus \{i\}$, and $\{i\}$ denote the cases of all units being perturbed, no unit being perturbed (i.e., normal example), all units excluding the unit $i$ being perturbed, and only the unit $i$ being perturbed, respectively. 
It was shown that interaction is negatively correlated with transferability~\cite{IRA_attack}: a large interaction indicates that the perturbation units need to work closely to jointly fool the surrogate model,
thus leading to low transferability, as a large interaction is more likely to be broken on target models. 

\subsection{Simultaneous Masking}
Simultaneous 
masking refers to the phenomenon 
that one faint but audible sound (the maskee) becomes inaudible 
in the presence of another simultaneously occurring louder audible sound (the masker)~\cite{auditorymasking,audio-watermark}. 
The masker introduces a curve of masking threshold which specifies the minimal sound pressure level of 
a tone to be human perceptible with respect to the tone frequency. In other words, any signal below this curve is inaudible to human. 
The masking threshold of a masker signal can be approximated using the psychoacoustic model~\cite{audio-watermark}. 

%% file: overview.tex
\section{Overview of \toolname}\label{sec:overview} 

\subsection{Objective and Design} 
Our goal is to protect the rights of songs by mitigating SVC-based song cover (\emph{prevention}) {no matter songs are used as the source and/or target singing voices in SVC}, but
without impacting the release, spread and enjoyment of songs (\emph{harmlessness}). 
These two objectives 
are achieved by \toolname. 

\begin{figure}[t]
    \centering
    \includegraphics[width=0.46\textwidth]{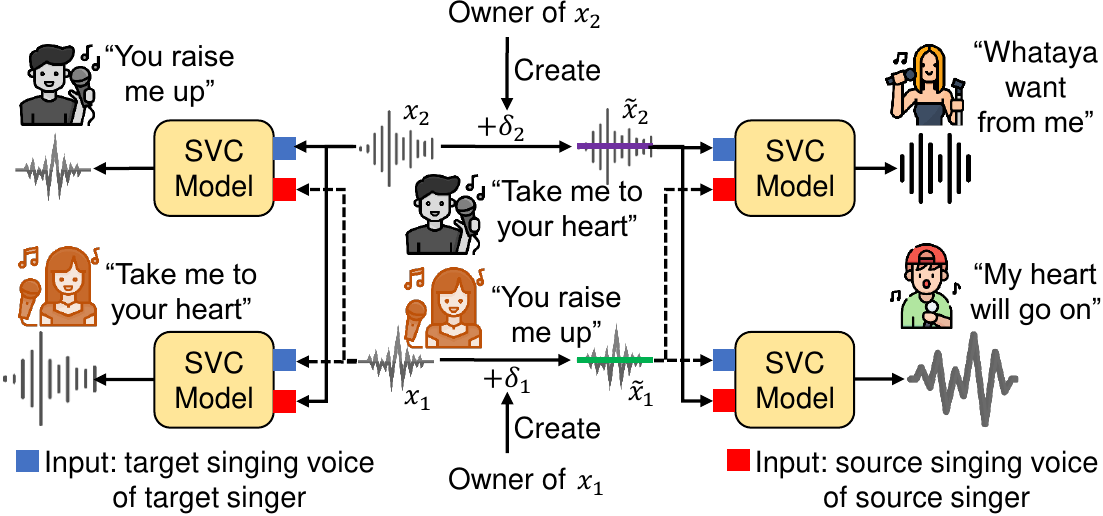}
    \caption{Overview of \toolname. Song owners apply \toolname to singing voices ($x_1$ and $x_2$) and obtain the protected counterparts  ($\tilde{x}_1$ and $\tilde{x}_2$) to prevent them from being used as source or target singing voices (dual prevention), by disrupting lyrics and singer identity in SVC-covered singing voices, respectively. 
    }
    \label{fig:overview-framework}
\end{figure}

\figurename~\ref{fig:overview-framework} depicts 
the overview of \toolname, where the left part shows the workflow of SVC without \toolname while 
the right part shows that song owners
create protected counterparts 
by adding perturbations with \toolname. When protected ones are used as the source and/or target singing voices in SVC,
the conversion fails to produce the expected one, achieving the prevention objective, while the  perturbations are inaudible by audiences,
achieving the harmlessness objective.

It is unknown to song owners
in advance if a singing voice will be used as the source or target singing voice in SVC, 
thus 
\toolname is designed to feature a dual prevention
by causing the following two disruptions in SVC-covered singing voices.

\begin{itemize}[leftmargin=*]
    \item {\bf Identity disruption}. 
    To prevent a singing voice from being used as the target singing voice in SVC, \toolname crafts the perturbation on the identity encoder so that the SVC-covered singing voice sounds unlike being covered by the target singer, 
    protecting both the performing and civil rights of the target singer. 
  \item {\bf Lyric disruption}. 
   To prevent a singing voice from being used as the source singing voice in SVC, 
\toolname crafts the perturbation on the lyric encoder so that the SVC-covered singing voice contains unclear and even distinct lyrics from the expected one, protecting the copyrights of the lyrics in the source singing voice.
\end{itemize}

\toolname directly protects the civil rights of singers and the copyrights of lyrics in a straightforward manner, while the copyrights of melodies and the copyrights to reproduce and distribute songs are indirectly protected by \toolname, as \toolname worsens the performance of singing voice conversion 
and thus discourages the release, distribution and spread of SVC-covered songs, and the usage of SVC. We will discuss possible solutions 
to directly protect more rights in \cref{sec:discussion}.  

We note that one may only want to directly protect the civil rights of singers (resp. copyrights of lyrics),
for which it suffices to prevent the singing voice from being used as target (resp. source) singing voice in SVC.
Thus, \toolname is designed to be configurable to provide a sole or dual prevention, by causing one of two disruptions or both.

\subsection{Threat Model}\label{sec:threa_model}
We discuss the threat model of the adversary and defender,
where the adversary can be neutral or malicious SVC users. 

\noindent {\bf Adversary's purpose.}
Neutral users use SVC for entertainment purposes, e.g.,
fans of a singer hope the singer covers some songs,  
and music enthusiasts who greatly admire the lyrics and melody of a song 
wish for it to be covered and spread widely.  
Malicious users gain improper benefits such as financial gain via SVC. 
For example, a company might use SVC to release records sung by a target singer, 
competing with the original record company. 
They may also use SVC to create singing voices with sensitive contents, 
for product promotion, advocacy, and so on. 
Both neutral and malicious users can cause right infringements, regardless of their purposes. 
 
\noindent {\bf Adversary's capacity.} 
{(1) {\it Singing voices:} We assume that the adversary can collect a few target singing voices
and a source singing voice, 
e.g., downloading or recording songs available on music platforms 
and then easily extracting singing voices from the songs. 
(2) {\it SVC models:} 
We assume that the adversary has access to a \emph{few-shot} SVC model, 
which requires much fewer resources (computation and the target singer's singing voices) 
than a non-few-shot model (cf.~\Cref{sec:background_svc}).  
This makes it accessible to a broader range of adversaries, producing more illegally covered songs. 
Preventing such adversaries is thus more urgent and expands the application of \toolname. 
}

\noindent {\bf Adversary's knowledge.} 
The adversary may be unaware of prevention or has complete knowledge of \toolname (cf.~\cref{sec:robustness}) under which the adversary may adopt adaptive strategies to bypass the prevention. 

\noindent {\bf Defender.}   
{(1) {\it Subject:} The song owners are the defenders. The composer, lyricist, and singer can be the defenders as well, e.g., when they are the same person such as online singers. 
(2) {\it Purpose:} By applying \toolname to their original clean singing voices, defenders can prevent their songs from being used as both target and source songs by disrupting the identity and/or lyrics of SVC-covered singing voices, 
protecting the civil rights of singers and/or the copyrights of songs. 
(3) {\it Knowledge of SVC models:} We first assume that the defender knows the identity and the lyric encoders of the SVC model adopted by the adversary. Later, we will relax this assumption in \cref{sec:exper_transfer}. 
(4) {\it Knowledge of target singers:} 
When applying \toolname to prevent a song from being used as a target song (target singer) by SVC, 
the target singer is the singer of the song, thus known to the defender.}

{\subsection{Practicality of \toolname}}
\noindent {\bf Platform difference.} 
{\toolname is applied by song owners before song release so that the same protected songs  
can be distributed across various platforms. 
Despite variations in storage or transmission methods (e.g., compression), 
\toolname is robust against transformations like compression (cf.~\cref{sec:robust_adaptive}).}

\noindent {\bf Unprotected songs.} 
{While adversaries may utilize songs covered by individuals (including the adversaries themselves) with excellent vocal skills as source singing voices, 
they still need songs sung by target singers as target singing voices to replicate their singing styles. 
Thus, at least the target singing voices are controlled by defenders and protected by \toolname, 
which causes at least identity disruption in SVC-covered singing voices. We consider this case in \Cref{sec:individual_effectiveness}.}
{In addition, though adversaries may have copies of some songs released prior to \toolname and thus they are not protected, \toolname remains effective for identity disruption even when the ratio of the protected target singing voices is small (cf.~\Cref{sec:impact_ratio}).}
{To enhance the practicality of \toolname in real-world usage and better protect copyrights and civil rights, non-technical strategies also can be adopted (cf. \cref{sec:discussion} for more discussions).}

%% file: methodology.tex
\section{Methodology of \toolname}\label{sec:methodology}
\begin{figure}
    \centering
    \includegraphics[width=0.5\textwidth]{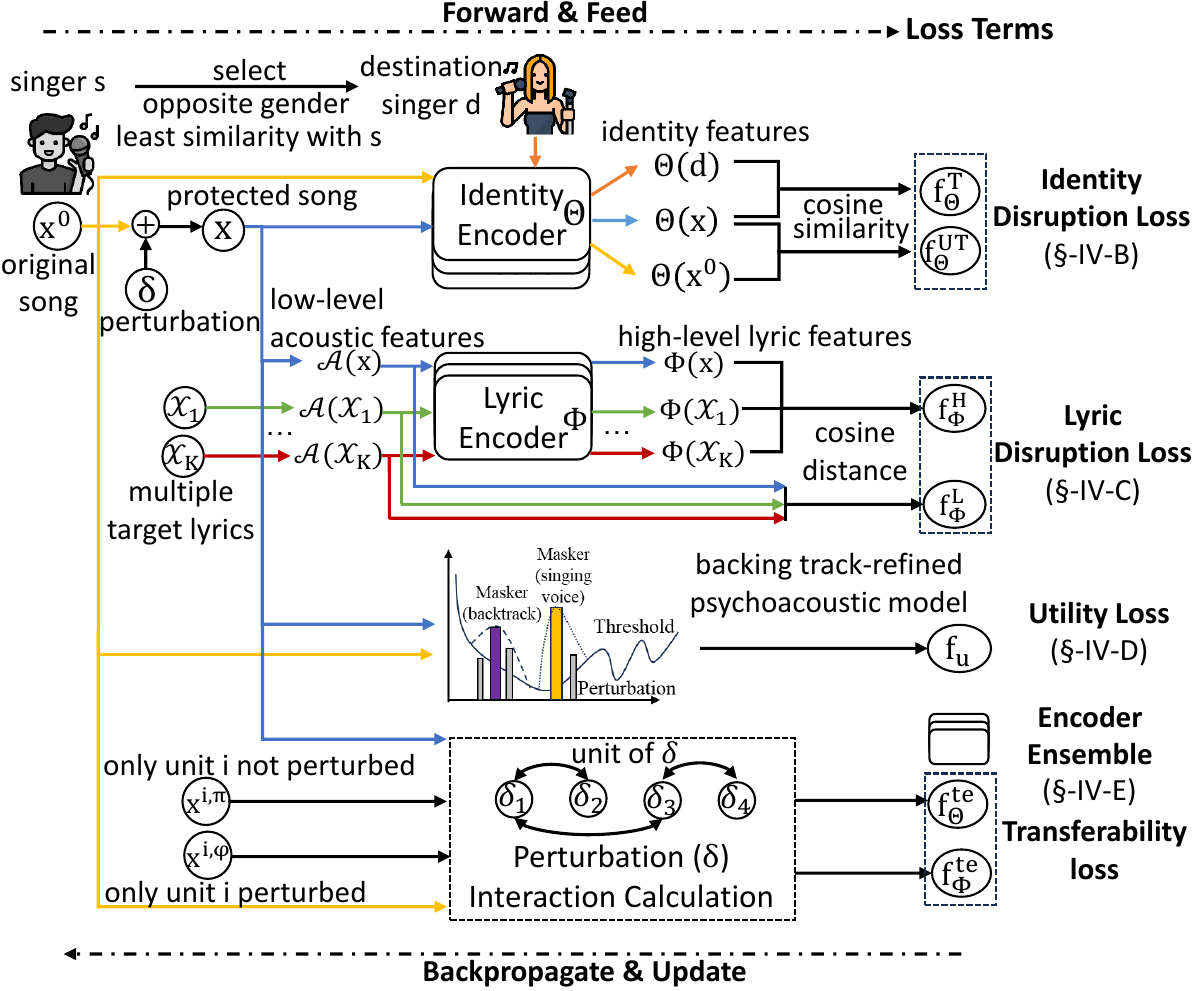}
    \caption{{Overview of the methodology of \toolname}}
    \label{fig:method_overview}
\end{figure}

{
\figurename~\ref{fig:method_overview} presents 
overview of our methodology of \toolname. 
We first formulate the optimization problem of crafting protected songs and then detail the loss functions designed for identity disruption, lyric disruption, utility, and transferability.}

\subsection{Problem Formulation}
Given a singing voice 
{$x^0\in \mathbb{R}^{1\times D}$ represented by a waveform with length $D$,} 
the identity encoder $\Theta$, and the lyric encoder $\Phi$, 
we attempt to craft a (protected) singing voice $x$ to disrupt the identity and lyrics of SVC-covered singing voices 
while preserving the quality of songs.   
Formally, we need to solve the following optimization problem: 
\begin{center}
$\min\limits_{x}\left(
\begin{array}{cc}
   f_{\Theta}(x,{x^0}) + f_{\Phi}(x) + 
   \lambda_u f_{u}(x,{x^0}) \\
    + \lambda_{\Theta}^{te} f_{\Theta}^{te}(x,{x^0}) +  \lambda_{\Phi}^{te} f_{\Phi}^{te}(x,{x^0})
\end{array}\right)$\\
$\text{subject to } x\in [-1, 1]$
\end{center}
where $f_{\Theta}$ and $f_{\Phi}$ are prevention losses for the identity and lyric disruptions; $f_{u}$ is the utility loss for harmlessness (i.e., song quality);  $f_{\Theta}^{te}$ and $f_{\Phi}^{te}$ are the transferability enhancement loss for the identity and lyric disruptions. The positive factors  
$\lambda_{u}$, $\lambda_{\Theta}^{te}$, and $\lambda_{\Phi}^{te}$ are used to control the impact of these losses on the perturbation ($x-x^0$).    
{Following~\cite{V-Cloak, FakeBob, FenceSitter, QFA2SR, AS2T, SpeakerGuard}, we normalize singing voices by dividing their magnitudes by the maximum value of the bit-width, to avoid causing potential overflow 
during optimization, so valid singing voices subject to $[-1, 1]$.}

\subsection{{Identity Disruption: Gender-Transformation Loss}}\label{sec:idloss}  
\smallskip \noindent {\bf Untargeted loss $f_\Theta^{\tt UT}$.}
{We attempt to cause SVC-covered singing voices to {\it \uline{sound unlike the singer of $x^0$}} (i.e., the target singer when the protected singing voice $x$ is used as the target singing voice by SVC). Since the identity information for SVC is provided by identity features, we achieve this purpose by ensuring that {\it \uline{the identity feature $\Theta(x)$ of the protected singing voice $x$ deviates from the original identity feature $\Theta(x^0)$}}. Formally, we minimize the following loss that quantifies the similarity between identity features $\Theta(x)$ and $\Theta(x^0)$:}
\[f_{\Theta}^{\tt UT}(x,{x^0}) = {\tt Sim}(\Theta(x), \Theta(x^0)) \]
where ${\tt UT}$ denotes untargeted and ${\tt Sim}(\cdot)$ is the similarity function.
In this work, we use the cosine similarity~\cite{COSS} {due to its bounded nature 
which results in bounded losses and therefore more stable optimization~\cite{bounded_loss_3, bounded_loss_1, bounded_loss_2}.}

\smallskip \noindent {\bf Targeted loss $f_\Theta^{\tt T}$.}
{
Humans can better perceive the vocal difference between opposite-gender singers than between same-gender singers. 
Hence, to enhance identity disruption, 
we 
cause SVC-covered singing voices to {\it \uline{sound like being covered by a singer with the {opposite} gender}} (called \emph{destination} singer) from the original singer (i.e., the singer of $x^0$).}

{
We design 
the following process to choose the {destination} singer
that {\it \uline{has the opposite gender and is the most objectively identity-dissimilar from the original singer}}. 
Firstly, we collect a set of auxiliary singers 
with the \emph{opposite} gender, each of which has a set of singing voices $\mathcal{V}_i$. 
Then we represent each auxiliary singer $i$ by the centroid identity feature 
$\Theta^c_{a, i}=\frac{1}{|\mathcal{V}_{i}|}\sum_{v\in \mathcal{V}_{i}}\Theta(v)$
and the original singer by $\Theta^c=\frac{1}{N}\sum_{i=1}^{N}\Theta(x^0_i)$ 
where $N$ is the number of singing voices. 
Finally, the destination singer is chosen as the auxiliary singer whose centroid identity feature $\Theta^c_{a, i}$ 
is the farthest from $\Theta^c$, i.e., $k=\argmin_{i}{\tt Sim}(\Theta^c_{a,i}, \Theta^c)$. 
Let $\Theta_{\tt des}^{c}$ denote $\Theta^c_{a,k}$.}
{
This process is characterized by 
(1) comprising both subjective perception (opposite-gender) and objective perception (least identity-similar) and (2) precise singer representation by the centroid embedding rather than a voice embedding.  
The defender can select auxiliary singers from open-source datasets without infringing their civil rights.}

{With the {destination} singer, we ensure that {\it \uline{the identity feature $\Theta(x)$ of the protected singing voice $x$ approaches that of the destination singer}} by minimizing the following loss: }
\[f_{\Theta}^{\tt T}(x) = - {\tt Sim}(\Theta(x), \Theta_{\tt des}^{c})\]  
where ${\tt T}$ denotes targeted.

\smallskip \noindent {\bf Final loss.} 
{Putting $f_{\Theta}^{\tt UT}$ and $f_{\Theta}^{\tt T}$ together, our final loss for identity disruption, called gender-transformation loss, is defined as:}
\[f_{\Theta}(x,{x^0}) = f_{\Theta}^{\tt UT}(x,{x^0})  + \lambda_{\Theta} f_{\Theta}^{\tt T}(x)\]
where $\lambda_{\Theta}>0$ is the loss balancing factor. 

We emphasize that {the two loss terms}, the selection of the least identity-similar and opposite-gender destination singer, and the representation of the destination singer by centroid embeddings, all contribute to identity disruption (cf.~\Cref{sec:exper_gt}).

\subsection{{Lyric Disruption: High/Low Hierarchy Multi-Target Loss}}\label{sec:lyric_disrupt_loss}

\smallskip \noindent {\bf High hierarchy loss.}
{Since lyric features provide the lyric information for SVC, 
to achieve lyric disruption, we ensure that the lyric feature $\Phi(x)$ of the protected singing voice $x$ differs from the original lyric feature $\Phi(x^0)$. 
Prior work~\cite{AdvDDoS} on adversarial attacks against speech-to-text tasks has shown that targeted attacks are more transferable than untargeted attacks regarding mistranscription. 
Inspired by this, we {\it \uline{choose a singing voice $\chi$ with different lyrics from $x^0$ and pull together the lyric feature $\Phi(x)$ and the lyric feature $\Phi(\chi)$ of $\chi$}} by minimizing the following designated loss:}
\[ f_{\Phi}^{H}(x) = {\tt Dist}(\Phi(x), \Phi(\chi)) \] 
{where ${\tt Dist}(\cdot)$ is the distance function, initialized by the cosine distance (i.e., $1 \text{ minus cosine similarity}$) in this work due to the same reason as in \cref{sec:idloss}.}

\smallskip \noindent {\bf Low hierarchy loss.}
{The loss $f_{\Phi}^{H}(x)$ only minimizes the distance between high-level lyric features $\Phi(x)$ and $\Phi(\chi)$. 
However, mainstream SVC models also rely on low-level acoustic features~\cite{acoustic-feature-li}, including handcrafted ones (e.g., filter Bank~\cite{FilterBanks}) 
and representations produced by shallow hidden layers (e.g., Hubert-based features~~\cite{HuBERT}),   
which are extracted from voice waveform and used to derive lyric features. 
Inspired by this, 
we hypothesize that {\it \uline{aligning 
the low-level acoustic features $\mathcal{A}(x)$ with $\mathcal{A}(\chi)$ 
can improve the alignment of high-level lyric features}}, thus enhancing the lyric disruption.}
{
Therefore, we define the following low hierarchy lyric disruption loss:
}
\[ f_{\Phi}^{L}(x) = {\tt Dist}(\mathcal{A}(x), \mathcal{A}(\chi)).\]

\smallskip \noindent {\bf Multiple targets.}
{Both $f_{\Phi}^{H}(x)$ and $f_{\Phi}^{L}(x)$ only utilize a single singing voice $\chi$ to provide target lyrics.  
{\it \uline{Due to the phoneme difference among different singing voices, given the protected singing voices $x$, the difficulty of optimizing $f_{\Phi}^{H}(x)$ and $f_{\Phi}^{L}(x)$ may vary with $\chi$.}} 
To tackle this issue, we propose to {\it \uline{enhance the effect of lyric disruption with multiple target lyrics}} by adapting $f_{\Phi}^{H}(x)$ and $f_{\Phi}^{L}(x)$ as follows:
\[\begin{array}{l}
     f_{\Phi}^{H}(x) = \frac{1}{K}\sum_{k=1}^{K}{\tt Dist}(\Phi(x), \Phi(\chi_k)) \\[1pt]
    f_{\Phi}^{L}(x) = \frac{1}{K}\sum_{k=1}^{K}{\tt Dist}(\mathcal{A}(x), \mathcal{A}(\chi_k))
\end{array}\]
where $\chi_1,\cdots,\chi_K$ are singing voices with distinct lyrics. 
}

\smallskip \noindent {\bf Final loss.}
{
Our final loss for lyric disruption, called high/low hierarchy multi-target loss, is formulated as follows:
\[
    f_{\Phi}(x)=\lambda_{\Phi}^{H}f_{\Phi}^{H}(x) + \lambda_{\Phi}^{L}f_{\Phi}^{L}(x)
\]
where $\lambda_{\Phi}^{H}$ and $\lambda_{\Phi}^{L}$ are positive balancing factors.
}

\subsection{{Utility: Backing Track-Refined Simultaneous Masking Loss}}\label{sec:utilityloss} 

\smallskip
\noindent {\bf Basic loss.} 
Since the original singing voice $x^0$ and the perturbation simultaneously occur   
when the singing voice $x$ is played, 
{the perturbation can be hidden with simultaneous masking.} 
Specifically, we {{\it \uline{treat $x^0$ as the masker and make the perturbation (maskee) inaudible  
by forcing it to fall under the masking threshold of the masker.}}}
Let $\theta_{a}\in \mathbb{R}^{T\times F}$ denote the masking threshold of the audio $a$ 
where $T$ is the number of frames (audio's short segments) and $F$ is the number of frequencies. 
Let $p_a\in \mathbb{R}^{T\times F}$ denote the log-magnitude power spectral density of the audio $a$.  
Formally, we minimize the following utility loss based on simultaneous masking:
\begin{center}
   $f_{u}(x,{x^0}) = \frac{1}{T\cdot F}\sum_{t=1}^T\sum_{k=1}^F\max\{0, p_{x-x^0}(t, k)-\theta_{x^0}(t, k)\}.$ 
\end{center}

\smallskip
\noindent {\bf Refined loss.} 
{\it \uline{A singing voice is typically accompanied by a backing track $\mathcal{M}$ in a different channel of a song.}}
Thus, we propose to {\it \uline{utilize the backing track as an additional masker}} to improve harmlessness: {\it \uline{the perturbation will not be {audible} as long as it is under one of the masking thresholds}} of the singing voice and the backing track.
The backing track-refined simultaneous masking utility loss is defined as: 
{\small
\begin{center}
   $    f_{u}(x,{x^0}) = \frac{1}{T\cdot F}\sum_{t=1}^T\sum_{k=1}^F\max\{0, p_{x-x^0}(t, k)-\theta_{x^0,\mathcal{M}}(t, k)\} $ 
\end{center}
}
    where $\theta_{x^0,\mathcal{M}}(t, k) = \max\{\theta_{x^0}(t, k), \theta_{\mathcal{M}}(t, k)\}$ is the joint masking threshold of the two maskers. 
Intuitively, minimizing the loss $f_{u}$ 
minimizes the density of the perturbation for each frame and frequency 
until it is no greater than one of the masking thresholds of the singing voice and the backing track. 

Remark that the refined loss $f_{u}$ is {tailored for songs} since it utilizes the unique backing tracks of songs that do not exist for ordinary speech voices. 
It adopts a simplified joint psychoacoustic model for the singing voice and backing track, as the modeling of simultaneous masking of multiple-channel signals is a very complex task~\cite{stuart1996psychoacoustics}. 
Despite being simplified, it can effectively improve harmlessness in our experiments (cf.~\Cref{sec:imper_exp}). More precise modeling is left as future work.

\subsection{{Transferability Enhancement}}\label{sec:tr-enloss}
\noindent {\bf Frame-level interaction reduction-based (FL-IR) loss $f_{\Theta}^{te}$\&$f_{\Phi}^{te}$.} 
Inspired by the {{\it \uline{negative correlation between transferability and interaction inside perturbations}}} (cf.~\cref{sec:adver_example}), we define $f_{\Theta}^{te}$ and $f_{\Phi}^{te}$ 
to enhance transferability by reducing interaction: 
\[\small
\begin{array}{l}
  f_{\Theta}^{te}(x,{x^0}) = \mathbb{E}_{i}
    \Big(
      f_\Theta^{\tt UT} (x, {x^0}) 
    + 1
    - f_\Theta^{\tt UT} (x^{i,\pi}, {x^0}) 
    - f_\Theta^{\tt UT} (x^{i,\varphi}, {x^0}) 
    \Big) \\[1pt]
    f_{\Phi}^{te}(x,{x^0}) = \mathbb{E}_{i}
    \Big(
    f_\Phi^H (x) 
    + f_\Phi^H (x^0) 
    - f_\Phi^H (x^{i,\pi}) 
    - f_\Phi^H (x^{i,\varphi}) 
    \Big) 
\end{array}\]
where {$1$ is $f_\Theta^{\tt UT} (x^0, {x^0})$ in defining $f_{\Theta}^{te}(x,{x^0})$}, 
$x^{i,\pi}$ is identical to $x$ except that its $i$-th unit is not perturbed and $x^{i,\varphi}$ is identical to $x^0$ except that its $i$-th unit is perturbed as $x$.

The computation of the FL-IR 
loss $f_{\Theta}^{te}(x)$ and $f_{\Phi}^{te}(x)$ involves iterating over all the sample points within a singing voice, which however, may contain numerous sample points due to the high sampling rate (e.g., 48KHz)~\cite{QFA2SR}, leading to costly and even intractable computational overhead. Observing that singing voices are split into multiple short fragments  (called frames) before being fed to SVC models, 
we address this challenge by calculating the losses at the frame level. Specifically, given the frame length $w_l$ and the frame shift $w_s$, we first decide the boundaries of each frame. The boundaries of the $i$-th frame are $i\times w_s$ and $i\times w_s + w_l$. We treat each frame as a whole, that is, all points within a frame are simultaneously perturbed or not perturbed. Then we compute the FL-IR loss by iterating over the frames instead of all the sample points, where $x^{i,\pi}$ becomes identical to $x$ except that all sample points within its $i$-th frame are not perturbed and $x^{i,\varphi}$ becomes identical to $x^0$ except that all sample points within its $i$-th frame are perturbed as $x$. We also approximate the expectation $\mathbb{E}$ by $R$ times random sampling to further reduce the overhead~\cite{IRA_attack}. 
 
\noindent {\bf Encoder ensemble}.
It is known that model ensemble can enhance transferability~\cite{QFA2SR, AntiFake, LCLS17}. 
Thus, we collect various identity/lyric encoders on which average losses $f_{\Theta}(x)$, $f_{\Phi}(x)$, $f_{\Theta}^{te}(x)$ and $f_{\Phi}^{te}(x)$ are computed.
{{\it \uline{It improves transferability from a different perspective than the FL-IR loss.}}}
The FL-IR loss reduces the interaction among  perturbation units which has a negative correlation with transferability, 
while encoder ensemble enforces that the features of the singing voice $x$ deviate enough from that of the original one $x_0$. 
Our experimental results confirm that both methods can improve transferability, and
their combination yields the best transferability 
(cf.~\cref{sec:exper_transfer}).

\subsection{Final Approach}\label{sec:final_app}
Finally, we solve the following optimization problem: 
\begin{center}
$\min\limits_{x}\left(
\begin{array}{cc}
   f_{\Theta}^{\tt UT}(x,{x^0}) + \lambda_{\Theta}f_{\Theta}^{\tt T}(x) + \lambda_{\Phi}^{H} f_{\Phi}^{H}(x) +   \lambda_{\Phi}^{L} f_{\Phi}^{L}(x) \\ 
   +  \lambda_u f_{u}(x,{x^0})+  \lambda_{\Theta}^{te} f_{\Theta}^{te}(x,{x^0}) +  \lambda_{\Phi}^{te} f_{\Phi}^{te}(x,{x^0}) 
\end{array}\right)$\\
$\text{subject to } x\in [-1, 1].$
\end{center}

Instead of manually setting the balance factors 
$\lambda_\Theta$, 
$\lambda_\Phi^{H}$, $\lambda_\Phi^{L}$,
$\lambda_u$, 
$\lambda_\Theta^{te}$, and $\lambda_{\Phi}^{te}$, 
we utilize automatic and dynamic loss balance 
by loss normalization~\cite{QFA2SR}, 
due to its advantage of nearly equally weighing different loss functions with different ranges and scales. 
Specifically, at each iteration of crafting the singing voice $x$, we normalize each loss $f_k$ by its mean $\mu_k$ and 
variance 
$\sigma_k$, i.e.,  $f'_k=\frac{1}{\sqrt{\sigma_k}}(f_k-\mu_k)$.
Both $\mu_k$ and $\sigma_k$ are loss-specific 
and iteratively updated via $\mu_k=\mu_k+\frac{1}{n}({f_k-\mu_k})$ and $\sigma_k=\sigma_k+\frac{1}{n}((f_k-\mu_k)^2-\sigma_k)$, 
where $n$ is the current iteration. 
Finally, the total loss function is defined as the sum of the normalized losses. 

 \begin{figure}[htp]\small\removelatexerror
\begin{algorithm*}[H]
      \caption{\toolname}
      \label{al:fianl_approach}
      \KwIn{original singing voice $x^0$; number of steps $N$; learning rate $\alpha$; identity encoder $\Theta$; lyric encoder $\Phi$; ${\tt protect\_target}$; ${\tt protect\_source}$; ${\tt transfer\_identity}$; ${\tt transfer\_lyric}$}
      \KwOut{singing voice $x$}
      $\text{Adam } \gets \text{ initialize Adam optimizer with } \alpha$\; 
      $K\gets 7$; $F\gets [f_\Theta^{\tt UT}, f_\Theta^{\tt T}, f_\Phi^{H}, f_\Phi^{L}, f_u, f_\Theta^{te}, f_\Phi^{te}]$\;
      \lFor{$k$ from $1$ to $K$} {$\mu_k\gets 0$; $\sigma_k\gets 1$}
      \For{$n$ from $1$ to $N$}{ \label{al_line:fianl_approach_2nd_loop}
          $f_{\tt total}\gets 0$\;
          \For{$k$ from $1$ to $K$}{ \label{al_line:fianl_approach_3rd_loop}
              $f\gets F_k$\; 
              \lIf{$f \in \{f_\Theta^{\tt UT},f_\Theta^{\tt T}\}\wedge{\tt protect\_target}={\tt False}$}{ \label{al_line:fianl_approach_protect_target} 
              \Continue
              } 
              \lIf{$f \in \{f_{\Phi}^{H},f_{\Phi}^{L}\}\wedge{\tt protect\_source}={\tt False}$}
              {\label{al_line:fianl_approach_protect_source} 
              \Continue
              }
              \lIf{$f =f_{\Theta}^{te}\wedge{\tt transfer\_identity}={\tt False}$}{ \label{al_line:fianl_approach_transfer_identity} 
              \Continue
              }
              \lIf{$f=f_{\Phi}^{te}\wedge{\tt transfer\_lyric}={\tt False}$}{ 
              \Continue \label{al_line:fianl_approach_transfer_lyric} 
              }
              $f_k\gets f(x^{n-1}, {x^0})$; 
              $\mu_k\gets \mu_k+\frac{f_k-\mu_k}{n}$\; \label{al_line:fianl_approach_compute_loss_1} 
              $\sigma_k\gets \sigma_k+\frac{1}{n}((f_k-\mu_k)^2-\sigma_k)$; $f_k\gets \frac{f_k-\mu_k}{\sqrt{\sigma_k}}$\; \label{al_line:fianl_approach_compute_loss_2} 
              $f_{\tt total} \gets f_{\tt total} + f_k$\; \label{al_line:fianl_approach_tol_loss} 
          } \label{al_line:fianl_approach_3rd_loop_end}
        $x^n \gets \text{Adam}(x^{n-1}, \nabla_{x^{n-1}}f_{\tt total})$\; \label{al_line:fianl_approach_update}
          $x^n \gets \max\{\min\{x^n,1\},-1\}$\; \label{al_line:fianl_approach_clip}
      } \label{al_line:fianl_approach_2nd_loop_end}
      \Return{$x^{N}$}
  \end{algorithm*}
\end{figure}

We minimize the loss by $N$-iteration gradient descent using the Adam optimizer (with learning rate $\alpha$). The overall algorithm 
is shown in Alg.~\ref{al:fianl_approach}. 
In each iteration (Lines~\ref{al_line:fianl_approach_2nd_loop}--\ref{al_line:fianl_approach_2nd_loop_end}), we iteratively (Lines~\ref{al_line:fianl_approach_3rd_loop}--~\ref{al_line:fianl_approach_3rd_loop_end}) compute each $f_k$ of 7 losses and normalize it using its mean $\mu_k$ and 
variance 
$\sigma_k$ (Lines~\ref{al_line:fianl_approach_compute_loss_1}--\ref{al_line:fianl_approach_compute_loss_2}). 
Remark that if encoder ensemble is enabled, $f_k$ is the average over all encoders.
We then compute the total loss $f_{\tt total}$ by summing the 7 normalized losses (Line~\ref{al_line:fianl_approach_tol_loss}), update the singing voice using the Adam optimizer and the gradient w.r.t. the total loss (Line~\ref{al_line:fianl_approach_update}), and clip it to be a valid singing voice (Line~\ref{al_line:fianl_approach_clip}). 
To be flexible, we provide the following flags:
${\tt protect\_target}$, ${\tt protect\_source}$, ${\tt transfer\_identity}$, and ${\tt transfer\_lyric}$. 
If the defender does not prevent singing voices from being used as target (resp. source) singing voices,  ${\tt protect\_target}$ (resp. ${\tt protect\_source}$) can be {\tt False}. Similarly, if the defender has access to the identity (resp. lyric) encoder of the SVC, ${\tt transfer\_identity}$ (resp. ${\tt transfer\_lyric}$) can be ${\tt False}$. When a flag is {\tt False}, \toolname will ignore 
the respective loss 
(Lines~\ref{al_line:fianl_approach_protect_target}-\ref{al_line:fianl_approach_transfer_lyric}).  
We also provide convergence analysis in \Cref{sec:convergence_analysis}.

Remark that except for the refined utility loss that utilizes unique elements of songs, the others could generalize to other domains. 
In particular, the loss $f_{\Theta}$ 
for identity disruption, and the FL-IR loss and encoder ensemble for complementarily enhancing transferability could be exploited to prevent ordinary speech voice conversion/synthesis to protect speaker identity. 
The high/low hierarchy multi-target loss could be used to enhance the generation of adversarial speech examples against speech-to-text models for malicious purposes. 
Also, the FL-IR loss could be utilized by attackers to strengthen transfer-based adversarial attacks against speech processing systems~\cite{QFA2SR}. 
We leave these as interesting future works.

%% file: evaluation.tex
\section{Evaluation}\label{sec:evaluate}

\subsection{Experimental Setup}\label{sec:steup}
\noindent {\bf Models.} 
{We adopt four recent promising SVC models with few-shot conversion capability: 
Lora-SVC~\cite{Lora-SVC}, Vits-SVC~\cite{Vits-SVC}, Grad-SVC~\cite{Grad-SVC}, and NeuCo-SVC~\cite{NeuCoSVC}. We also consider one non-few-shot SVC model StarGANv2-SVC~\cite{StarGANv2_VC}. As shown in \tablename~\ref{tab:detail_svc_model}, they are diverse in information disentanglement method, identity, lyric, and pitch encoders, the decoder, and the sampling rate of singing voices. Since
we target few-shot SVC models (cf.~\Cref{sec:threa_model}),
this section only considers 4 few-shot models
while StarGANv2-SVC is evaluated in \Cref{sec:results_non_few_shot}.}

To evaluate the transferability of \toolname, 
we consider another 8 distinct identity encoders: X-vectors (XV)~\cite{X-Vector}, 
ECAPA-TDNN (ECAPA)~\cite{ECAPA-TDNN}, ResNet18 for identification (Res18-I)~\cite{resnet18, Autospeech_github}, ResNet34 for identification (Res34-I)~\cite{resnet34, Autospeech_github}, ResNet34 for verification (Res34-V)~\cite{resnet34, Autospeech_github}, AutoSpeech (Auto)~\cite{auto-speech}, 
ResNetSE34V2 (Res-SE)~\cite{he2016deep}, and VGGVox-40 (VGG)~\cite{nagrani2017voxceleb};
and 
another 5 distinct lyric encoders: Whisper-Tiny~\cite{Whisper}, Whisper-Base~\cite{Whisper}, Whisper-Small~\cite{Whisper}, Wav2vec2~\cite{wav2vec2}, and Decoar2~\cite{decoar2}. 
These 13 encoders are different from those used in all SVC models. 

\begin{table}[t]
    \centering\setlength\tabcolsep{1pt} 
    \caption{{Details of singing voice conversion (SVC) models.}}
     \resizebox{0.49\textwidth}{!}{
    \begin{threeparttable}
    \begin{tabular}{c|c|c|c|c|c|c|c}
    \hline
          {\bf Dis.$^\ddag$} & {\bf Model} & \makecell[c]{{\bf Few-} \\ {\bf Shot?}} & \makecell[c]{{\bf Identity} \\ {\bf Encoder}} & \makecell[c]{{\bf Lyric} \\ {\bf Encoder}} & \makecell[c]{{\bf Pitch} \\ {\bf Encoder}} & \multicolumn{1}{c|}{\makecell[c]{{\bf Decoder$^\natural$} \\ \boldmath{}{($G$ / $G+V$)}\unboldmath{}}} & \makecell[c]{{\bf Sample} \\ {\bf Rate}} \\ \hline
         \multirow{3}[8]{*}{\rotatebox{90}{\bf Explict}} & \makecell[c]{{\bf Lora} \\ {\bf -SVC}} &  \cmark & LSTM~\cite{GE2E} & \makecell[c]{Whisper\\ -Medium~\cite{Whisper}} &  WORLD~\cite{WORLD} & \multicolumn{1}{c|}{BigVGAN$^\sharp$~\cite{BigVGAN}} & 16kHz \\ \cline{2-8}
        & \makecell[c]{{\bf Vits} \\ {\bf -SVC}} & \cmark & LSTM~\cite{GE2E} & \makecell[c]{Whisper\\-Large~\cite{Whisper} \\ \& Hubert~\cite{HuBERT}} &  Crepe~\cite{Crepe} & \multicolumn{1}{c|}{BigVGAN$^\sharp$~\cite{BigVGAN}} & 32kHz \\ \cline{2-8}
        & \makecell[c]{{\bf Grad} \\ {\bf -SVC}} & \cmark & LSTM~\cite{GE2E} & Hubert~\cite{HuBERT} & Praat~\cite{praat} & \multicolumn{1}{c|}{\makecell[c]{Diffusion~\cite{Grad-TTS}\\+BigVGAN$^\sharp$~\cite{BigVGAN}}} & 32kHz\\ 
    \hline
      \multirow{1}{*}{\rotatebox{90}{\bf Implicit}} & \makecell[c]{{\bf NeuCo} \\ {\bf -SVC}} & \cmark & \makecell[c]{WavLM \\ -Large~\cite{WavLM}} & \makecell[c]{WavLM \\ -Large~\cite{WavLM}} & \makecell[c]{pYIN~\cite{PYIN} \&
     \\ REAPER~\cite{reaper}} & \multicolumn{1}{c|}{\makecell[c]{FiLM UNet~\cite{FiLM_Unet}}} & 24kHz \\ \cline{2-8}
      & \makecell[c]{{\bf StarGANv2} \\ {\bf -SVC}} & \xmark & \makecell[c]{Style \\ Encoder~\cite{StarGANv2_VC}} & \makecell[c]{VGG \\ -BLSTM~\cite{VGG_BLSTM}} & \makecell[c]{JDCNet~\cite{JDCNet}} & \multicolumn{1}{c|}{\makecell[c]{StarGANv2~\cite{starganv2}\\+ParallelWaveGAN~\cite{ParallelWavegan}}} & 24kHz \\ 
     \hline
    \end{tabular}
    \begin{tablenotes}
    \item (1) 
    {\small $\ddag$: Dis. is short for information disentanglement. 
    (2) $\natural$: Decoder may directly produce waveforms with a generator ($G$) or first use $G$ to produce acoustic features and utilize vocoders $V$ to synthesize waveforms ($G+V$). 
     (3) $\sharp$: Their specific architectures and parameters are different.}
    \end{tablenotes}
    \end{threeparttable}
    }
    \label{tab:detail_svc_model}
\end{table}

\noindent {\bf Datasets.} 
We use two datasets: OpenSinger~\cite{OpenSinger} and  NUS-48E~\cite{NUS-48E}, 
{whose attributes are shown in \tablename~\ref{tab:dataset_attirbute}.}

{We select target singers, target and source singing voices as follows. 
Let $m$ denote the number of singers in a dataset. 
Firstly, we regard each singer as target singer and randomly select $t$ singing voices sung by the singer as target singing voices and randomly select $s$ singing voices from other singers as the source singing voices, leading to $p=m\times s$ pairs of target singer and source singing voice. 
Then, we run the SVC model and choose 2,000 pairs out of $p$ pairs with top identity similarity, 
following the practice of previous works~\cite{Attack-VC, AntiFake}.  
The rationale behind this selection is that the SVC model performs better on these selected pairs, 
thus they are more necessary to be protected than the others. 
For OpenSinger, 
$m$=76, $t$=10, $s$=100, $p$=7,600. 
For NUS-48E with a smaller volume, 
$m$=12, $t$=4, $s$=200, $p$=2,400.  
The number of pairs is large enough to cover all singers and songs for both datasets.} 

Since both datasets do not contain any backing tracks, 
for each singing voice, we randomly crop the backing track ``Amazing Grace'' to match the length of each singing voice. 
For the loss $f_\Theta$ (cf.~\cref{sec:idloss}), all the singers with the opposite gender of the target singer are used as auxiliary singers. 
For the loss $f_\Phi$ (cf.~\cref{sec:lyric_disrupt_loss}), 
$\chi_{k\in [1,K]}$ are selected from all singing voices with different lyrics from the source singing voice, and $K$ is set to 10 after investigation.

\begin{table}[]
    \centering\setlength\tabcolsep{2pt}
    \caption{{The attributes of datasets.}}
    \resizebox{0.49\textwidth}{!}{
    \begin{threeparttable}
    \begin{tabular}{c|c|c|c|c|c|c|c|c}
    \hline
         \makecell[c]{{\bf Data}\\{\bf set}} &  {\bf Language} & {\bf \#Accent} & \makecell[c]{{\bf Voice} \\ {\bf Type}} & \makecell[c]{{\bf Tempo} \\ {\bf (bpm$^\S$)}} & {\bf Pitch} & {\bf \#Singers$^\dag$} & \makecell[c]{{\bf \#Songs$^\natural$}} & {\bf \#Pairs} \\
         \hline
         \makecell[c]{Open \\ Singer} & Chinese & NA$^\ddag$ & NA & NA & \makecell[c]{280.4 \\ $\pm$ 94.6} & \makecell[c]{76 \\ (48F, 28M)} & 363 & 2,000 \\ \hline
         \makecell[c]{NUS\\ -48E} & English & \makecell[c]{7} & \makecell[c]{5$^\flat$} & \makecell[c]{68 \\ $\sim$ \\ 150} & NA & \makecell[c]{12 \\ (6F, 6M)} & 20 & 2,000 \\
         \hline
    \end{tabular}
    \begin{tablenotes}
        \item (1) NA$^\ddag$ means that the respective metadata is not available. (2) bpm$^\S$ is short for beats per minutes. (3) \#Singers$^\dag$ and \#Songs$^\natural$ denote the number of singers and songs of the dataset and our selected pairs of target singers and source singing voices cover all singers and songs. (4) ``(xF, yM)'' denotes x female and y male singers. (5) $\flat$: Soprano, Alto, Tenor, Baritone, and Bass. 
    \end{tablenotes}
    \end{threeparttable}
    }
    \label{tab:dataset_attirbute}
\end{table}

\noindent {\bf Metrics.} Besides human study in \cref{sec:human_study} as subjective evaluation metrics, 
we use the following objective metrics. 
    {(1) \emph{Identity similarity (IS)}: cosine similarity between the centroid identity feature of the target singer and the identity feature of the SVC-covered output. It measures how well SVC models imitate the timbre of the target singer.}
    We extract identity features with the Resnet18 for verification (Res18-V) model~\cite{resnet18, Autospeech_github} differing from the other encoders. 

    {(2) \emph{Lyric word error rate (WER)} measures the lyric differences between the SVC-covered output and the source singing voice, i.e., 
    the error that SVC models commit in retaining lyrics. }
    ${\tt WER} = \frac{D+I+S}{N}$    
    where $N$ is the number of words in the source singing voice, 
    and $D$, $I$, and $S$ are the numbers of deletions, insertions, and substitutions, 
    respectively. We use the speech-to-text model Conformer~\cite{Conformer} to recognize lyrics.
    
   {(3) \emph{Success reduction rate (SRR)} is the reduction of SVC success rate after applying prevention, including SRR for target identity imitation (SRR-I), 
   SRR for source lyric preservation (SRR-L), and overall SRR (SRR-T):}

   \begin{center}
$\begin{array}{l}
    \text{SRR-I} = 
    \frac{\sum_{i=1}^{Q}\mathbb{I}(\text{IS}(y_i)\geq \xi_I) - \sum_{i=1}^{Q}\mathbb{I}(\text{IS}(\hat{y}_i) \geq \xi_I)}{Q}\\
    \text{SRR-L} = 
\frac{\sum_{i=1}^{Q}\mathbb{I}(\text{WER}(y_i)\leq \xi_L) - \sum_{i=1}^{Q}\mathbb{I}(\text{WER}(\hat{y}_i) \leq \xi_L)}{Q}\\
    \text{SRR-T} =
            \frac{  \left(\begin{array}{c}
            \sum_{i=1}^{Q}\mathbb{I}(\text{IS}(y_i)\geq \xi_I \bigwedge \text{WER}(y_i)\leq \xi_L) \\
            - \\
            \sum_{i=1}^{Q}\mathbb{I}(\text{IS}(\hat{y}_i) \geq \xi_I \bigwedge \text{WER}(\hat{y}_i) \leq \xi_L)\end{array}\right)}{Q}
\end{array}$
\end{center}
    where $y_i$ and $\hat{y}_i$ for $i=1,\cdots, Q$ are the undefended and defended SVC-covered outputs, respectively; 
    $\xi_I$ and $\xi_L$ are the thresholds for deciding the success of SVC w.r.t. identity 
    imitation 
    and lyric preservation, respectively;
    and $\mathbb{I}$ is the indicator function. We set $\xi_I=0.41$ (the same as \cite{QFA2SR}) and  $\xi_L$ to the average WER of undefended SVC-covered outputs. 
    
    {The higher (resp. lower) WER, SRR-I, SRR-L, and SRR-T (resp. IS) are, 
    the more effective a prevention method is.}
    
    (4) \emph{Signal-to-noise ratio (SNR)~\cite{SpeakerGuard}}, $SNR=10\log_{10}\frac{P_x}{P_\delta},$ is widely used to measure 
    the imperceptibility of voice perturbations, where $P_x$ and $P_\delta$ are the power of the original singing voice $x$ and the perturbation $\delta$, respectively.

    (5) \emph{Perceptual evaluation of speech quality (PESQ)~\cite{rix2001perceptual}} is an objective perceptual metric that simulates the human auditory system~\cite{xiang2017digital}, ranging from -0.5 to 4.5.  
    
    To compute SNR and PESQ for a stereo song where one channel is the singing voice and the other is the backing track, we merge the song into a mono audio using the ``pydub'' package~\cite{pydub}.
    {Remark that higher SNR and PESQ indicate better imperceptibility and thus better harmlessness of prevention.}

\noindent {\bf Baselines.}
{\toolname is the first for preventing SVC, so we select from \tablename~\ref{tab:related_work} the closest baselines (targeting voice modality and generative models) for comparison, AttackVC and AntiFake (using its best target-based scheme~\cite{AntiFake}),  
which are designed to prevent ordinary voice conversion or synthesis. VSMask is not considered since it is not publicly available.} 

\noindent {\bf Experimental design.} 
We first evaluate the dual prevention effectiveness of \toolname (i.e., disrupting both identity and lyrics in SVC-covered singing voices), 
assuming the defender is aware of the identity and lyrics encoders of adversaries.
Next, we relax this assumption by evaluating the transferability of \toolname to unknown SVC models 
{and analyze the efficiency of \toolname}.
Finally, we subjectively evaluate \toolname via human study 
and evaluate the robustness of \toolname in over-the-air scenario and against adaptive adversaries.

\subsection{Dual Prevention of \toolname}
\label{sec:overall_dual}

\noindent {\bf Setting.} 
To evaluate the dual prevention of \toolname, 
we set ${\tt protect\_target}$ and ${\tt protect\_source}$ to {\tt True}, 
the initial learning rate $\alpha$=0.001 for the Adam optimizer, 
and the number of iterations $N$=1,000. 
The results are shown in  \tablename~\ref{tab:overall_performance}.

\noindent {\bf Results of identity disruption.}
{With each of \toolname, AntiFake, and AttackVC, the identity similarity of the defended SVC-covered outputs $\tilde{y}$ is lower compared to the undefended SVC-covered outputs $y$, indicating that all of them disrupt the identity of SVC-covered outputs away from target singers. 
However, \toolname achieves much lower identity similarity and much higher SRR-I than baselines,  
regardless of datasets and SVC models, demonstrating that {\it \uline{\toolname is significantly more effective than baselines for the prevention of SVC regarding identity disruption.}}} 
This is probably because (1) AttackVC perturbs acoustic features and uses the Griffin-Lim algorithm~\cite{GriffinL} to reconstruct voices, which is a lossy procedure that may interrupt the perturbation~\cite{SpeakerGuard}. 
It is evidenced by the much higher SRR-I of AttackVC-W, a modified version that directly perturbs voices. 
(2) AttackVC randomly chooses the destination speaker from some opposite-gender speakers, while \toolname selects the opposite-gender singer having the least identity similarity with the target singer. 
(3) AntiFake only penalizes the distance of embeddings between the protected voice and the destination speaker while \toolname additionally penalizes the similarity of embeddings between the protected and original voices (i.e., $f_\Theta^{\tt UT}$, cf.~\cref{sec:idloss}). 
(4) Both AttackVC and AntiFake represent the destination speaker by a voice embedding, while \toolname uses the centroid of multiple voice embeddings. 
The ablation study reported in  \Cref{sec:exper_gt} justifies 
the reasons (2)--(4).

\begin{table}[t]
\centering\setlength\tabcolsep{1pt}
  \centering
  \caption{{Comparison of prevention effectiveness and harmlessness between \toolname and baselines.}}
  \resizebox{0.49\textwidth}{!}{
  \begin{threeparttable}      
    \begin{tabular}{c|c|c|c|c|c|c|c|c|c|c|c|c|c}
    \hline
    \multicolumn{1}{c|}{\multirow{4}{*}{\textbf{Dataset}}} & \multicolumn{1}{c|}{\multirow{4}{*}{\makecell[c]{\textbf{SVC} \\ {\bf Model}}}} & \multirow{4}{*}{\textbf{Approach}} & \multicolumn{7}{c|}{\textbf{Prevention Effectiveness}} & \multicolumn{4}{c}{\textbf{Harmlessness}} \\
\cline{4-14}          &       & \multicolumn{1}{c|}{} & \multicolumn{2}{c|}{\multirow{2}{*}{\makecell[c]{\textbf{Identity} \\ {\bf Similarity} $\downarrow$}}} & \multicolumn{2}{c|}{\multirow{2}{*}{\makecell[c]{\textbf{Lyric} \\ {\bf WER (\%)} $\uparrow$ }}} & \multicolumn{3}{c|}{\multirow{2}{*}{\textbf{SRR (\%)} $\uparrow$}} & \multicolumn{2}{c|}{\multirow{2}{*}{\textbf{SNR (dB)} $\uparrow$}} & \multicolumn{2}{c}{\multirow{2}{*}{\textbf{PESQ} $\uparrow$}} \\
          &       & \multicolumn{1}{c|}{} & \multicolumn{2}{c|}{} & \multicolumn{2}{c|}{} & \multicolumn{3}{c|}{} &  \multicolumn{2}{c|}{} & \multicolumn{2}{c}{} \\
\cline{4-14}          &       & \multicolumn{1}{c|}{} & \boldmath{}\textbf{$y$}\unboldmath{} &  \boldmath{}\textbf{$\tilde{y}$} & \boldmath{}\textbf{$y$}\unboldmath{} & \boldmath{}\textbf{$\tilde{y}$}\unboldmath{} & \multicolumn{1}{c|}{\textbf{SRR-I}} & \multicolumn{1}{c|}{\textbf{SRR-L}} & \multicolumn{1}{c|}{\textbf{SRR-T}} & \boldmath{}\textbf{$\tilde{\mathcal{I}}$}\unboldmath{} & \boldmath{}\textbf{$\tilde{\mathcal{L}}$}\unboldmath{} & \boldmath{}\textbf{$\tilde{\mathcal{I}}$}\unboldmath{} & \boldmath{}\textbf{$\tilde{\mathcal{L}}$}\unboldmath{} \\
    \hline
    \multicolumn{1}{c|}{\multirow{13}{*}{\makecell[c]{\textbf{Open} \\ {\bf Singer}}}} & \multicolumn{1}{c|}{\multirow{4}{*}{\makecell[c]{\textbf{Lora} \\ {\bf -SVC}}}} & \textbf{AntiFake} & \multirow{4}{*}{0.54} & 0.15  & \multirow{4}{*}{13.9} & 13.2  & 65.5  & 9.8   & 68.0    & 24.6  & 26.6  & 3.1   & 3.3 \\
          &       & \textbf{AttackVC} &       & 0.55  &       & 13.9  & 0.1   & 9.3   & 9.4   & -5.0    & -4.5  & 2.0     & 2.3 \\
          &       & \textbf{AttackVC-W} &       & 0.24  &       & 13.1  & 41.4  & 8.8   & 44.5  & 10.1  & 12.4  & 1.3   & 1.4 \\
          &       & \textbf{\toolname} &       & \textbf{0.05} &       & \textbf{76.1} & \textbf{88.1} & \textbf{92.2} & \textbf{99.3} & \textbf{26.5} & \textbf{30.6} & \textbf{3.9} & \textbf{4.2} \\
\cline{2-14}          &  \multicolumn{1}{c|}{\multirow{3}{*}{\makecell[c]{\textbf{Vits} \\ {\bf -SVC}}}} & \textbf{AntiFake} & \multirow{3}{*}{0.51} & 0.15  & \multirow{3}{*}{14.9} & 15.2  & 71.1  & 11.2  & 73.6  & 24.5  & 25.5  & 3.0     & 3.1 \\
          &       & \textbf{AttackVC-W} &       & 0.26  &       & 14.7  & 40.1  & 9.5   & 44.8  & 10.2  & 11.1  & 1.4   & 1.7 \\
          &       & \textbf{\toolname} &       & \textbf{0.09} &       & \textbf{90.4} & \textbf{82.6} & \textbf{92.7} & \textbf{99.1} & \textbf{26.3} & \textbf{27.5} & \textbf{3.9} & \textbf{4.0} \\
\cline{2-14}          &  \multicolumn{1}{c|}{\multirow{3}{*}{\makecell[c]{\textbf{Grad} \\ {\bf -SVC}}}} & \textbf{AntiFake} & \multirow{3}{*}{0.48} & 0.17  & \multirow{3}{*}{32.1} & 31.4  & 77.8  & 8.9   & 78.9  & 24.7  & 24.8  & 3.2   & 3.1 \\
         &       & \textbf{AttackVC-W} &       & 0.23  &       & 30.9  & 59.2  & 8.2   & 62.1  & 10.0    & 10.3  & 1.5   & 1.7 \\
         &       & \textbf{\toolname} &       & \textbf{0.11} &       & \textbf{103.6} & \textbf{85.2} & \textbf{94.6} & \textbf{99.4} & \textbf{26.6} & \textbf{27.7} & \textbf{4.1} & \textbf{4.0} \\
\cline{2-14}          & \multicolumn{1}{c|}{\multirow{3}{*}{\makecell[c]{\textbf{NeuCo} \\ {\bf -SVC}}}} & \textbf{AntiFake} & \multirow{3}{*}{0.65} & 0.33  & \multirow{3}{*}{18.1} & 20.8  & 70.7  & 14.1  & 72.2  & 18.8  & 19.2  & 2.2   & 2.4 \\
          &       & \textbf{AttackVC-W} &       & 0.28  &       & 20.1  & 78.5  & 13.0    & 79.3  & 10.7  & 11.2  & 1.4   & 1.6 \\
         &       & \textbf{\toolname} &       & \textbf{0.22} &       & \textbf{86.5} & \textbf{88.8} & \textbf{90.4} & \textbf{98.9} & \textbf{27.6} & \textbf{28.3} & \textbf{3.8} & \textbf{4.1} \\
    \hline
    \multicolumn{1}{c|}{\multirow{12}{*}{\makecell[c]{\textbf{NUS} \\ {\bf -48E}}}} & \multicolumn{1}{c|}{\multirow{3}{*}{\makecell[c]{\textbf{Lora} \\ {\bf -SVC}}}} & \textbf{AntiFake} & \multirow{3}{*}{0.47} & 0.22  & \multirow{3}{*}{23.3} & 22.0    & 70.8  & 5.7   & 72.4  & 26.5  & 26.5  & 3.2   & 3.0 \\
        &       & \textbf{AttackVC-W} &       & 0.25  &       & 23.1  & 79.2  & 7.3   & 80.4  & 6.2   & 6.1   & 1.3   & 1.2 \\
         &       & \textbf{\toolname} &       & \textbf{0.12} &       & \textbf{79.9} & \textbf{87.7} & \textbf{93.6} & \textbf{98.7} & \textbf{32.4} & \textbf{28.3} & \textbf{4.4} & \textbf{4.3} \\
\cline{2-14}          & \multicolumn{1}{c|}{\multirow{3}{*}{\makecell[c]{\textbf{Vits} \\ {\bf -SVC}}}} & \textbf{AntiFake} & \multirow{3}{*}{0.48} & 0.19  & \multirow{3}{*}{18.4} & 19.4  & 75.8  & 10.3  & 78.3  & 26.3  & 24.2  & 3.3   & 2.0 \\
          &       & \textbf{AttackVC-W} &       & 0.25  &       & 19.3  & 69.1  & 10.8  & 72.6  & 6.2   & 5.9   & 1.5   & 1.3 \\
         &       & \textbf{\toolname} &       & \textbf{0.12} &       & \textbf{78.4} & \textbf{80.1} & \textbf{92.5} & \textbf{98.0} & \textbf{32.1} & \textbf{25.7} & \textbf{4.4} & \textbf{4.2} \\
\cline{2-14}          & \multicolumn{1}{c|}{\multirow{3}{*}{\makecell[c]{\textbf{Grad} \\ {\bf -SVC}}}} & \textbf{AntiFake} & \multirow{3}{*}{0.45} & 0.24  & \multirow{3}{*}{41.1} & 41.2  & 74.8  & 7.8   & 78.6  & 26.0    & 23.3  & 3.4   & 3.2 \\
          &       & \textbf{AttackVC-W} &       & 0.24  &       & 43.6  & 69.7  & 10.9  & 73.9  & 6.0     & 5.8   & 1.5   & 1.5 \\
          &       & \textbf{\toolname} &       & \textbf{0.16} &       & \textbf{94.5} & \textbf{83.4} & \textbf{91.7} & \textbf{97.3} & \textbf{31.8} & \textbf{26.4} & \textbf{4.3} & \textbf{4.2} \\
\cline{2-14}          & \multicolumn{1}{c|}{\multirow{3}{*}{\makecell[c]{\textbf{NeuCo} \\ {\bf -SVC}}}} & \textbf{AntiFake} & \multirow{3}{*}{0.59} & 0.24  & \multirow{3}{*}{22.6} & 22.7  & 79.0    & 10.4  & 79.1  & 14.9  & 16.9  & 1.9   & 2.2 \\
         &       & \textbf{AttackVC-W} &       & 0.22  &       & 21.7  & 81.9  & 8.2   & 82.4  & 6.8   & 7.3   & 1.3   & 1.7 \\
         &       & \textbf{\toolname} &       & \textbf{0.16} &       & \textbf{76.6} & \textbf{86.9} & \textbf{94.4} & \textbf{98.9} & \textbf{26.4} & \textbf{27.3} & \textbf{3.7} & \textbf{4.3} \\
    \hline
    \end{tabular}%

    \begin{tablenotes}
        \item 
        {\small (1) Acronyms refer to \tablename~\ref{tab:abbr} and \cref{sec:steup}-``Metrics''. (2) For each combination of datasets and models, the best results among all prevention approaches are highlighted in {\bf bold}. (3) $\uparrow$: the higher, the more effective or harmless the approach is. (4) $\downarrow$: the lower, the more effective the approach is. (5) AttackVC is only considered on OpenSinger and Lora-SVC since it is much less effective than its variants AttackVC-W.}
    \end{tablenotes}
  \end{threeparttable}
    }
  \label{tab:overall_performance}
\end{table}

\noindent {\bf Results of lyric disruption.}
{With \toolname, the lyric WER of $\tilde{y}$ is 53\%-75\% higher 
than that of $y$.
On each combination of SVC models and datasets, \toolname achieves 
more than 90\% overall SVC success reduction rate (SRR-T), higher than both SRR-I and SRR-L. 
In comparison, the SRR-T of AttackVC and AntiFake is nearly identical to SRR-I, 
and the WER of $\tilde{y}$ is very close to that of $y$. 
These demonstrate that {\it \uline{both baselines cannot disrupt lyrics, 
while \toolname is effective in the dual prevention of SVC by both identity and lyric disruptions.}}
}

\noindent {\bf Results of harmlessness.}
{The SNR and PESQ of protected singing voices crafted by \toolname exceed {25 dB} and {3.7}, respectively, higher than that of protected singing voices crafted by the two baselines, especially for PESQ, indicating that {\it \uline{\toolname outperforms both baselines regarding harmlessness and perturbations' side-effect on songs.}}}
This is attributed to our refined simultaneous masking loss utilizing backing tracks as additional maskers (cf. \Cref{sec:imper_exp} for justification).

\noindent {\bf Ablation study.} 
We conduct ablation study to evaluate: (E1) the impact of the ratio of protected target singing voices;
(E2) the effectiveness of \toolname w.r.t. singer genders and song genres; 
(E3) the single prevention of \toolname 
for disrupting lyric or identity but not both;
{(E4) the effectiveness of the gender-transformation loss and high/low hierarchy multi-target loss;}
and (E5) the effectiveness of the refined utility loss. 
The results show that: 
(R1) \toolname is not impacted by the ratio for disrupting lyric,
and is still effective for disrupting identity even only a small fraction of target singing voices are protected and becomes more effective
when increasing the ratio;
(R2) \toolname exhibits  universality across different singer genders and song genres;
(R3) \toolname is still effective for single prevention 
of disrupting  identity or lyric;
{(R4) 
our gender-transformation loss outperforms the loss without or with only the loss term $f_\Theta^{\tt UT}$, the loss randomly selecting a destination singer with the opposite gender, and the loss representing the destination singer by a voice embedding.
Our high/low hierarchy multi-target loss achieves better lyric disruption than the low hierarchy loss, high hierarchy loss, 
and the loss without multiple targets;} and
(R5) the refined utility loss with backing tracks as additional maskers achieves better harmlessness 
than the basic utility loss solely using the singing voice as the masker.   
More details refer to \Cref{sec:impact_ratio}-\Cref{sec:imper_exp}.

We will mainly consider the OpenSinger dataset and Lora-SVC, 
as they generally achieve the best SVC in \tablename~\ref{tab:overall_performance}.

\subsection{Transferability of \toolname}
\label{sec:exper_transfer}

\begin{figure}[t]
\centering
    \begin{subfigure}{0.235\textwidth}
\includegraphics[width=1.\textwidth,height=.8\textwidth]{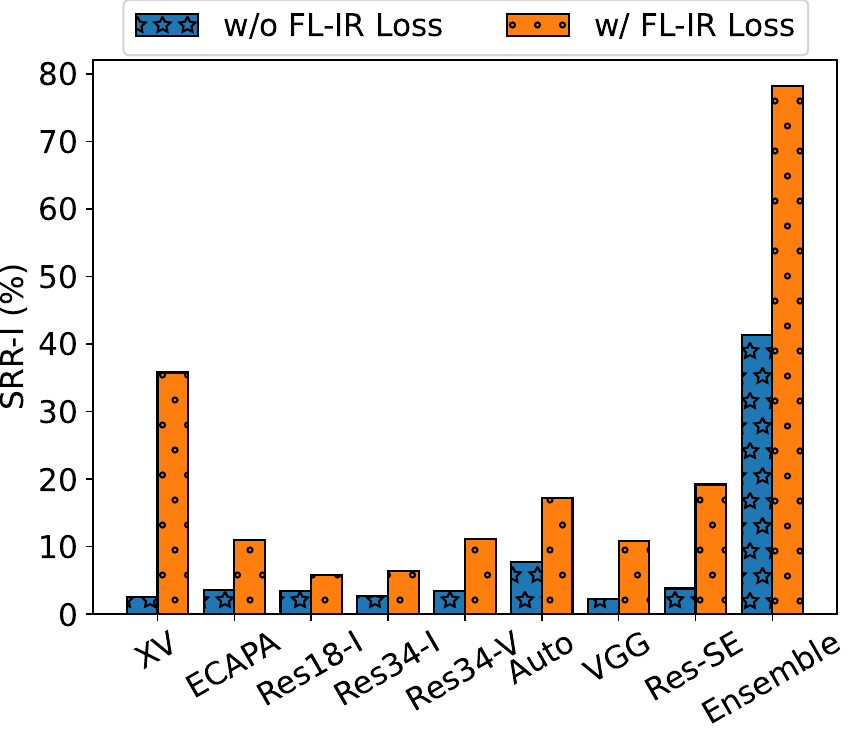}
    \caption{Identity disruption}
    \label{fig:transfer_identity_lora_open}
    \end{subfigure}
    \begin{subfigure}{0.235\textwidth}
\includegraphics[width=1.\textwidth,height=.8\textwidth]{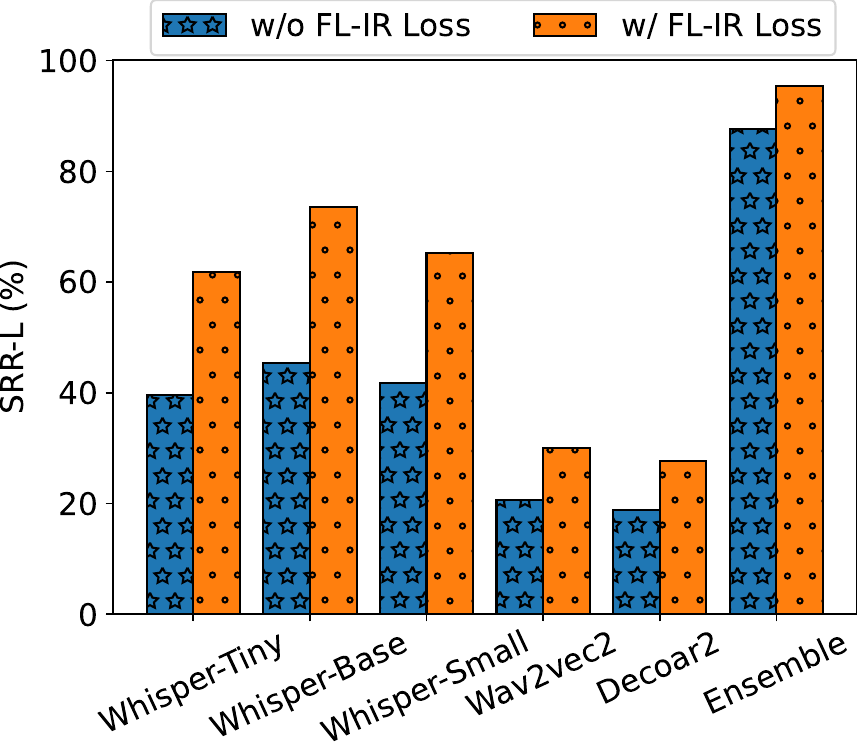}
    \caption{Lyric disruption}
    \label{fig:transfer_lyric_lora_open}
    \end{subfigure}
    \caption{Transferability of \toolname.}
    \label{transfer_identity_lyric_lora_open}
\end{figure}
\begin{figure}
    \centering
    \includegraphics[width=0.3\textwidth]{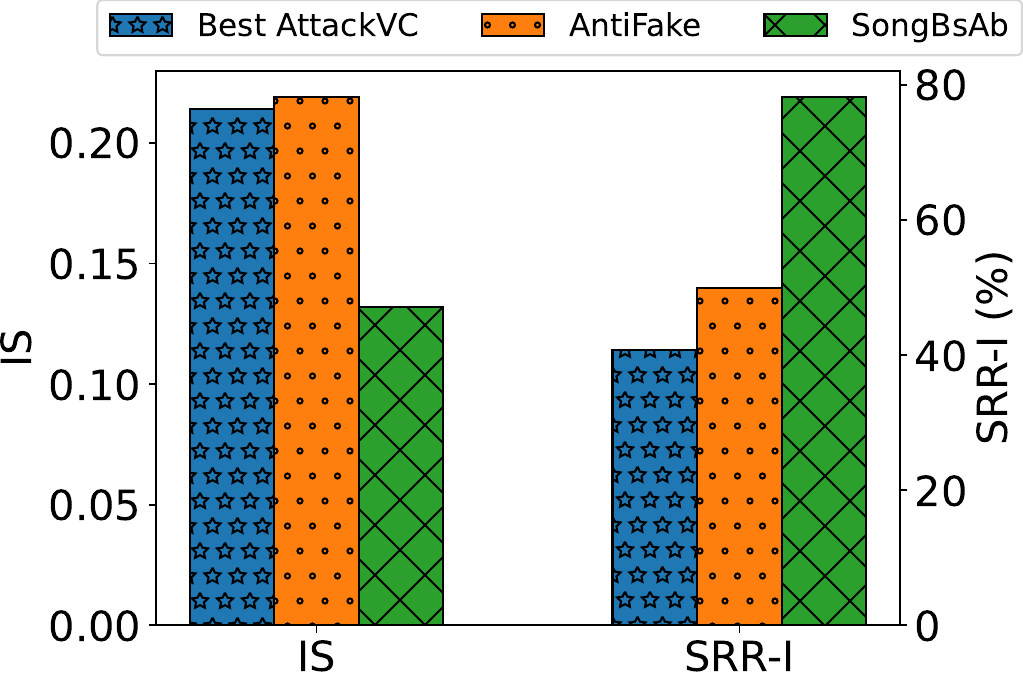}
    \caption{Comparison of transferability for identity disruption in terms of identity similarity. AttackVC uses a single encoder, and Best AttackVC means the best result among all encoders.}
    \label{fig:transfer_comp}
\end{figure}

We evaluate the transferability 
for disrupting identity and lyric separately to avoid interference. For the FL-IR loss, we set $R=32$, $w_l=w_s=\frac{L}{200}$ where $L$ is the number of sample points of a singing voice. The impact of $w_l$ and $w_s$ is evaluated in \Cref{sec:frame_length_exper}. 
Remark that we also evaluate the transferability via human study in \Cref{sec:human_study}.

\noindent {\bf Transferability for identity disruption.} 
Each of the 8 identity encoders (cf.~\cref{sec:steup})
is used to craft protected target singing voices. 
The results are shown in \figurename~\ref{fig:transfer_identity_lora_open}. 
SRR-I 
is largely improved after applying either our FL-IR loss (regardless of the identity encoder)
or encoder ensemble. 
Applying both the FL-IR loss and encoder ensemble
yields the best transferability,
{\it \uline{confirming the effectiveness and complementarity of the FL-IR loss and encoder ensemble in boosting transferability for identity disruption.}}
According to \figurename~\ref{fig:transfer_comp}, 
\toolname achieves lower identity similarity (IS) and higher SRR-I than both AttackVC and AntiFake, 
{\it \uline{indicating the superiority of \toolname over baselines for transferring to unknown identity encoders.}}
This is because AttackVC and AntiFake are not incorporated with the encoder ensemble 
or the FL-IR loss.

\noindent {\bf Transferability for lyric disruption.} 
Each of the 5 lyric encoders (cf.~\cref{sec:steup}) is used to craft protected source singing voices. 
The results are shown in \figurename~\ref{fig:transfer_lyric_lora_open}. 
We can observe that {\it \uline{\toolname has inherent transferability 
for lyric disruption,
and the FL-IR loss and encoder ensemble can further enhance the transferability 
for lyric disruption.}}
More results of transferability with different SVC models
and metrics are reported in \Cref{sec:more_results_transfer}, from which 
the same conclusions can be drawn.

\begin{table}[t]
    \centering\setlength\tabcolsep{1pt} 
    \caption{Pairs of singing voices for human study task 1.}
    \label{tab:task1}
   \scalebox{0.95}{
   \begin{tabular}{c|c|c}  \hline
   {\bf Pair Name} & {\bf No.} & {\bf Description} \\\hline
     {\bf Normal} & 9 & \makecell[l]{A pair consists of two original singing voices from\\ a target singer and a source singer, respectively.} \\ \hline

     \makecell[c]{{\bf Undefended}\\{\bf Output}} & 9 & \makecell[l]{A pair is built by replacing the source singer's voice in a \\ Normal pair with its
    SVC-covered output using the identity \\ of corresponding target singer,  where \toolname is disabled.} \\   \hline

    \makecell[c]{{\bf Defended}\\{\bf Output}} & 5 & \makecell[l]{The SVC-covered output in each of 5 randomly selected\\
    Undefended Output pairs is replaced with another\\
     SVC-covered output, where \toolname is enabled\\
    and uses the same identity encoder as the SVC model.}\\   \hline

    \makecell[c]{{\bf Defended}\\{\bf Output}\\{\bf (Transfer)}} & 5 &  \makecell[l]{They are built the same as Defended Output pairs \\ except that \toolname uses different identity encoders \\ from the SVC model.}  \\   \hline

    {\bf Prot} & 4 & \makecell[l]{The target singing voices protected by \toolname\\ for building 5 Defended Output pairs (the duplicated\\ one is removed), with their original counterparts.} \\   \hline

     \makecell[c]{{\bf Prot}\\{\bf (Transfer)}} & 5 & \makecell[l]{The target singing voices protected by \toolname\\ for building 5 Defended Output (Transfer) pairs, \\ with their original counterparts.}  \\   \hline

     {\bf Special} & 3 & \makecell[l]{Each pair consists of two original singing voices from two  \\singers with opposite genders. 
 If a participant fails to choose\\{\it different} for any of them,  we exclude all his/her submissions.} \\   \hline
    \end{tabular}}
\end{table}

\subsection{{Run Time and Efficiency Analysis}}\label{sec:efficiency}
{The optimization process of \toolname took an average of 287 seconds (0.287 seconds per iteration $\times$ 1000 iterations) using an NVIDIA RTX 2080Ti GPU, comparable to the baselines.
Given the relatively short runtime and that \toolname is applied offline before song release, thus holding a minimal real-time requirement and a high tolerance for runtime, 
we regard \toolname as a computationally efficient toolkit.}

\subsection{Human Study}\label{sec:human_study}
{To further confirm the effectiveness and harmlessness of \toolname in practice, 
we conduct a human study as subjective evaluation metrics.
The human study was approved by the Institutional Review Board (IRB) of our institutes.
We design the following 3 tasks for human study in the form of questionnaires on Credamo~\cite{credamo}, an online opinion research questionnaire completion platform. 
Here we report the results on the Chinese dataset OpenSinger 
while the results on the English dataset NUS-48E are similar (cf. \Cref{sec:humn_study_english}).}

\noindent {\bf Low-quality answers filtering.} 
{We set up special questions as concentration tests. 
Each task contains 3 different questions inserted at random positions, and each question is designed to be trivial and tailored for each task. Details refer to \tablename~\ref{tab:task1}, \tablename~\ref{tab:task2}, and Task-3.}

\noindent {\bf Participants.} 
{We recruited 120 participants (after filtering) for each task, and each participant can only participate in one task, resulting in 360 participants. 
We 
restricted participants to be within China. Overall, the participants come from 27 provinces and 112 cities in China, offering a reasonable representation.}

\noindent {\bf Spent Time.} 
{Participants have adequate time to review each sample and complete the whole task without any time restriction. 
Statistically, they spent $20\pm 9$, $17\pm 9$, and $9\pm 6$ minutes for Task 1, Task 2, and Task 3, respectively. In contrast, filtered participants by special questions spent $8\pm 5$, $7\pm 3$, and $6\pm 3$ minutes, 
indicating a positive correlation between spent time and answer quality. }

\begin{figure*}\centering
\begin{subfigure}{0.325\textwidth}
        \centering
       \includegraphics[width=.95\textwidth,height=.65\textwidth]{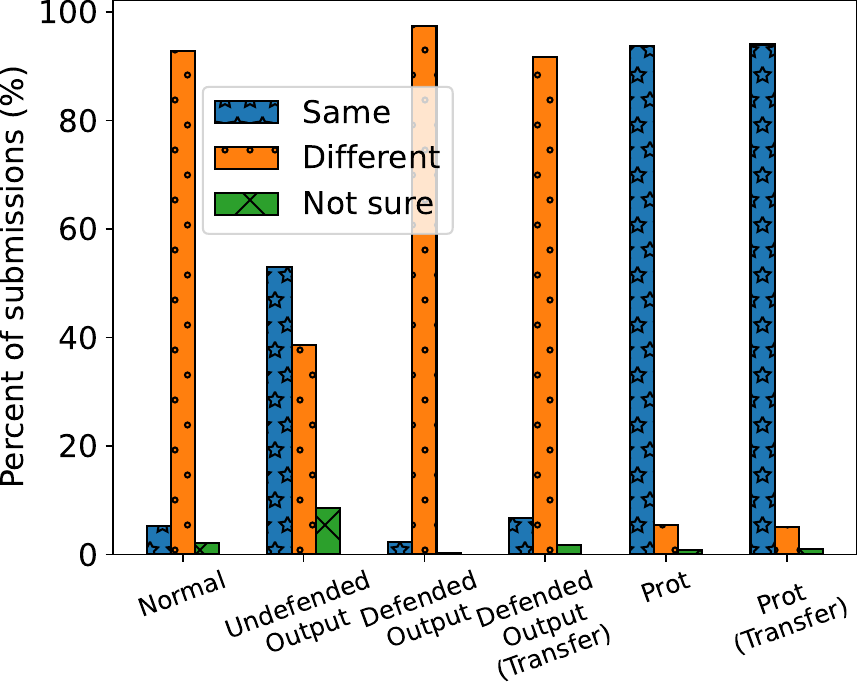}
\caption{Identify Singer}\label{fig:human_study_ID}
\end{subfigure}\quad
\begin{subfigure}{0.325\textwidth}
        \centering
\includegraphics[width=.9\textwidth,height=.65\textwidth]{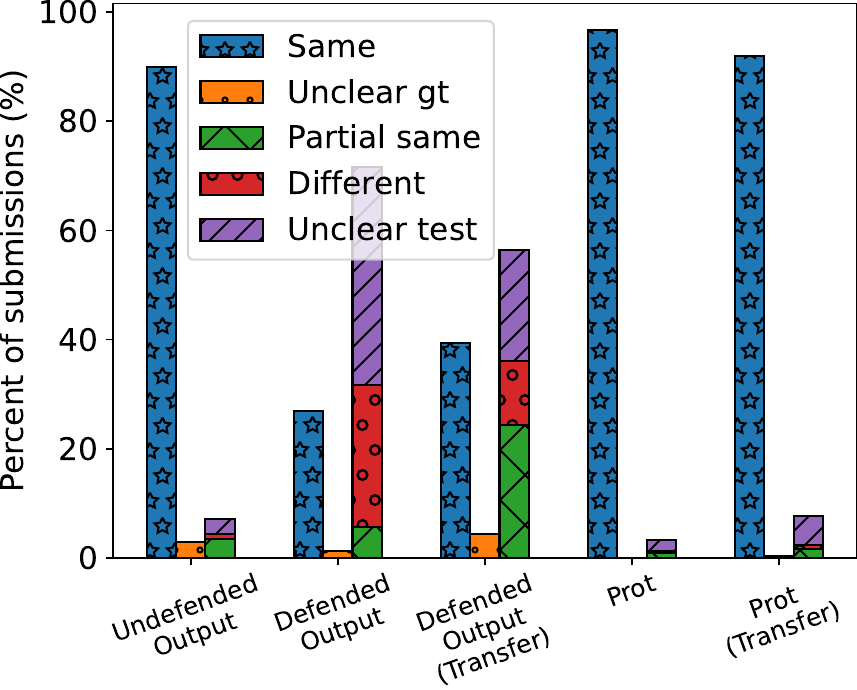}     
\caption{Identify Lyric}\label{fig:human_study_LD}
\end{subfigure}\quad
\begin{subfigure}{0.31\textwidth}
        \centering
       \includegraphics[width=.95\textwidth,height=.65\textwidth]{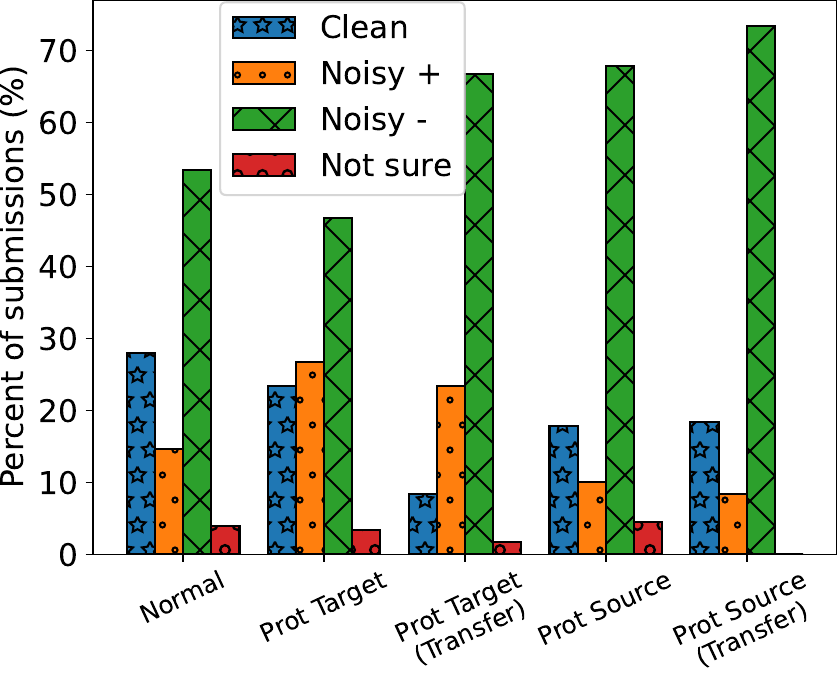}
\caption{Clean or Noisy}\label{fig:human_study_clean_noisy}
\end{subfigure}
\caption{{Results of human study. ``Noise +'' and ``Noise -'' denote the answers ``{\it noisy w/ influence}'' and ``{\it noisy w/o influence}''.}}\label{fig:human_study}
\end{figure*}

\begin{table}[t]
    \centering\setlength\tabcolsep{1pt} 
    \caption{Pairs of singing voices for human study task 2.}
    \label{tab:task2}
   \scalebox{0.95}{
   \begin{tabular}{c|c|c}  \hline
   {\bf Pair Name} & {\bf No.} & {\bf Description} \\\hline
    \makecell[c]{{\bf Undefended} \\ {\bf Output}}  & 10 &  
    \makecell[l]{A ground-truth voice is an original source singing voice and 
    \\  its test voice is an SVC-covered output using the lyrics\\  of the ground-truth voice, where \toolname is disabled.} \\ \hline
    \makecell[c]{{\bf Defended} \\ {\bf Output}} & 5 & 
    \makecell[l]{The SVC-covered output in each of 5 randomly selected \\ 
    Undefended Output pairs is replaced by
    another SVC-covered \\ output, where \toolname is enabled 
    and uses the same lyric \\ encoder as the SVC model.} \\ \hline
    \makecell[c]{{\bf Defended} \\ {\bf Output} \\ {\bf (Transfer)}} & 5 & \makecell[l]{They are built
    the same as Defended Output pairs except that \\ \toolname uses different 
    lyric encoders from the SVC model.} \\ \hline 
    {\bf Prot} & 5 & \makecell[l]{The source singing voices protected by \toolname\\ for building 5 Defended Output pairs, \\ with their original counterparts as ground-truth voices.}  \\ \hline
     \makecell[c]{{\bf Prot} \\ {\bf (Transfer)}} & 5 & \makecell[l]{The source singing voices protected by \toolname\\ for building 5 Defended Output (Transfer) pairs, \\ with their original counterparts as ground-truth voices.}  \\ \hline
    {\bf Special}  & 3 & \makecell[l]{Each pair consists of two original singing voices in Chinese  \\ and English, respectively. 
If a participant fails to choose \\ {\it different} for
any of them, we exclude all his/her submissions.} \\\hline
    \end{tabular}
    }
\end{table}

\noindent {\bf Task 1: identify singer.} 
To evaluate \toolname's effectiveness in disrupting identity, 
participants are asked to tell whether each pair of singing voices is sung 
by the same singer, with options: {\it same}, {\it different}, or {\it not sure}. 
We randomly created 37 pairs; see \tablename~\ref{tab:task1} for details.

The results are shown in \figurename~\ref{fig:human_study_ID}. 
Over 
{92\%} of participants choose {\it different} for Normal pairs, confirming the quality of submissions. 
By contrast, most participants choose {\it same} for Undefended Output pairs, demonstrating the identity conversion capacity of SVC models. 
Remark that it is reasonable that a few participants choose {\it different} for Undefended Output pairs, as when humans consecutively listen to the undefended SVC-covered and original target singing voices, 
they are conservative in considering them as sung by the same singer, 
consistent with previous human studies~\cite{FakeBob,AS2T}.
Remarkably, 
{97\%} and {91\%} 
of participants choose {\it different} for 
Defended Output and Defended Output (Transfer) pairs, 
{58\%} and {52\%} 
higher than that of the undefended counterparts, respectively. It indicates that \toolname is very effective for disrupting the target singer's identity in SVC-covered singing voices, even
using different identity encoders. More than 
{93\%} 
of participants 
choose {\it same} for Prot and  Prot (Transfer) pairs,
confirming the harmlessness of \toolname on preserving the singer's identity in the protected singing voices. 

\noindent {\bf Task 2: identify lyric.} To evaluate the effectiveness 
of \toolname for disrupting lyric, participants 
are asked to tell if a ground-truth voice and a test voice contain the same lyrics. We provide 5 options: 
{\it same}, {\it partially same}, {\it different}, {\it unclear ground-truth}, and {\it unclear test},
where the first three options denote intelligible lyrics, and ``unclear'' in other options means that lyrics are too vague to be recognized, i.e., unintelligible. We randomly build 30 pairs; see \tablename~\ref{tab:task2} for details. 

The results are shown in \figurename~\ref{fig:human_study_LD}. 
{Nearly 90\%} of participants
choose {\it same} for Undefended Output pairs, confirming the capacity of
SVC models for preserving the lyrics.
By contrast,  
more than 
{71\%} 
and {56\%} of participants believe that the SVC-covered singing voices 
in Defended Output and Defended Output (Transfer) pairs contain either unclear or (partially) different lyrics from the ground-truth ones, much higher than that of the undefended counterparts 
{(7\%)}.  
It confirms the effectiveness of \toolname for disrupting lyrics in SVC-covered singing voices. Moreover, over 
{92\%} 
of participants choose {\it same} for Prot and Prot (Transfer) pairs, indicating the harmlessness of \toolname on preserving the lyrics in the protected singing voices. 

\noindent {\bf Task 3: clean or noisy.} The above two tasks have confirmed the harmlessness of \toolname on preserving the identity and lyrics in protected singing voices, respectively. This task performs a much stricter study by asking participants if a given song contains any background noise and if so, how the noise influences their enjoyment of the song, provided with 4 options: {\it clean}, {\it noisy w/ influence}, {\it noisy w/o influence}, and {\it not sure}. 
We randomly select 5 normal songs, 
5 protected source songs and 5 protected target songs.
Among 5 target songs, 3 (resp. 2) songs are crafted using the same (resp. different) identity encoders than the SVC model, denoted by ``Prot Target'' (resp. ``Prot Target (Transfer)''). Similarly, the 5 source songs consist of 3 ``Prot Source'' songs and 2 ``Prot Source (Transfer)'' songs.
We remark that these songs contain the backing tracks since in practice, singing voices are usually accompanied by backing tracks. We additionally insert 3 silent audios with zero magnitude as the concentration test. If a participant didn't choose  {\it clean} or {\it not sure} for any of silent audios, we exclude all his/her submissions. 

The results are shown in \figurename~\ref{fig:human_study_clean_noisy}. 
Although the number of {\it clean} answers of four types of adversarial songs decreases compared to the normal songs, 
a large majority of them do not influence the perception and enjoyment. 
This demonstrates that \toolname can maintain the song quality and enjoyment of protected songs in practice. 

\begin{figure*}\centering
\begin{subfigure}{0.19\textwidth}
        \centering
       \includegraphics[width=1\textwidth]{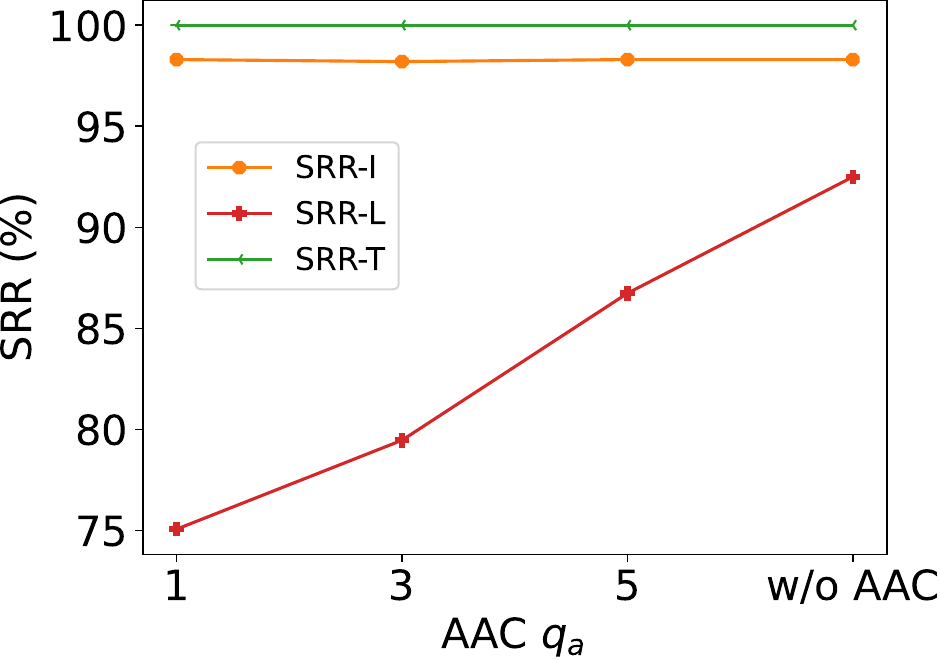}
\caption{AAC}\label{fig:robustness_AAC}
\end{subfigure} 
\begin{subfigure}{0.182\textwidth}
        \centering
       \includegraphics[width=1\textwidth]{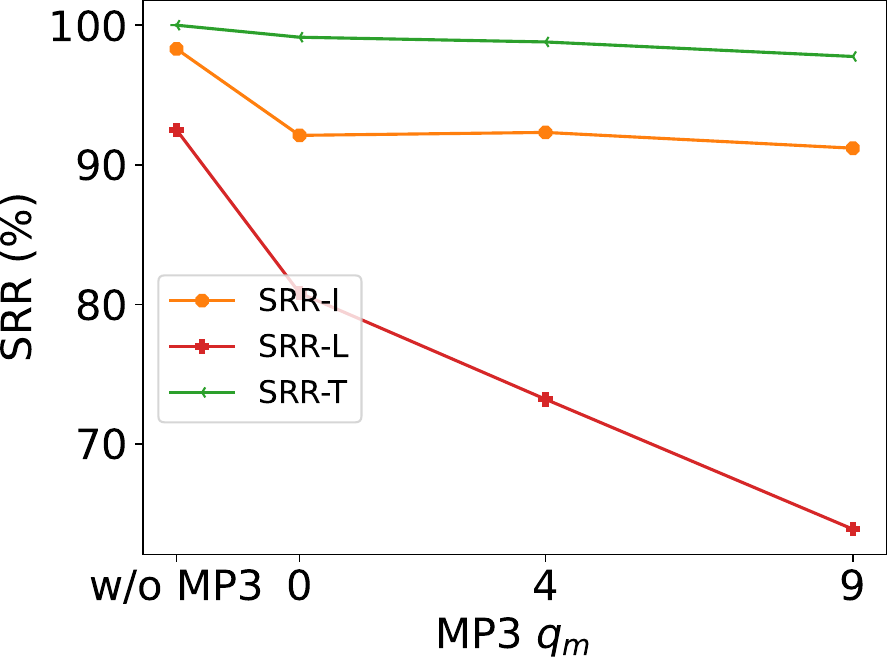}
\caption{MP3}\label{fig:robustness_MP3}
\end{subfigure}
\begin{subfigure}{0.19\textwidth}
        \centering
       \includegraphics[width=1\textwidth]{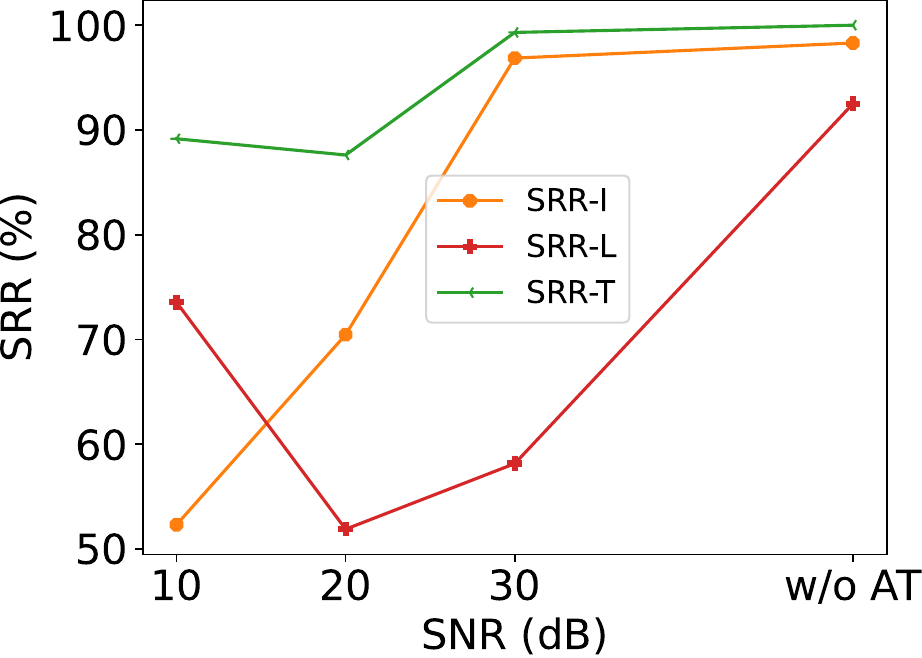}
\caption{AT}\label{fig:robustness_AT}
\end{subfigure}
\begin{subfigure}{0.17\textwidth}
        \centering
       \includegraphics[width=1\textwidth]{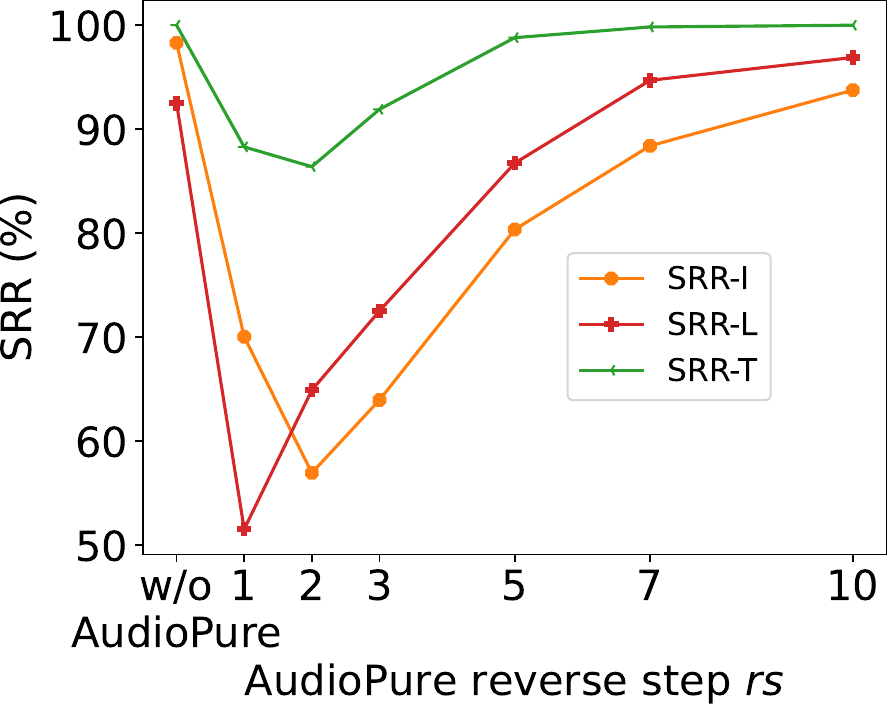}
\caption{AudioPure}\label{fig:robustness_AudioPure}
\end{subfigure}\quad
\begin{subfigure}{0.225\textwidth}
        \centering
    \includegraphics[width=1\textwidth]{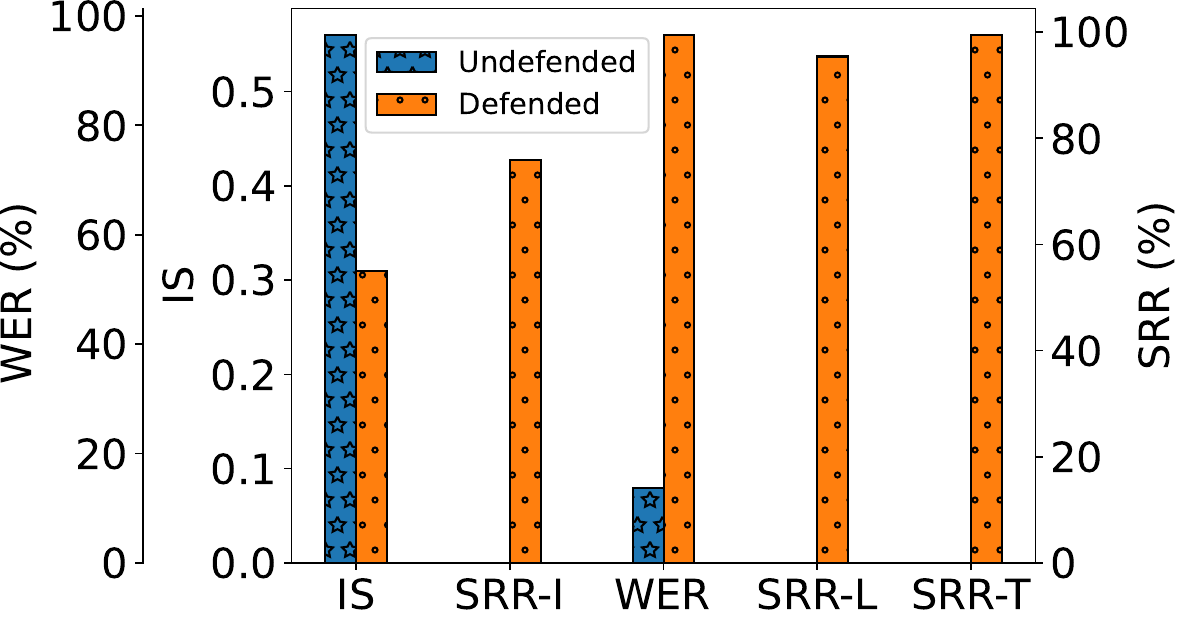}
    \caption{Optimization-based}
    \label{fig:opt_adv}
\end{subfigure}
\caption{Robustness of \toolname against transformation-based and optimization-based adaptive adversaries}\label{fig:robustness}
\end{figure*}

\subsection{Robustness of \toolname}
\label{sec:robustness}

\subsubsection{Robustness against Adaptive Adversaries}\label{sec:robust_adaptive}
Here we consider adaptive adversaries who know and attempt to bypass \toolname. 
We design {three} types of adversaries, i.e., transformation-based and optimization-based adversaries by modifying singing voices,  
{and {fine-tuning-based} adversaries by modifying SVC models.} 

\noindent  
{\bf Transformation-based adversaries.}  
The adversary 
pre-processes protected singing voices via some transformations before
SVC. 
We consider three typical methods in the audio domain: AAC compression (AAC)~\cite{SpeakerGuard}, MP3 compression (MP3)~\cite{SpeakerGuard}, and Audio Turbulence (AT)~\cite{yuan2018commandersong}, and one recent advanced method AudioPure~\cite{AudioPure}.
AAC and MP3 perform different speech compression schemes controlled by the compression quality parameters $q_a$ and $q_m$, respectively. 
AT adds white Gaussian noise to each voice within a pre-defined SNR limit (cf.~\cref{sec:steup}).
We set $q_a$ of AAC as 1, 3 and 5,  $q_m$ of MP3 as 0, 4 and 9, and the SNR of AT as 10, 20 and 30 dB, following the setting in \cite{SpeakerGuard}. AudioPure first adds noise to the input voice and then runs a reverse process with $rs$ reverse steps to recover the purified voice from the noisy one. We set $rs$ to 1, 2, 3, 5, 7, and 10, the same as \cite{AudioPure}. 

The results are shown in \figurename~\ref{fig:robustness_AAC}--\figurename~\ref{fig:robustness_AudioPure}, 
where the SVC success reduction rate (SRR) decreases compared to no transformation, 
indicating the reduction of the prevention effect of \toolname.  
However, regardless of the pre-processing methods and their specific parameters, 
the SRR-I, SRR-L, and SRR-T are larger than 52\%, 51\%, and 87\%, respectively. 
These demonstrate the robustness of \toolname against transformation-based adversaries. 
{There are two main reasons. Firstly, they transform singing voices without any guidance, so although interfering with perturbations, they may just push the target and source singing voices towards another different singer and lyrics rather than the target singer and source lyrics. 
Thus, the SVC success rate is still low. 
Secondly, transformations have side-effects on the quality of singing voices~\cite{QFA2SR, SpeakerGuard,du2020sirenattack} by injecting noises (AT and AudioPure) or lossily compressing (AAC and MP3). SVC models accepting low-quality inputs tend to produce poor-quality outputs with low target identity similarity and high source lyric WER.}

\noindent {\bf Optimization-based adversaries.} 
The adversary tries to restore the features of the expected lyrics and the target singer's identity 
from protected singing voices by applying 
\toolname in a ``reverse'' direction. 
The key challenge is to determine the ``reverse'' direction. 
For lyrics,  
the adversary can use a Text-to-Speech tool (we use the iFlytek TTS~\cite{iFlytek_tts}) 
to craft a voice with the expected lyrics, which replaces the voice $\chi$ 
in $f_{\Phi}$ (cf.~\cref{sec:lyric_disrupt_loss}) as the ``reverse'' direction. 
However, since adversaries cannot acquire the target singer's clear singing voices (otherwise they can be directly used for SVC), the ``reverse'' direction for the target singer is unknown. 
We instead increase adversaries' capacity by assuming that they can probe queries to a speaker recognition system enrolled by the target singer and use recognition scores to guide the optimization. 
Specifically, we use the speaker recognition system XV (cf. \cref{{sec:steup}}) and 
natural evolution strategy (NES) to estimate the gradient of the loss $f_{\Theta}^{\tt T}$. 
The same as \cite{FakeBob, QFA2SR}, the parameter {\tt samples\_per\_draw} and the number of iterations of NES are set to 50 and 1,000, respectively, thus the number of total queries is $50\times 1,000=50,000$.

The results are shown in \figurename~\ref{fig:opt_adv}, where IS denotes the (cosine) identity similarity. 
\toolname can still reduce the overall singing voice conversion success rate by 99\% (SRR-T is 99\%), 
indicating that the optimization-based adversary 
fails to circumvent \toolname on almost all the protected singing voices. 
{The reason is two-fold. Firstly, while adversaries may obtain a relatively precise reverse direction for expected lyrics, the estimated gradient by NES is less informative than the exact gradient, preventing an accurate reverse direction for target identity. 
Secondly, although the optimization-based adversaries outperform the 
transformation-based ones
by using guidelines, 
they still suffer from degrading input quality since perturbations are applied twice to singing voices. 
}

\noindent {\bf Fine-tuning-based adversaries.} 
{This adversary fine-tunes the identity and lyric encoders, aimed to produce identity and lyric features for the protected singing voices that are close to that of original singing voices. 
We design two different fine-tuning approaches: $\arg\min_{\vartheta}f_1$ and $\arg\min_{\vartheta}f_1+f_2$, where 
$\vartheta\in\{\Theta, \Phi\}$ is the encoder, 
$f_1={\tt Dist}(\vartheta(x^0), \vartheta(x))$, 
and $f_2$ is the encoder's original training loss, which intends to preserve the functionality of recognizing singers and lyrics of the identity and lyric encoders, respectively. 
With the fine-tuned identity and lyric encoders, the adversary may or may not fine-tune the decoder to align with encoders' modification, leading to four types of adversaries. For both encoders and the decoder, we adopt their respective official training settings,  
since they have been optimized and tailored towards more effective training. 
}

The results are shown in \tablename~\ref{tab:fine_tune_adversary}. 
Unsurprisingly,
regardless of the loss for fine-tuning encoders, additional fine-tuning of the decoder further reduces the effectiveness of \toolname compared to only fine-tuning encoders. This is because the fine-tuned decoder aligns with the modified feature space of fine-tuned encoders.
When the decoder is not fine-tuned, $f_1+f_2$ is less effective in bypassing \toolname than $f_1$. 
The reason is that $f_2$ introduces larger modifications to the feature space for preserving the functionality destroyed by $f_1$ which forcefully pulls together two distinct features, while the un-fine-tuned decoder is trained to cooperate well with the unmodified feature space. 
When the decoder is fine-tuned, $f_1+f_2$ exhibits a larger impact on \toolname than $f_1$, due to the more functional feature space achieved by $f_1+f_2$, based on which the decoder is more effective for SVC. 

Under the strongest adversary, \toolname still achieves over 68\% SRR, indicating that fine-tuning adversaries 
cannot render \toolname to be ineffective. 
This is due to 
the continuous and large input and output spaces of generative SVC models, 
so even less effective perturbation in inputs still directly affects the covered songs. 
It is consistent with the finding that adversarial training has an upper bound of defeating adversarial examples~\cite{madry2017towards}. 

\begin{table}
    \centering
     \begin{minipage}{0.24\textwidth}
    \centering
     \caption{{Robustness of \toolname against fine-tuning adversaries in terms of SRR}}
    \label{tab:fine_tune_adversary}
    \resizebox{1\textwidth}{!}{
     \begin{threeparttable}
    \begin{tabular}{c|c|c}
    \hline
        \diagbox{\bf E}{\bf D} &  \makecell[c]{{\bf w/o FT}} &  \makecell[c]{{\bf FT}} \\ \hline
        \makecell[c]{{\bf w/o FT}} & \multicolumn{2}{c}{100\%} \\ \hline
        \makecell[c]{{\bf FT } (${f_1}$)} & 83.7\% & 76.6\% \\ \hline
        \makecell[c]{{\bf FT} (${f_1}$+${f_2}$)} & 91.2\% & 68.9\% \\ \hline
    \end{tabular}
    
    \begin{tablenotes}
        \item E: Encoders; D: Decoders 
        \item FT: Fine-tuning
    \end{tablenotes}
    \end{threeparttable}
    }
    \end{minipage}
    \quad
    \begin{minipage}{0.22\textwidth}
    \centering
     \caption{Over-the-air robustness of \toolname in terms of SRR}
    \label{tab:over_the_air}
    \resizebox{1\textwidth}{!}{
     \begin{threeparttable}
    \begin{tabular}{c|c|c}
    \hline
       \diagbox{\bf L}{\bf M}  &  \makecell[c]{{\bf iPhone}} &   {\bf OPPO} \\ \hline
        {\bf JBL} & 90\% & 90\% \\ \hline
        {\bf Xiaodu} &  92\% & 91\% \\ \hline
        \makecell[c]{{\bf iPad}} & 89\% & 85\% \\ \hline
    \end{tabular}
    \begin{tablenotes}
        \item L: Loudspeakers
        \item M: Microphones
    \end{tablenotes}
    \end{threeparttable}
    }
    \end{minipage}
\end{table}

\subsubsection{Over-the-air robustness of \toolname}
The adversary may obtain singing voices by recording them
using microphones, 
during which perturbations used for prevention may be disrupted~\cite{AS2T, QFA2SR}.
We evaluate the robustness of \toolname by playing singing voices via 3 loudspeakers (JBL clip3 portable loudspeaker, Xiaodu smart speaker, and iPad Pro 10.5)
and recording the air channel-transmitted singing voices using 2 microphones (iOS iPhone 15 Plus and Android OPPO), leading to 6 diverse combinations of hardware settings. 
We randomly select 100 pairs of target singers and protected source singing voices with 3 protected target singing voices per target singer. 
To ensure the quality of recorded singing voices for SVC~\cite{FakeBob, AdvPulse, QIN2023100098, Imperio, Metamorph}, 
we conduct experiments in a relatively quiet room with air-conditioner noise, 
and set the distance between microphones and loudspeakers to 2 meters.
The results are shown in {\tablename~\ref{tab:over_the_air}}. 
Though the effectiveness of \toolname varies slightly with loudspeakers and microphones, 
\toolname achieves at least 85\% SRR regardless of devices, 
confirming the over-the-air robustness of \toolname. 
{Over-the-air transmissions introduce hardware distortion, ambient noise, and reverberation to perturbations~\cite{AS2T, Metamorph}, which can be regarded as transformations, so \toolname's over-the-air robustness shares the reason with robustness against transformation-based adversaries.}

%% file: discussion.tex
\section{Discussion}\label{sec:discussion}
In this section, we discuss the limitations of \toolname,  insights, and potential future works motivated by this work.

\noindent {\bf Copyrights of melodies.} 
\toolname directly protects the civil rights of target singers and the copyright of lyrics, 
but only indirectly safeguards melody copyright 
by discouraging the sharing of SVC-covered songs. 
To directly disrupt the melody, we crafted perturbations by maximizing the root mean square error of pitch features between the protected and original singing voices. 
However, significant deviation in pitch features leads to noticeable changes in melody, likely due to the tight mapping between singing voices and low-dimension pitch features. 
Future work should study extracting pitch features and the correlation with singing voices to enable pitch disruption in SVC while preserving melody in the protected songs.

\noindent 
{\bf Real-world usage of \toolname.} 
While we have confirmed the effectiveness of \toolname to protect both Chinese and English songs 
with different singer genders (cf.~\Cref{sec:gender}), {accents, and voice types (cf.~\tablename~\ref{tab:dataset_attirbute})}, 
and different song genres (cf.~\Cref{sec:genre}), {tempos, and pitches (cf.~\tablename~\ref{tab:dataset_attirbute}),} 
songs can be in other languages and diverse in other aspects, e.g., 
degree of pitch fluctuations and instrument types and loudness. 
We do not consider these factors due to the unavailability of suitable datasets, 
which should be examined in future to better understand the applicability of \toolname in the real world.

\noindent {\bf Arm race between adaptive adversaries and defenders.}
The defender, aware of transformations in \cref{sec:robust_adaptive}, can apply them to intermediate singing voices at each iteration such that the protected singing voices gain sufficient robustness against the transformations~\cite{DBLP:conf/icml/AthalyeC018, DBLP:conf/nips/ZimmermannBTC22}. 
However, this cannot protect previously released songs that have been kept 
by adversaries~\cite{honig2024adversarial}.
Adversaries might also try to bypass \toolname with a binary detector to identify and discard protected singing voices. However, this detector will inevitably produce false negatives, 
allowing \toolname to succeed, even with a small proportion of protected voices (cf.~\Cref{sec:impact_ratio}). Additionally, the defender can counter this detection by incorporating the detector's outputs into the loss to deceive both the encoders and the detector~\cite{DBLP:conf/ccs/Carlini017,DBLP:journals/corr/abs-1902-06705,tramer2020adaptive}.

\noindent {\bf Harmlessness of perturbations.}
We propose using the backing track as an additional masker, enhancing the hiding capacity of perturbations compared to previous methods that relied solely on the voice as the masker. This indicates that harmlessness can be improved by utilizing unique elements of singing voices versus ordinary speech. We believe this insight is valuable for future research. For example, different backing tracks may have varying masking capacities, so future studies could explore this correlation to identify or design backing tracks that more effectively conceal perturbations.
{
While \toolname restricts perturbations to stay below the hearing threshold for \emph{all} frequencies, an alternative is to position perturbations outside the human hearing range (20-20 kHz~\cite{Fundamentals_of_Acoustics}) using methods like the ultrasound transformation model~\cite{VRIFLE} or imposing larger penalties on audible frequency bands~\cite{AntiFake}. This approach eliminates the constraint on perturbation magnitude~\cite{VRIFLE}, but it may also reduce the frequency information available for optimizing prevention losses. We leave this as interesting future work.}

\noindent {\bf Other song cover techniques.} 
Other techniques for automated song covers include singing voice synthesis (SVS)~\cite{DiffSinger}, which uses musical scores with lyrics and the target singer's voices to generate a performance as if sung by the target. 
The main difference between SVC and SVS is how melody and lyric information are provided. 
While our identity disruption methods can be adapted for SVS, 
future work should focus on lyric and melody disruption for SVS, 
potentially using adversarial examples from natural language processing~\cite{bad_characters}.

\noindent {\bf Non-technical efforts for rights protection.}
{
Complex music copyrights and civil rights protection also require collaborative non-technical efforts. 
Firstly, applying and enforcing legal frameworks is challenging due to outdated definitions of infringement in the generative AI era and difficulties in cross-jurisdictional protection. Regulatory bodies must address these issues by collaborating with industry stakeholders to update requirements and promote international treaties.
To enhance the effectiveness of \toolname, song owners should minimize unprotected songs. For individual-covered songs, they need to conduct regular inspections and monitoring on various platforms to combat unlicensed covers. 
For songs released before \toolname, they should ``patch'' them on all controlled platforms and fight piracy and the secondary distribution that leads to ``unpatchable'' songs.}

\noindent {\bf Impact on authorized SVC.}
\toolname does not hinder authorized SVC, as song owners can maintain both unaltered and perturbed versions of a song, providing the unaltered version to authorized SVC entities. 
To prevent leaks of unaltered songs that could undermine \toolname, 
song owners can embed entity-specific watermarks into songs for traceability, 
allowing them to identify the source of leakage and seek compensation.

%% file: conclusion.tex
\section{Conclusion}
We have proposed \toolname, the first proactive approach that can be utlized by song owners to mitigate SVC-based illegal song covers for protecting their copyright and singers' civil rights. \mbox{\toolname} features a dual prevention, preventing singing voices from being used as the source and target singing voices in SVC, by perturbing singing voices prior to their release 
with a gender-transformation loss and a high/low hierarchy multi-target loss; preserves the quality of singing voices with a refined 
simultaneous masking 
loss; exhibits strong transferability to unknown SVC models with a transferability enhancement loss and encoder ensemble; 
and possesses robustness in over-the-air scenario and against adaptive adversaries. 
We make the first significant step towards coping with illegal automated song covers. 
Our open-source code, audio samples, and discussions on future works can foster researchers in exploring this direction further.

%% file: appendix.tex
\appendix

\begin{table*}
  \centering\setlength\tabcolsep{1pt}  
   \caption{Comparison between \toolname and related works}
    \resizebox{1\textwidth}{!}{
    \begin{threeparttable}
    \begin{tabular}{|c|c|c|c|c|c|}
    \hline
          & \textbf{Target Model} & \textbf{Purpose} & \textbf{Harmlessness} & {\bf Transfer $\uparrow$} & \textbf{Application} \\
    \hline
    \textbf{Unlearnable}~\cite{Unlearnable} & \multirow{2}{*}{\makecell[c]{image recognition \\ {(image$^\sharp$ \& D$^\natural$)}}} & \multirow{2}{*}{\makecell[c]{making data \\ unlearnable}} & \multirow{5}[2]{*}{$L_\infty$ norm} & \multirow{9}[3]{*}{\xmark} & \multirow{2}{*}{\makecell[c]{preventing unauthorized data \\ exploitation for training}} \\ \cline{1-1}
    \textbf{Robust Unlearnable}~\cite{Robust_Unlearnable} &  &  &  &  & \\ \cline{1-3}\cline{6-6}
    \textbf{$^\S$Glaze}~\cite{Glaze} & \multirow{2}{*}{\makecell[c]{text-to-image \\ {(image \& G$^\ddag$)}}} & {style} &  & & \multirow{2}{*}{\makecell[c]{protecting copyrights \\ of artworks}} \\
\cline{1-1}    \textbf{$^\S$MIST}~\cite{MIST} &       &  disruption &    &   &       \\
\cline{1-3}\cline{6-6}    \textbf{UnGANable}~\cite{UnGANable} & \makecell[c]{GAN-based face manipulator \\ {(image \& G)}} & \multirow{6}{*}{\makecell[c]{identity \\ disruption}} &   &     & \multirow{6}[3]{*}{\makecell[c]{preventing abuse \\ of biometric data}} \\
\cline{1-2}\cline{4-4}    \textbf{V-cloak}~\cite{V-Cloak} & \multirow{2}{*}{\makecell[c]{speaker  recognition \\ {(voice$^\dag$ \& D)}}} &       & {psychoacoustics model} &    &    \\
\cline{1-1}\cline{4-4}    \textbf{VoiceCloak}~\cite{VoiceCloak} &      &       & \multirow{3}{*}{$L_\infty$ norm} &      &  \\
\cline{1-2}    \textbf{$^*$AttackVC}~\cite{Attack-VC} & \multirow{3}{*}{\makecell[c]{speech voice \\ conversion/synthesis \\ {(voice \& G)}}} &       &       &   &     \\
\cline{1-1}    \textbf{$^*$VSMask}~\cite{VSMask} &     &       &       &  & \\
    \cline{1-1}\cline{4-5} \textbf{$^*$AntiFake}~\cite{AntiFake} &  & & \makecell[c]{frequency penalty \& SNR} & encoder ensemble & \\
    \hline
    \makecell[c]{{\bf Our work} \\ (\textbf{\toolname})} & \makecell[c]{singing voice conversion \\ {(voice \& G)}} & \makecell[c]{identity disruption \\ and \\ lyric disruption} & \makecell[c]{psychoacoustics model\\ (with backing tracks)}  & \makecell[c]{FL-IR loss \\ and  \\ encoder ensemble} & 
    \makecell[c]{protecting rights of\\ singers and lyrics (direct), \\ and melodies (indirect)}
    \\ \hline
    \end{tabular}
     \begin{tablenotes}
    \item {(1) ``Transfer $\uparrow$'': transferability enhancement; image$^\sharp$/voice$^\dag$: image/voice modality; D$^\natural$/G$^\ddag$: discriminative/generative models. 
    (2) $\S$/$*$: achieving purposes via artist/speaker style transfer, which is analogous to each other, 
    so their prevention techniques are generally the same, involving pulling artist or speaker style features towards targets. 
    (3) We also experimentally compare \toolname with the closet works AttackVC and AntiFake, all of which are of voice modality and target generative models (VSMask is not considered since it is unavailable).}
    \end{tablenotes}
    \end{threeparttable}
    }
  \label{tab:related_work}
\end{table*}

\subsection{Summary of Related Works that Exploiting Adversarial Examples for Good}\label{sec:summary_works}
\tablename~\ref{tab:related_work} summarizes related works that exploit adversarial examples for beneficial purposes, comparing prior works with our proposed \toolname. More detailed discussion refers to \cref{sec:adver_example}. 
Remarkably, we use the term ``harmlessness'' instead of ``imperceptibility'' in this work when exploiting adversarial examples for good.  This is because although protected samples (e.g., voices, images) are imperceptible 
to the adversary when they have not been used by the adversary in conversion/synthesis/training,
the adversary will largely become aware of the protection/prevention when the unexpected output is produced by conversion/synthesis/training.

\subsection{{Loss Convergence Analysis}}\label{sec:convergence_analysis}
{There is little theoretical understanding of the non-convex optimization problems involved in deep learning~\cite{dl_book}, 
hindering us from providing theoretical analysis or proof of the convergence of our optimization. 
Therefore, we instead analyze it empirically in the following two reasonable ways.}

\begin{figure*}
    \centering
    \begin{subfigure}{0.3\textwidth}
    \centering
        \includegraphics[width=.96\textwidth]{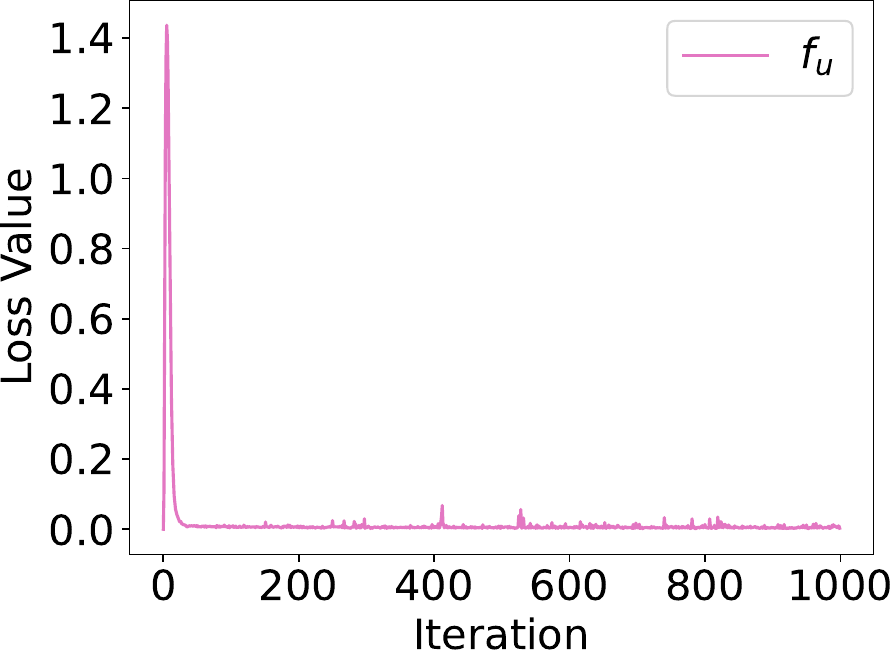}
        \caption{Utility loss $f_u$}
    \end{subfigure}
         \hspace{-0.2cm} 
    \begin{subfigure}{0.3\textwidth}
    \centering
        \includegraphics[width=1\textwidth]{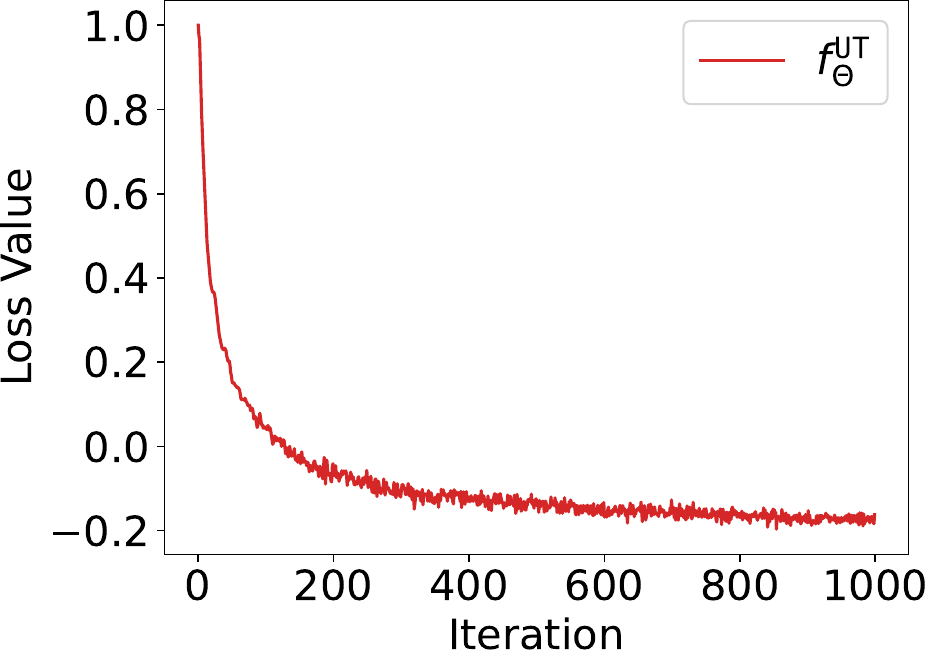}
        \caption{Untargeted identity disruption loss $f_\Theta^{\tt UT}$}
    \end{subfigure}
         \hspace{-0.2cm} 
    \begin{subfigure}{0.3\textwidth}
    \centering
        \includegraphics[width=1\textwidth]{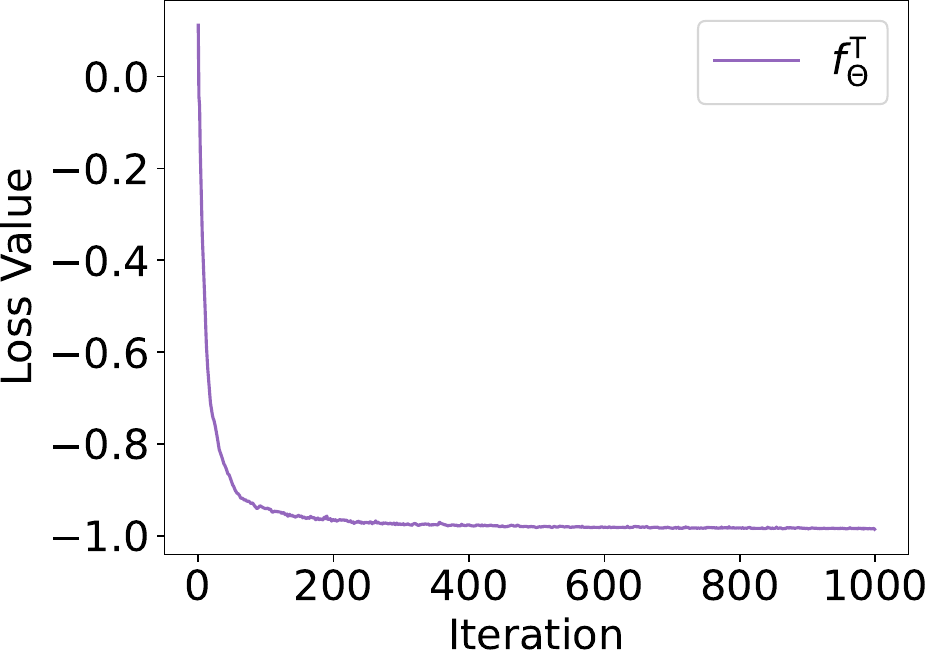}
        \caption{Targeted identity disruption loss $f_\Theta^{\tt T}$}
    \end{subfigure}
         \hspace{-0.2cm} 
     \begin{subfigure}{0.235\textwidth}
     \centering
        \includegraphics[width=1\textwidth]{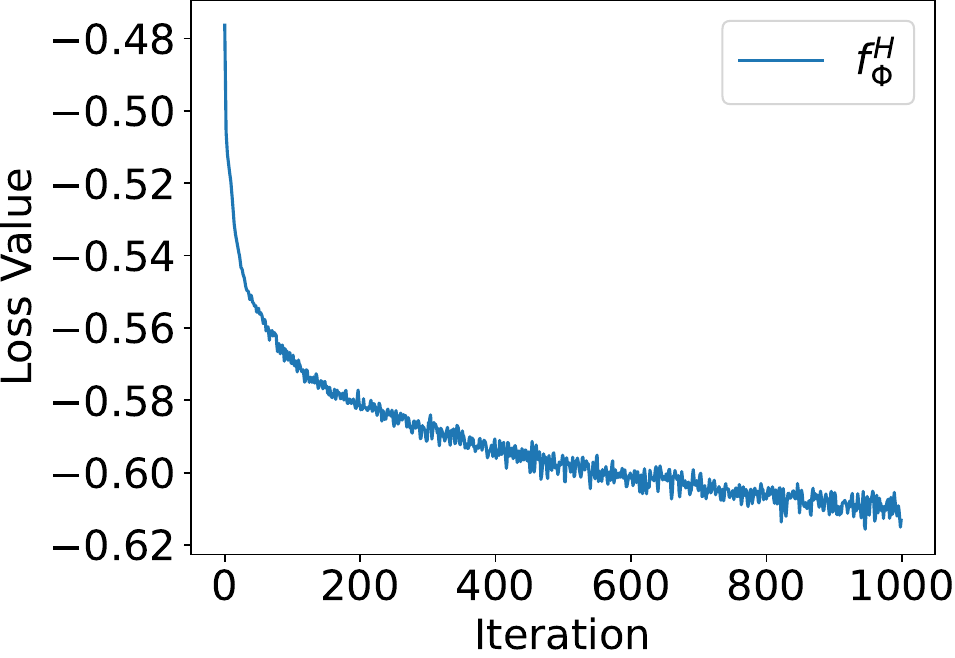}
        \caption{High-hierarchy lyric\\ disruption loss $f_\Phi^{H}$}
    \end{subfigure}
     \begin{subfigure}{0.22\textwidth}
     \centering
        \includegraphics[width=1\textwidth]{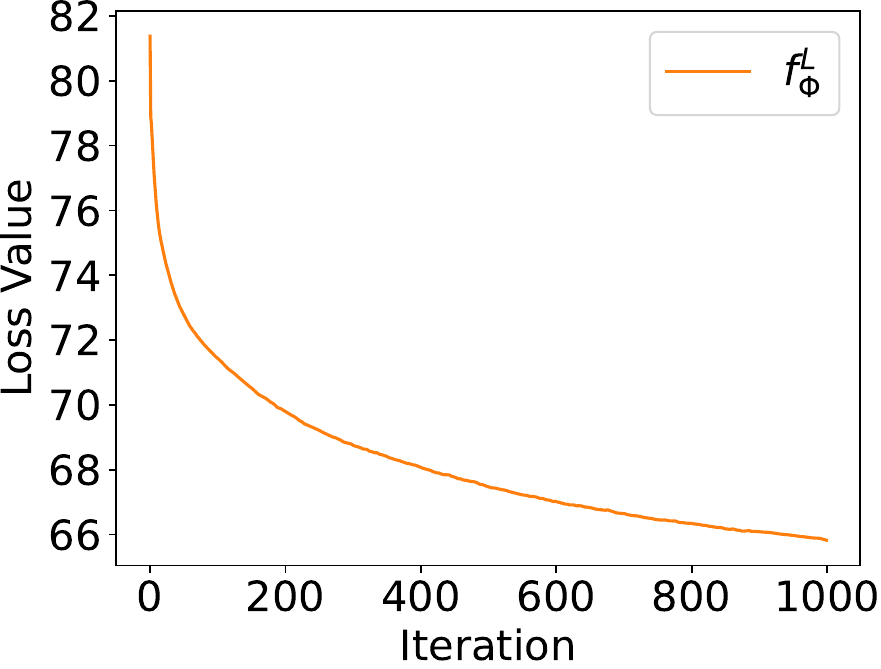}
         \caption{Low-hierarchy lyric\\ disruption loss $f_\Phi^{L}$}
    \end{subfigure}
     \begin{subfigure}{0.235\textwidth}
     \centering
        \includegraphics[width=1\textwidth]{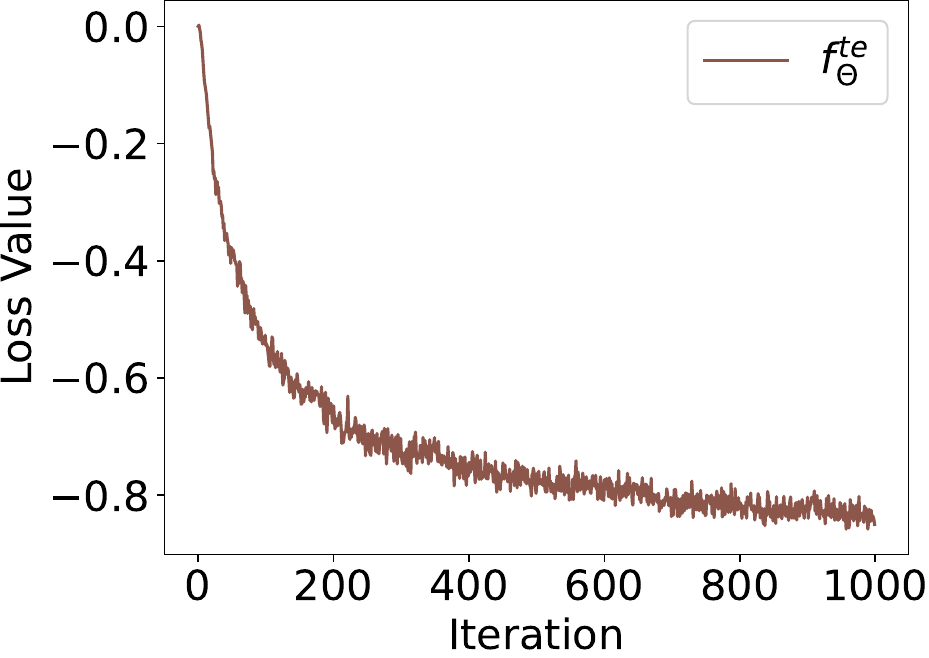}
        \caption{FL-IR loss $f_\Theta^{te}$ for  identity\\ disruption}
    \end{subfigure}
     \begin{subfigure}{0.24\textwidth}
     \centering
         \includegraphics[width=1\textwidth]{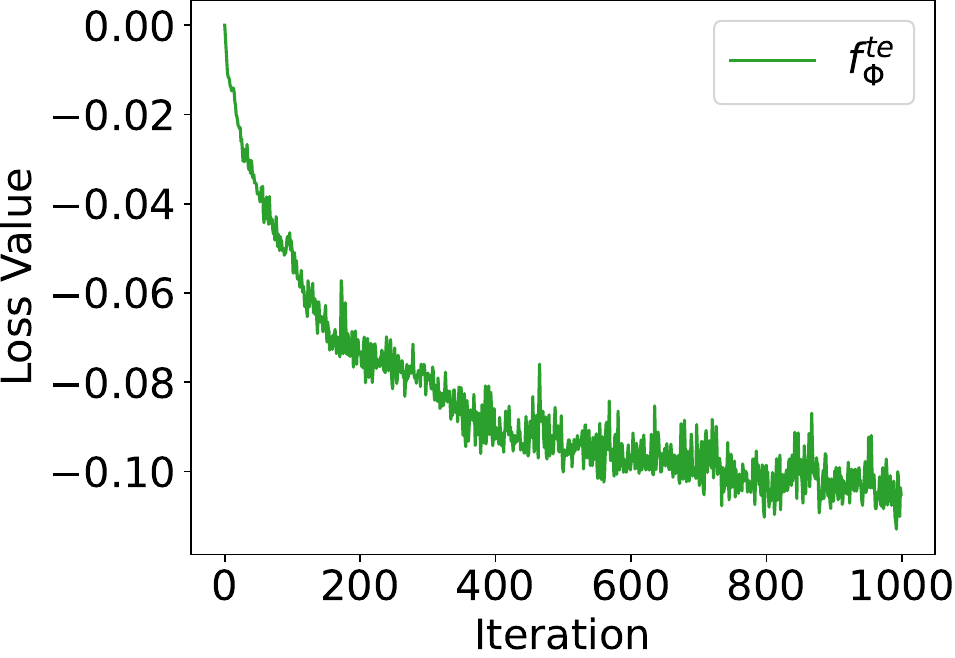}
        \caption{FL-IR Loss $f_\Phi^{te}$ for lyric\\ disruption}
    \end{subfigure}
    \caption{{The loss value w.r.t. the iteration during optimization. Remark that the optimal value of the utility loss $f_u$ is ``0''.}}
    \label{fig:loss_vs_iter}
\end{figure*}

\noindent {\bf Illustration.} 
{Following previous works~\cite{AS2T, QFA2SR, SpeakerGuard, UnGANable, Unlearnable, Robust_Unlearnable}, 
we empirically analyze the convergence by illustrating each loss’s change over iteration in \figurename~\ref{fig:loss_vs_iter}.  
It is easy to observe that each of the seven losses gradually converges. 
We attribute this to the better convergence property of Adam equipped with an adaptive learning rate compared to some other optimizers~\cite{ADAM, AS2T, CW}. 
We interestingly find that the utility loss $f_u$ increases suddenly at the beginning and then shortly starts decreasing close to the ideal value ``0''. 
This is reasonable since initially, all losses are large, so it is difficult to fix $f_u$ while minimizing other losses. Instead, the optimization may first sacrifice $f_u$ by allowing a large perturbation to minimize other losses. When the other losses decrease to some extent, it is possible to further reduce them with smaller perturbations, hence $f_u$ also decreases. }

\noindent {\bf Loss change.} 
{The computation of our loss goes through neural networks. 
The nonlinearity of neural networks causes neural network-based losses to become non-convex,
introducing challenges to optimize them~\cite{dl_book}. 
Therefore, it is a common practice to settle for finding and accepting a solution 
as long as it corresponds to a value of loss that is low enough for the focused task, 
instead of struggling for strict convergence~\cite{dl_book}. 
Inspired by this, we report the change of each loss between the start and the end of iterations in \tablename~\ref{tab:loss_change}. 
Except for the utility loss $f_u$ with the ideal value of  ``0'', the other losses decrease obviously. We attribute this to the adopted normalization-based dynamic loss balance strategy (cf.~\Cref{sec:final_app}).}

\begin{table}[]
    \centering\setlength\tabcolsep{4pt}
    \caption{{The average of the change of each loss between the start and the end of iterations.}} 
    \scalebox{1.}{
    \begin{tabular}{c|c|c|c|c|c|c|c}
    \hline
         & \boldmath{}$f_\Theta^{\tt UT}$\unboldmath{} & \boldmath{}$f_\Theta^{\tt T}$\unboldmath{} & \boldmath{}$f_\Phi^{H}$\unboldmath{} & \boldmath{}$f_\Phi^{L}$\unboldmath{} & \boldmath{}$f_u$\unboldmath{} 
         & \boldmath{}$f_\Theta^{te}$\unboldmath{} & \boldmath{}$f_\Phi^{te}$\unboldmath{} \\ \hline
         {\bf Start} & 1.0 & 0.05 & -0.466 & 94.734 & 0 & 0 & 0 \\ \hline
         {\bf End} &  -0.277 & -0.978 & -0.669 & 78.04 & 0.056 & -1.163 & -0.153 \\ \hline
         {\bf Decrease} & 1.277 & 1.029 & 0.202 & 16.693 & -0.056 & 1.163 & 0.153 \\ \hline
    \end{tabular}
    }
    \label{tab:loss_change}
\end{table}

\subsection{{Effectiveness of \toolname on the Non-Few-Shot SVC Model StarGANv2}}\label{sec:results_non_few_shot}
{
Recall that we focus on few-shot SVC models, which require much fewer computational resources and target singers' voices than non-few-shot models 
(cf. \Cref{sec:threa_model}), allowing for direct use by individuals without any training process and making them accessible to a broader range of adversaries, 
thus expanding the potential applications of \toolname.
We have confirmed the effectiveness of \toolname on four different few-shot SVC models. 
Here, we evaluate \toolname against adversaries who have sufficient singing voices of target singers and computational resources such that they can train from scratch or fine-tune SVC models with these singing voices. 
}

\begin{figure}[t]
    \centering
    \includegraphics[width=0.44\textwidth]{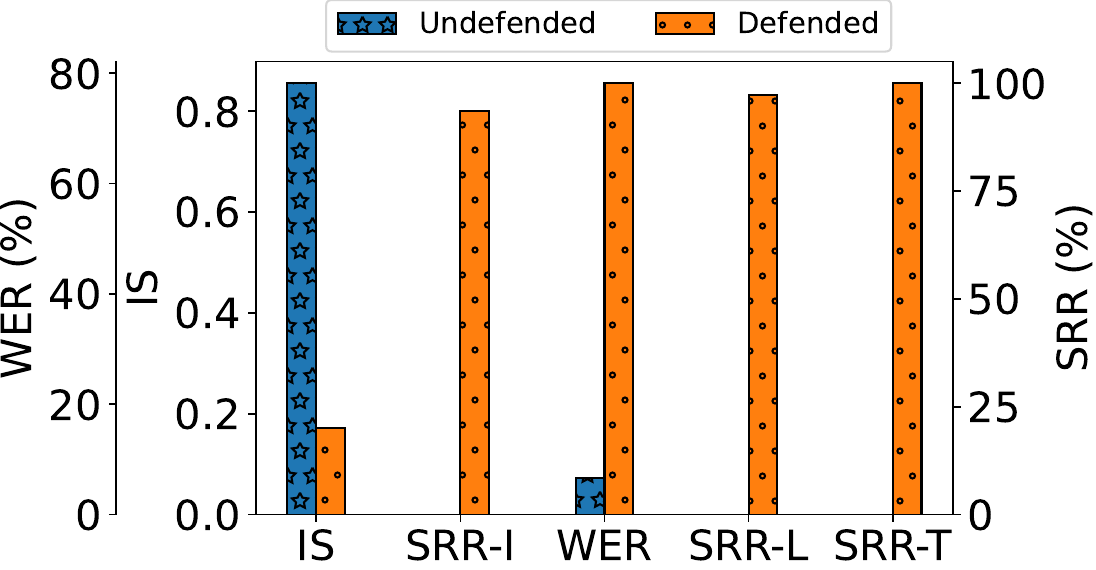}
    \caption{{The prevention effectiveness of \toolname against the non-few-shot model StarGANv2.}}
    \label{fig:exper_non_few_shot}
\end{figure}

{Specifically, we consider the promising non-few-shot model StarGANv2~\cite{StarGANv2_VC} 
(details refer to \tablename~\ref{tab:detail_svc_model}) 
and use its style encoder inference mode (in contrast to the mapping network mode). 
Although the released pre-trained model is trained only using ordinary voices, 
it exhibits emerging capacities including generalizing to singing conversion~\cite{StarGANv2_VC}. 
Therefore, we fine-tune the model respectively for each of the 12 singers in the English dataset NUS-48E instead of training from scratch. 
Note that StarGANv2 does not support Chinese.
Other experimental settings are the same as in \Cref{sec:steup}.}

{The results are shown in \figurename~\ref{fig:exper_non_few_shot}. 
The identity similarity (resp. lyric WER) of SVC-covered singing voices is much lower (resp. higher) than that of undefended output singing voices. 
Overall, \toolname is able to achieve nearly 100\% SRR-I, SRR-L, and SRR-T. 
These results demonstrate the effectiveness of \toolname against this non-few-shot model. 
We leave the evaluation of the prevention effectiveness of \toolname on more non-few-shot SVC models as future work. 
}

\begin{figure}[t]
    \centering
   \includegraphics[width=0.3\textwidth]{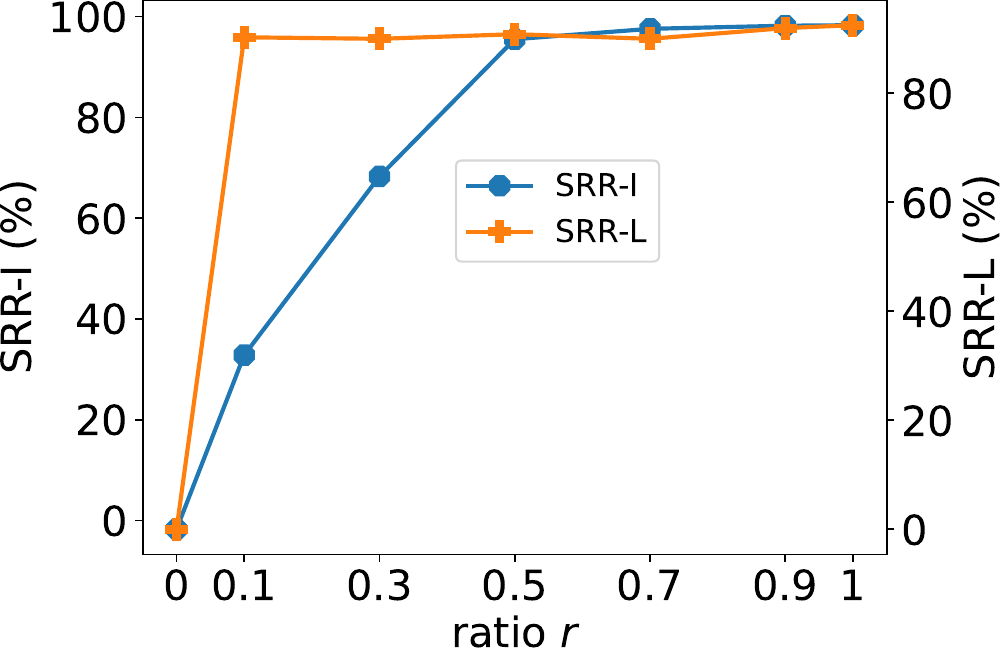}
\caption{Effectiveness of \toolname w.r.t. the ratio $r$ of the protected target singing voices, where $r=0$ disables \toolname.}
    \label{fig:ratio}
\end{figure} 
    
\subsection{Impact of the Ratio of Protected Target Singing Voices}
\label{sec:impact_ratio}
Recall that SVC models can take multiple target singing voices as input to better characterize the identity feature of the target singer. Previously,  all the 10 target singing voices are protected by \toolname for each target singer. Denote by $r$ the ratio between the number of protected target singing voices and the total number of target singing voices. Here we evaluate the impact of $r$ on the prevention effectiveness by setting $r=0,0.1,0.3,0.5,0.7,0.9$ and $1$. 

The results are depicted in \figurename~\ref{fig:ratio}. 
The ratio $r$ has no impact on the lyric disruption when $r>0$, as the lyric feature is extracted from the source singing voice. In contrast, the effectiveness of \toolname for identity disruption increases with the ratio. It is because the final identity feature is the centroid aggregation of all the target singing voices, and a large ratio is more likely to push the aggregated identity feature away from the original aggregated identity feature of the target singer. 

Remarkably, the SRR-I is always larger than 32\% when $r>0$,
demonstrating that \toolname can take effect even when only a small fraction of target singing voices are protected by \toolname. 

\begin{figure}[t] 
        \centering
    \begin{subfigure}{0.235\textwidth}
        \centering
        \includegraphics[width=0.98\textwidth]{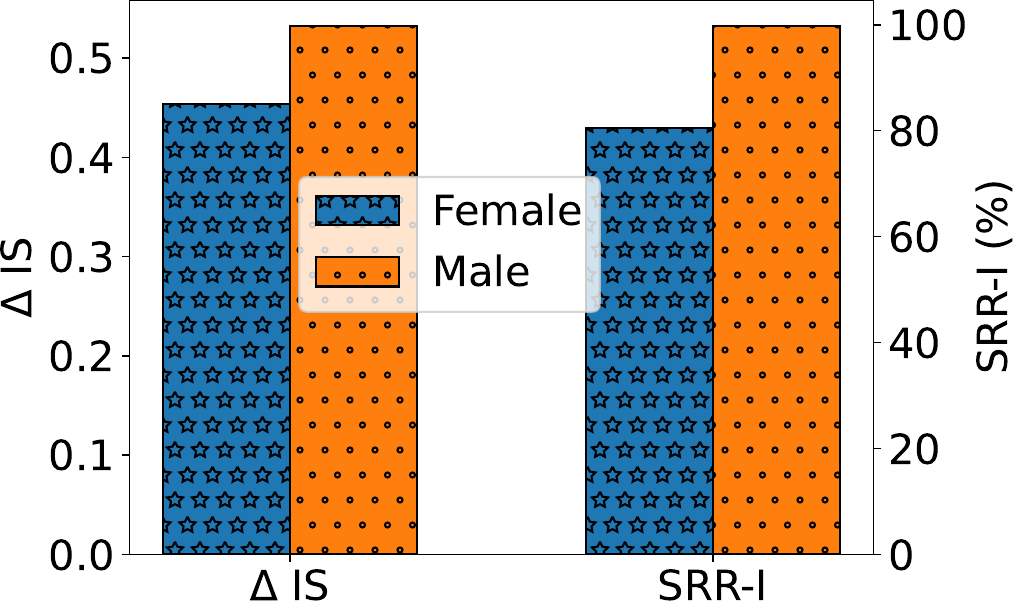}
        \caption{Identity disruption}
    \label{fig:gender_ID}
    \end{subfigure}
    \begin{subfigure}{0.235\textwidth}
        \centering
        \includegraphics[width=0.98\textwidth]{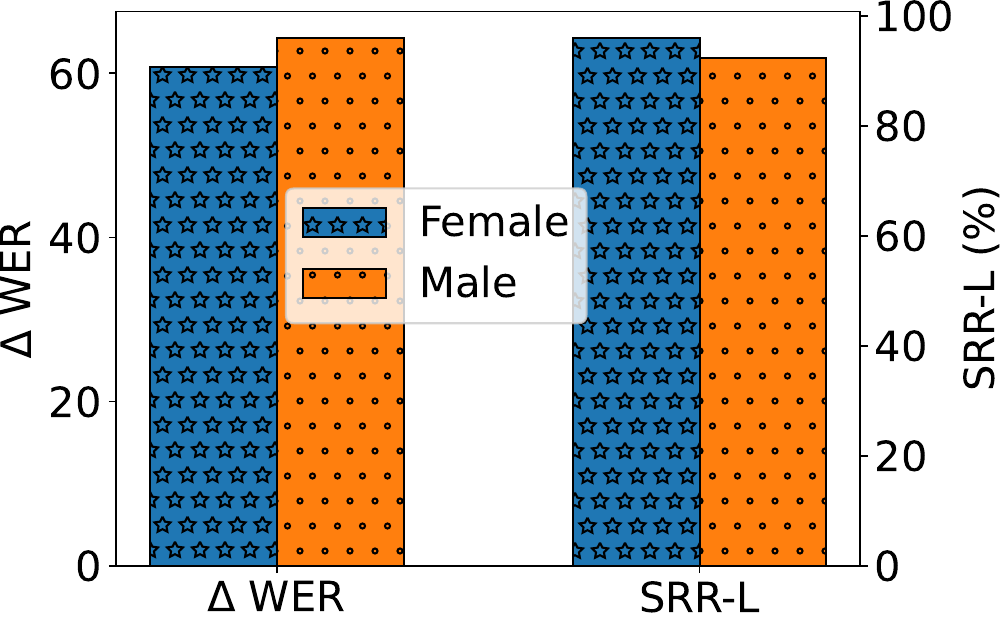}
        \caption{Lyric disruption}
    \label{fig:gender_LD}
    \end{subfigure}
    \caption{Effectiveness of \toolname w.r.t. singer gender.}
    \label{fig:gender}
\end{figure}

\begin{figure}[t]
\centering
    \begin{subfigure}{0.23\textwidth}
        \centering
        \includegraphics[width=1.05\textwidth]{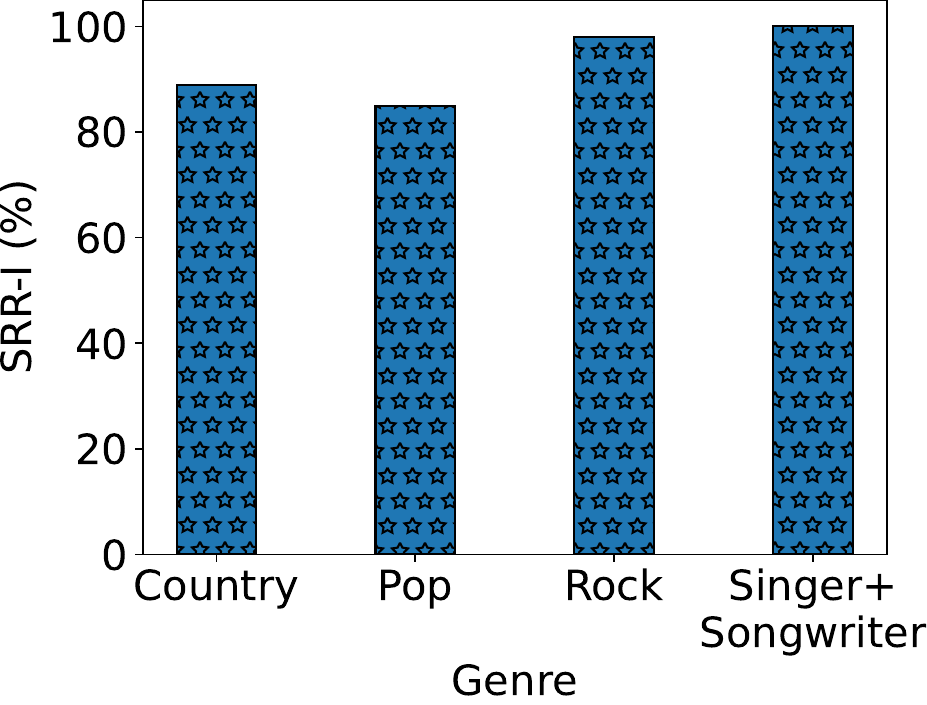}
        \caption{Identity disruption}
    \label{fig:genre_ID}
    \end{subfigure}
    \begin{subfigure}{0.23\textwidth}
        \centering
        \includegraphics[width=1\textwidth]{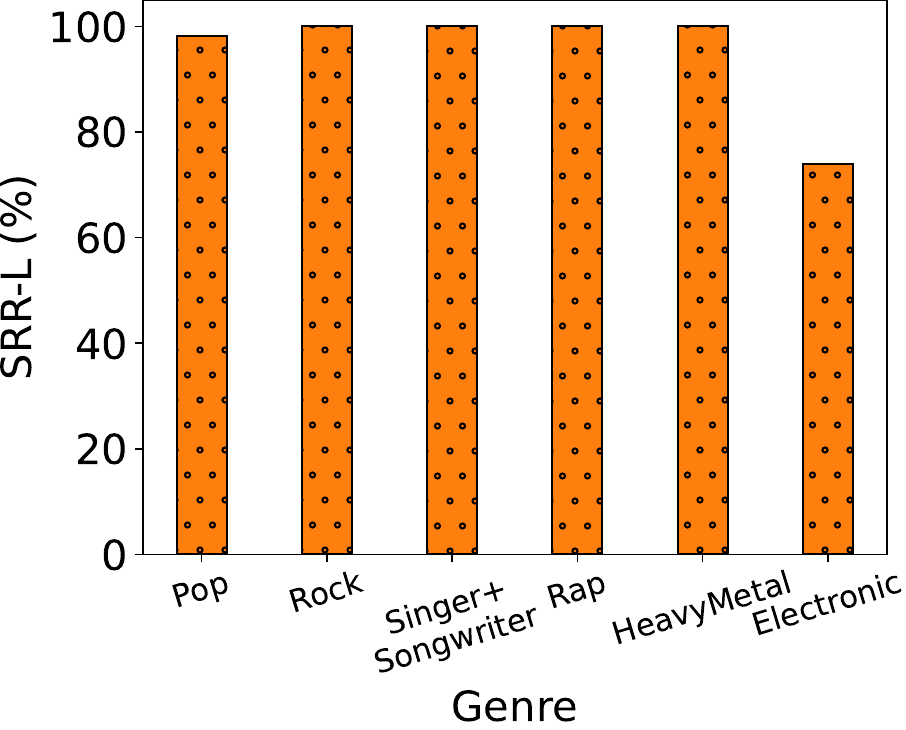}
        \caption{Lyric disruption}
    \label{fig:genre_LD}
    \end{subfigure}
    \caption{Effectiveness of \toolname w.r.t. song genres.}
    \label{fig:genre}
\end{figure}

\subsection{Effectiveness of \toolname w.r.t. Singer Gender}\label{sec:gender}
We compare the prevention effectiveness of \toolname between female and male singers. 
\figurename~\ref{fig:gender_ID} shows the SRR-I and the decrease of identity similarity (IS) compared to undefended SVC-covered singing voices
for identity disruption, while \figurename~\ref{fig:gender_LD} shows
the SRR-L and the increase of WER compared to undefended SVC-covered singing voices for lyric disruption. 

In general, \toolname achieves 
comparable 
SRR-I and SRR-L for
male singers and female singers, 
but exhibits larger changes of identity similarity and larger changes of WER for male singers than for female singers,
indicating that \toolname can better protect the civil rights of male target singers 
and better protect the copyrights of lyrics for source singing voices with male singers.  
Nevertheless, \toolname achieves 
over 80\% 
SRR-I and SRR-L for both genders, 
demonstrating the universality of \toolname for different genders. 

\subsection{Effectiveness of \toolname w.r.t. Song Genres}\label{sec:genre}
To evaluate the effectiveness of \toolname for protecting the copyrights of songs with different song genres, 
we choose songs from the dataset MUSDB18~\cite{musdb18, musdb18-hq}. Specifically, for identity disruption, we consider 4 song genres:  
{\sf Country, Pop, Rock,} and {\sf Singer+Songwriter},
while for lyric disruption, we consider 6 song genres: {\sf Pop, Rock, Singer+Songwriter, Electronic, HeavyMetal,} and {\sf Rap}, covering 7 song genres.
We do not consider all the 7 genres for each disruption type because the undefended SVC-covered singing voices have very low identity similarity or very high lyric WER under other not-considered genres. 

The results are shown in \figurename~\ref{fig:genre}. 
Regarding the identity disruption, \toolname reduces the SVC success rate by 85\%-100\%, 
with the lowest SRR-I for the {\sf Pop} song genre. 
Regarding the lyric disruption, \toolname achieves 74\% SRR-L for the 
{\sf Electronic} song genre and nearly 100\% SRR-L for the other song genres. 
These results demonstrate the universality of \toolname for songs with different song genres.

\subsection{Effectiveness of \toolname for Single Prevention}
\label{sec:individual_effectiveness}

We have showed that
\toolname is very effective in disrupting both the target singer's identity and the source lyrics
in SVC-covered singing voices when both
the source and target singing voices of SVC are protected by \toolname (cf.  \cref{sec:overall_dual}).
We now evaluate the effectiveness of \toolname for single prevention, namely, either the target singing voices (for identity disruption) or the source singing voices (for lyric disruption) are protected by \toolname, but not both.

\begin{figure*}\centering
\begin{subfigure}{0.71\textwidth}
        \centering
       \includegraphics[width=1\textwidth]{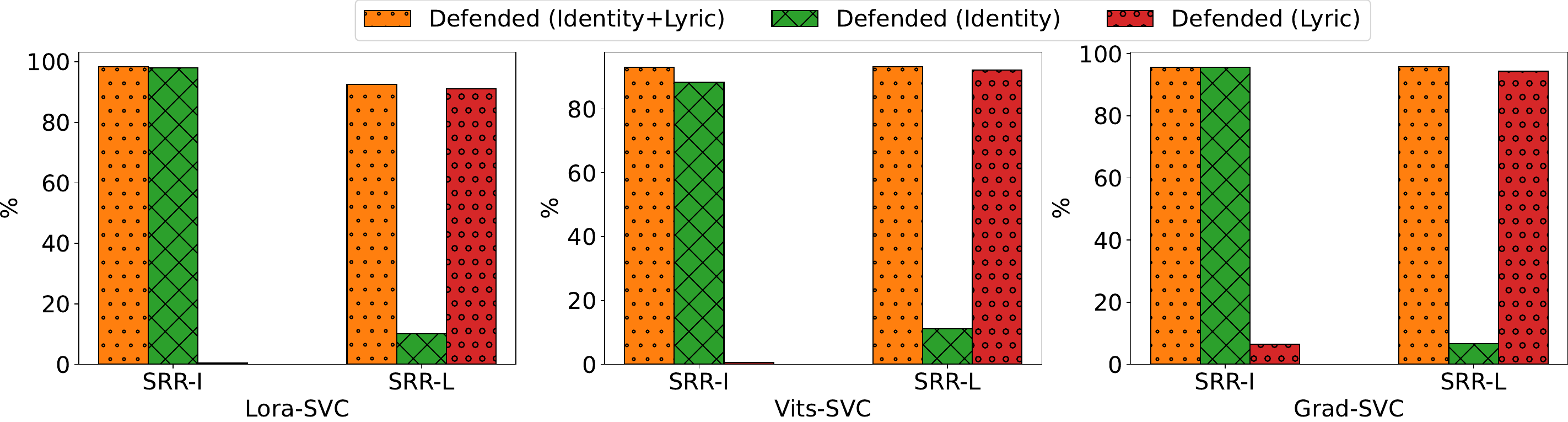} \vspace{-4mm}
\label{fig:ablation_dual_OpenSinger}
\caption{The dataset OpenSinger}
\end{subfigure}
\begin{subfigure}{0.71\textwidth}
        \centering
       \includegraphics[width=1\textwidth]{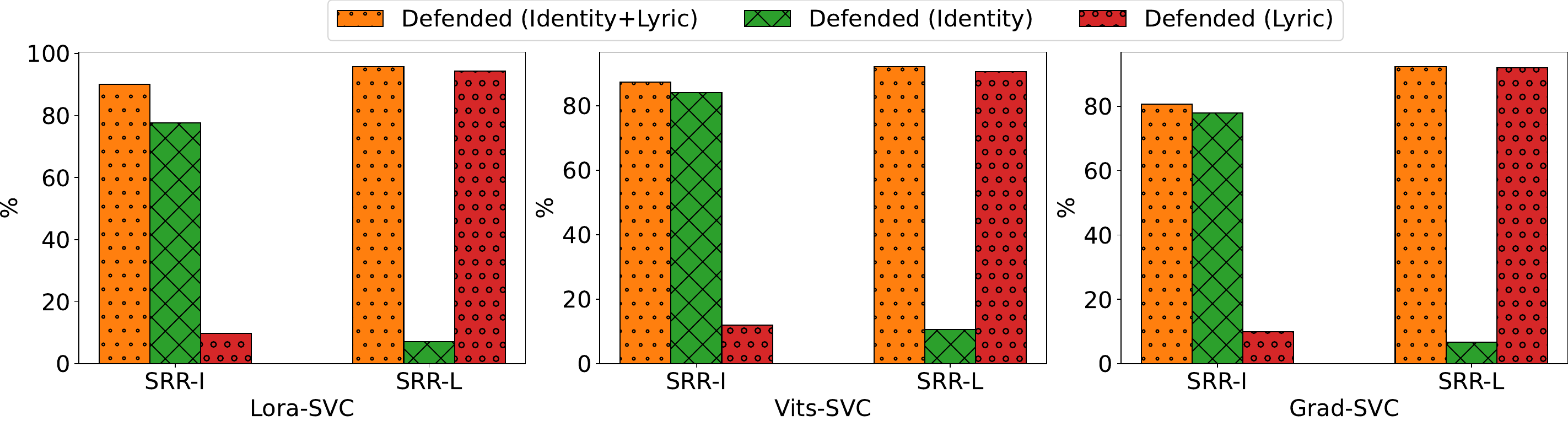} \vspace{-4mm}
\label{fig:ablation_dual_NUS_CMS_48} 
\caption{The dataset NUS-48E}
\end{subfigure}
\caption{Effectiveness of \toolname for single prevention in terms of SVC success reduction rate (SRR).} 
    \label{fig:ablation_dual_more}
\end{figure*}

The results in terms of SVC success reduction rate (SRR) are shown in \figurename~\ref{fig:ablation_dual_more}. 
For both Chinese and English datasets and 3 SVC models, when the target singing voices are protected by \toolname but the source singing voices are not, i.e., Defended (Identity), the SRR-I is close to that of Defended (Identity+Lyric) and the SRR-L is close to 0\%. Similarly, when the source singing voices are protected by \toolname but the target singing voices are not, i.e., Defended (Lyric), the SRR-L is close to that of Defended (Identity+Lyric) and the SRR-I is close to 0\%. 

\begin{figure*}\centering
\begin{subfigure}{0.72\textwidth}
        \centering
       \includegraphics[width=1\textwidth]{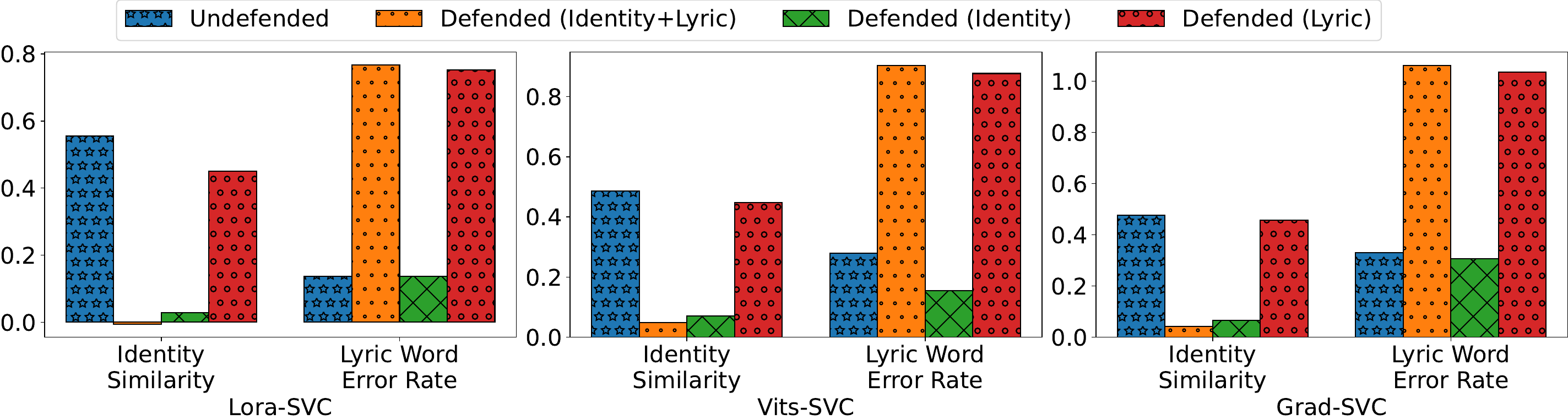}
\label{fig:ablation_dual_OpenSinger_2}\vspace{-4mm}
\caption{The dataset OpenSinger}
\end{subfigure}
\begin{subfigure}{0.72\textwidth}
        \centering
       \includegraphics[width=1\textwidth]{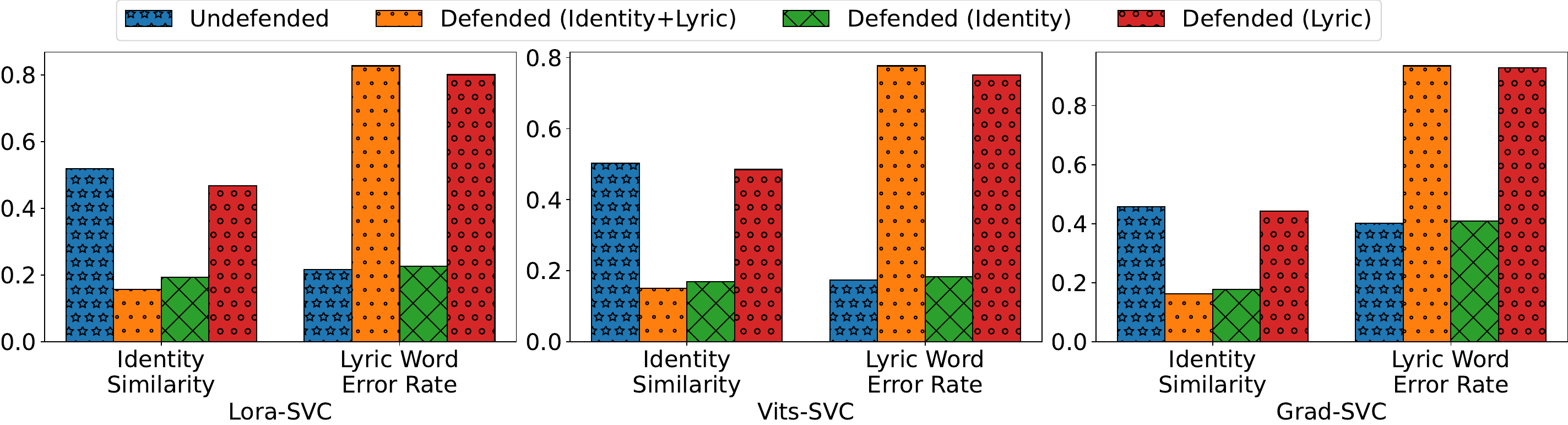}
\label{fig:ablation_dual_NUS_CMS_48_2}\vspace{-4mm}
\caption{The dataset NUS-48E}
\end{subfigure}
\caption{Effectiveness of \toolname for single prevention in terms of identity similarity and lyric WER.}
    \label{fig:ablation_dual_2}
\end{figure*}

The results in terms of identity similarity and lyric WER are shown in \figurename~\ref{fig:ablation_dual_2}. 
For both datasets and 3 SVC models, when the target singing voices are protected by \toolname but the source singing voices are not, i.e., Defended (Identity),
the identity similarity of output singing voices is close to that of Defended (Identity+Lyric), while the lyric WER is close to that of undefended output singing voices.
Similarly, when the source singing voices are protected by \toolname but the target singing voices are not, i.e., Defended (Lyric),  the lyric WER of output singing voices is close to that of Defended (Identity+Lyric), while the identity similarity is close to that of undefended output singing voices.

The above results confirm the effectiveness of \toolname for the single prevention, causing either identity or lyric disruption but not both, in all the three terms.

\subsection{Effectiveness of the Gender-transformation Loss $f_{\Theta}$ for Identity Disruption}\label{sec:exper_gt}

\begin{figure*}[t]
\centering
     \begin{minipage}{0.45\textwidth}
        \centering
     \includegraphics[width=.76\textwidth]{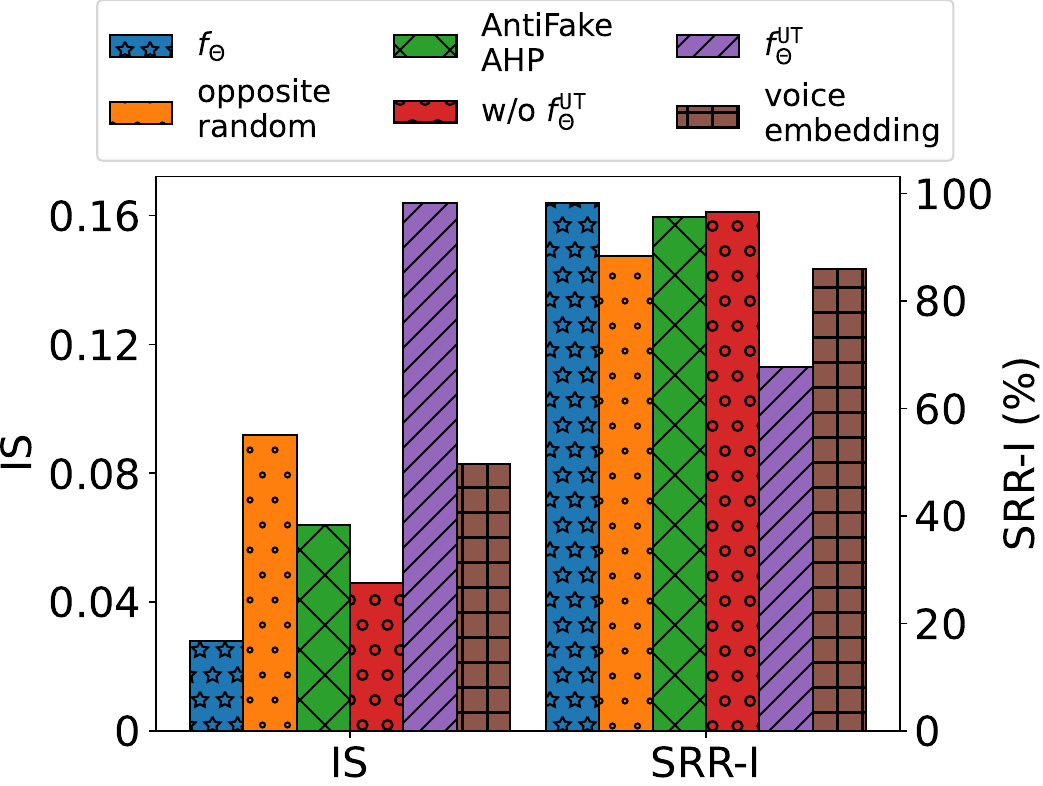}
    \caption{Effectiveness of $f_{\Theta}$ for identity disruption.}
    \label{fig:comp_gt_rd}
\end{minipage}
   \begin{minipage}{0.45\textwidth}
    \centering    
    \includegraphics[width=.76\textwidth]{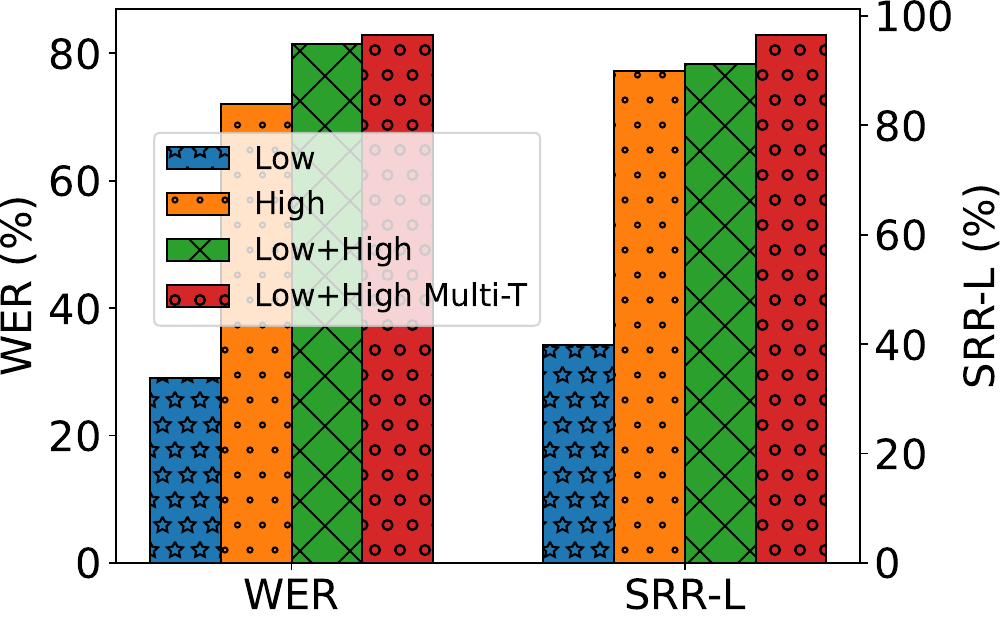}
    \caption{Effectiveness of $f_{\Phi}$ for lyric disruption.}
    \label{fig:low_high_multi_target}
\end{minipage}
\end{figure*} 

We evaluate the effectiveness of our gender-transformation loss $f_{\Theta}$ for causing identity disruption (cf.~\cref{sec:idloss}) by comparing the effectiveness of $f_\Theta$ with the following variants: 
(1) adopting the method of selecting destination speakers in AttackVC and AntiFake, i.e., random selection among opposite-gender speakers and AHP, 
respectively; (2) eliminating or only using the loss term $f_\Theta^{\tt UT}$; (3) adopting the method of representing the destination speaker in AttackVC and AntiFake, i.e., by a single voice embedding. 
The results are shown in \figurename~\ref{fig:transfer_comp}. 
These variants are strictly inferior to $f_\Theta$ in terms of both identity similarity and SRR-I, indicating the contribution of each component of $f_\Theta$ to identity disruption.  
This is probably because: (1) 
the method of \toolname for determining destination singers comprises both objective perception (i.e., least identity-similarity) and subjective perception (i.e., opposite-gender singers), which is more effective than the two variant methods; (2) the loss term $f_\Theta^{\tt UT}$ pushes the embedding of the protected singing voice away from the original embedding, improving the degree of embedding deviations; (3) Centroid embeddings can more precisely represent destination singers, thus contributing to a more accurate optimization direction. 

\subsection{Effectiveness of the Low/High Hierarchy Multi-target Loss $f_{\Phi}$ for Lyric Disruption}\label{sec:exper_low_high_multi_target}
We compare the effectiveness for lyric disruption among low hierarchy loss, high hierarchy loss, high/low hierarchy loss, and high/low hierarchy multi-target loss $f_{\Phi}$ (cf.~\cref{sec:lyric_disrupt_loss}). 

The results are shown in \figurename~\ref{fig:low_high_multi_target}. 
The high/low hierarchy (multi-target) loss achieves higher WER and SRR-L than both the low hierarchy loss and the high hierarchy loss, indicating the necessity and effectiveness of optimizing both the low-level acoustic features and high-level lyric features. The high/low hierarchy multi-target loss achieves the highest WER and SRR-L, indicating the effectiveness of using multiple target lyrics, i.e., multiple singing voices with distinct lyrics.

\subsection{Effectiveness of the Utility Loss $f_{u}$ for Imperceptibility}\label{sec:imper_exp}
We demonstrate the effectiveness of our utility loss $f_{u}$
by comparing SNR and PESQ of the protected singing voices 
using the basic utility loss and the refined one (cf.~\cref{sec:utilityloss}).
We compute the SNR and PESQ of the protected singing voices before and after adding the backing tracks. 
We note that the backing tracks are still used in optimization when the refined utility loss is used.

\begin{figure*}
    \centering  
\begin{subfigure}{0.48\textwidth}
       \includegraphics[width=.44\textwidth]{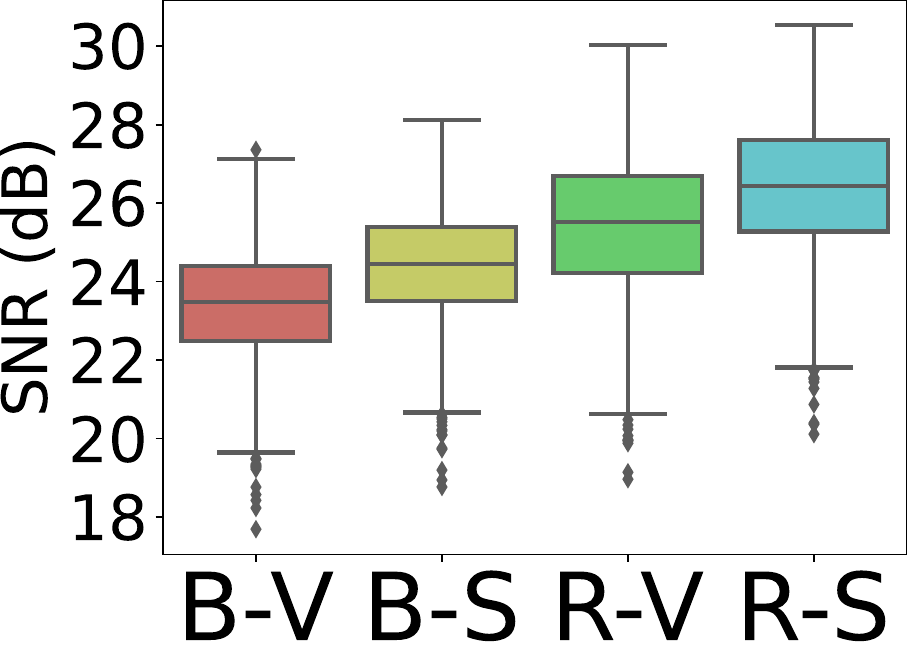}
       \includegraphics[width=.44\textwidth]{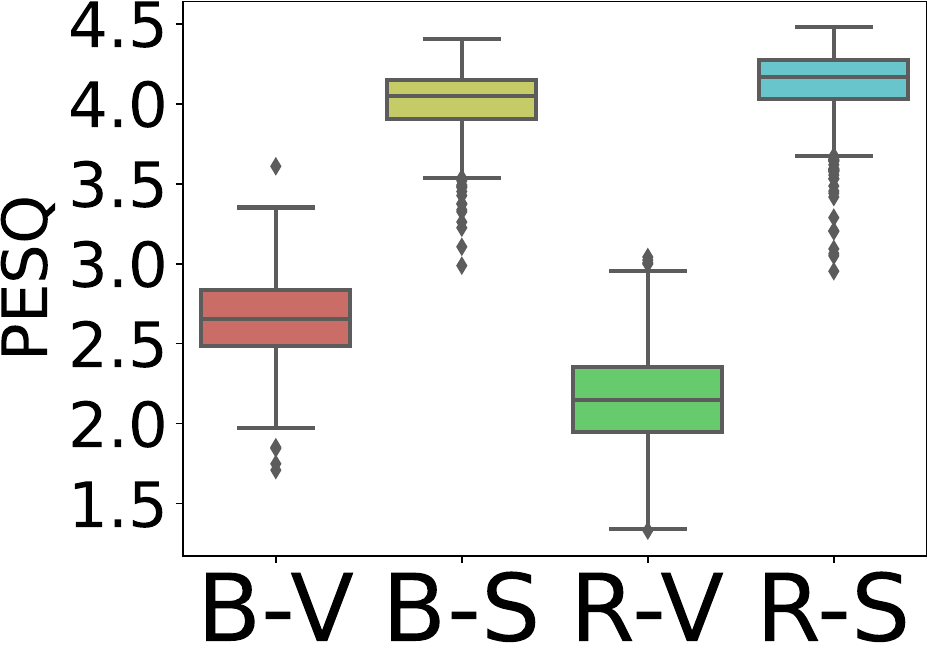}
\caption{Results on target singing voices.} 
    \label{fig:refined_imper_loss_target}
    \end{subfigure} 
\begin{subfigure}{0.48\textwidth}
      \centering
       \includegraphics[width=.43\textwidth]{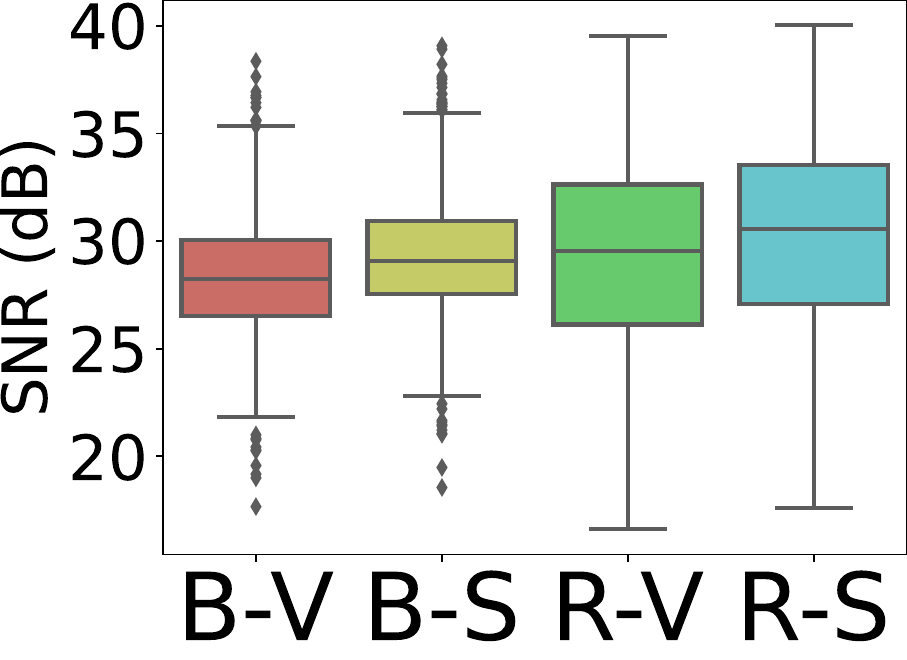}
       \includegraphics[width=.43\textwidth]{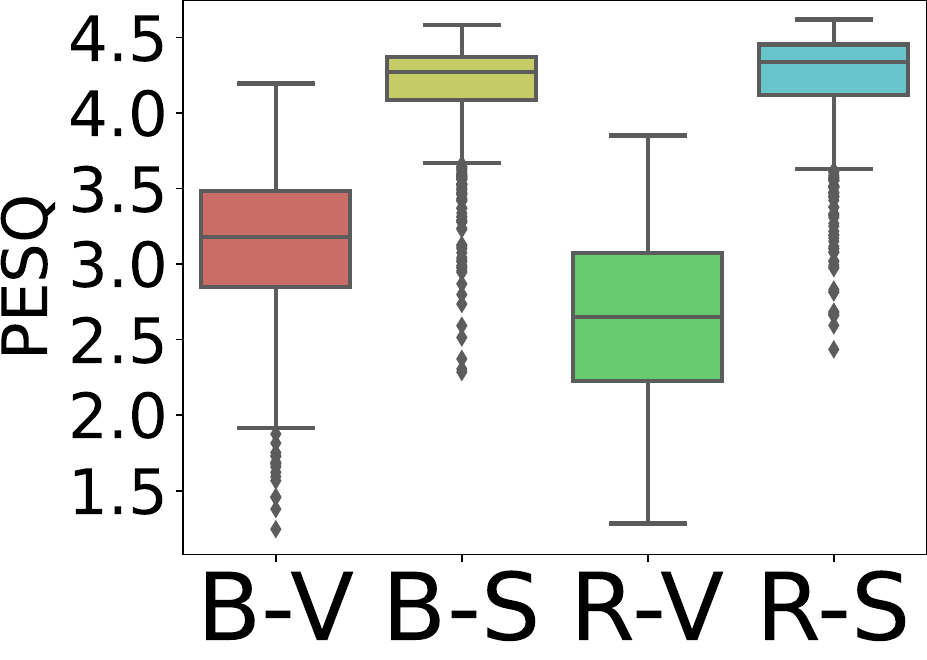}
\caption{Results on source singing voices.} 
    \label{fig:refined_imper_loss_source}
    \end{subfigure}
    \caption{Effectiveness of the utility loss $f_{u}$ for improving the harmlessness of the protected singing voices, where
``B-V'' and ``R-V'' denote the results of using the basic utility loss and refined utility loss, respectively. ``B-S'' and ``R-S'' are the results of songs obtained by adding backing tracks to singing voices of ``B-V'' and ``R-V'', respectively.} 
\end{figure*}

The results are depicted in \figurename~\ref{fig:refined_imper_loss_target} and \figurename~\ref{fig:refined_imper_loss_source} for protecting target singing
voices and source singing
voices, respectively. 
The SNR and PESQ of songs crafted by \toolname with the refined utility loss (i.e., ``R-S'') are higher than that of songs crafted by \toolname
with the basic utility loss (i.e., ``B-S''), indicating that the refined utility loss incorporating the backing track as an additional masker can better hide perturbations. We also find that the songs (i.e., ``R-S'') have much higher SNR and PESQ than the singing voices (i.e., ``R-V'') when both using the refined utility loss. For example, on target singing voices, ``R-V'' has a PESQ of 2.1 while ``R-S'' has a PESQ of over 4.0,
confirming the large capacity of backing tracks for hiding  perturbations;
and \toolname achieves a median SNR of 26.3 dB and a median PESQ of 4.1 (recall that the upper bound of PESQ is 4.5), showcasing the effectiveness of the refined utility loss. 

\begin{figure*}
    \centering 
\begin{subfigure}{0.4\textwidth}
        \centering
       \includegraphics[width=0.95\textwidth]{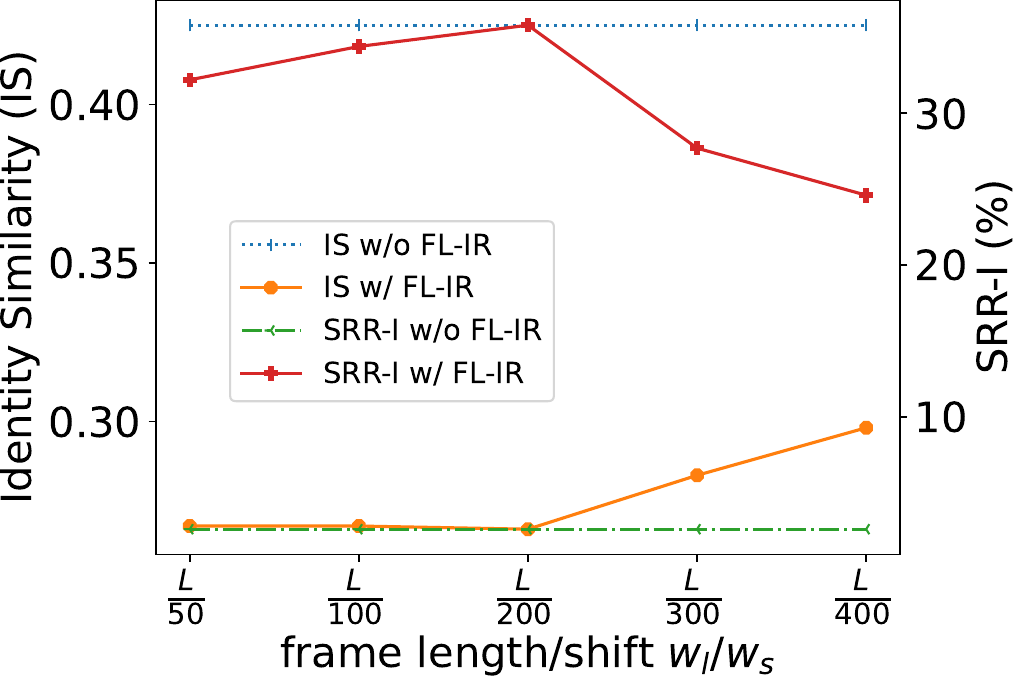} 
        \caption{XV}
        \label{fig:frame_length_xv}
\end{subfigure}
\qquad
\begin{subfigure}{0.4\textwidth}
        \centering
       \includegraphics[width=0.95\textwidth]{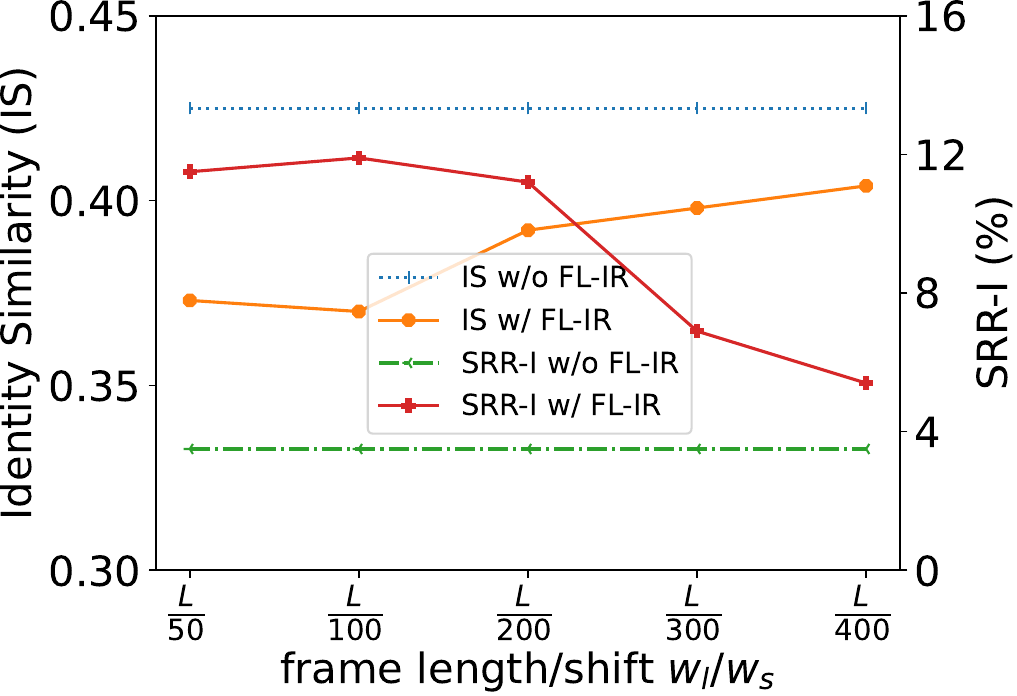}
        \caption{Res34-V}
        \label{fig:frame_length_res34_veri}
\end{subfigure}
\caption{Effectiveness of the FL-IR loss w.r.t. frame length.}
\label{fig:frame_length}
\end{figure*}

\begin{figure*}\centering
\begin{subfigure}{0.28\textwidth}
        \centering
       \includegraphics[width=.95\textwidth]{figure/transfer_identity_lora_svc_ens_srr_2.pdf}
\label{fig:transfer_identit_lora_svc}
\caption{Lora-SVC}
\end{subfigure}\quad
\begin{subfigure}{0.28\textwidth}
        \centering
       \includegraphics[width=.95\textwidth]{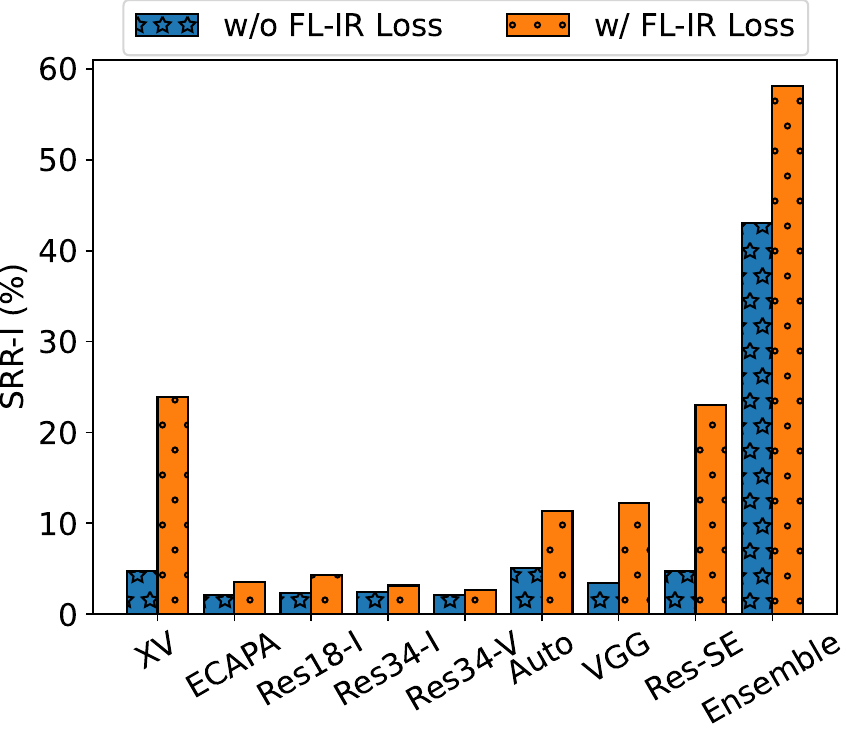}
\label{fig:transfer_identit_vits_svc}
\caption{Vits-SVC}
\end{subfigure}\quad
\begin{subfigure}{0.28\textwidth}
        \centering
       \includegraphics[width=.95\textwidth]{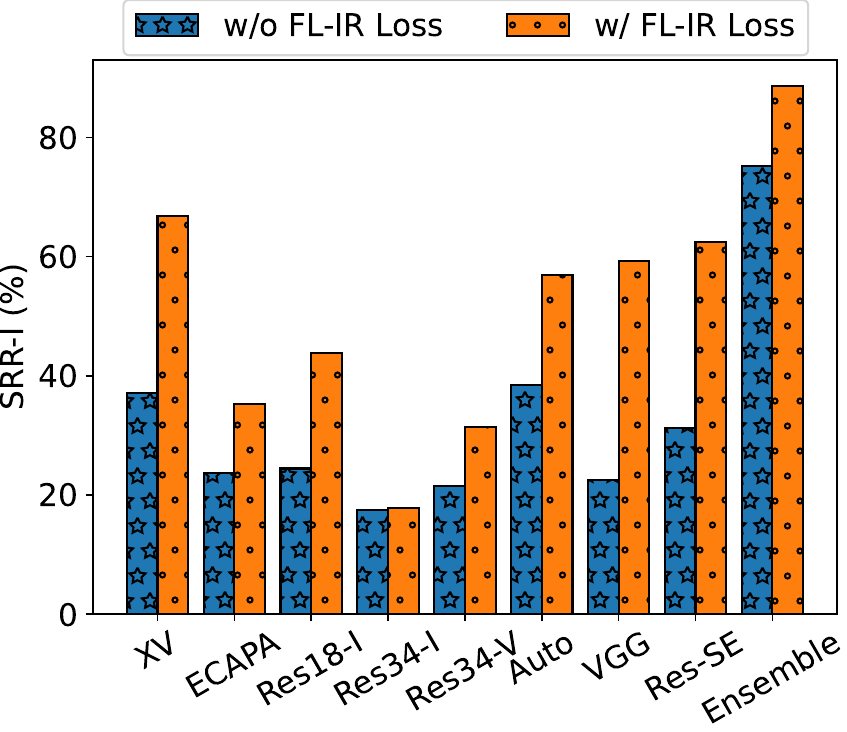}
\label{fig:transfer_identit_grad_svc}
\caption{Grad-SVC}
\end{subfigure}
\caption{Transferability of \toolname for causing identity disruption in terms of success reduction rate (SRR).  
}
    \label{fig:transfer_identity}
\end{figure*}

\begin{figure*}\centering
\begin{subfigure}{0.28\textwidth}
        \centering
       \includegraphics[width=.95\textwidth]{figure/transfer_lyric_lora_svc_ens_srr_2.pdf} 
\label{fig:transfer_lyric_lora_svc}
\caption{Lora-SVC}
\end{subfigure}\quad
\begin{subfigure}{0.28\textwidth}
        \centering
       \includegraphics[width=.95\textwidth]{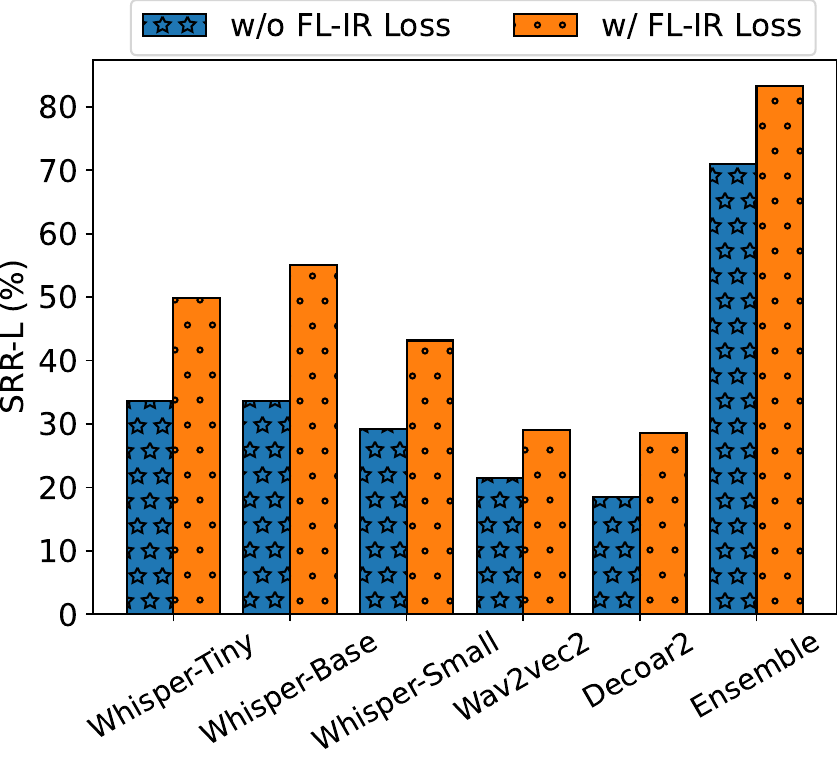}
\label{fig:transfer_lyric_vits_svc}
\caption{Vits-SVC}
\end{subfigure}\quad
\begin{subfigure}{0.28\textwidth}
        \centering
       \includegraphics[width=.95\textwidth]{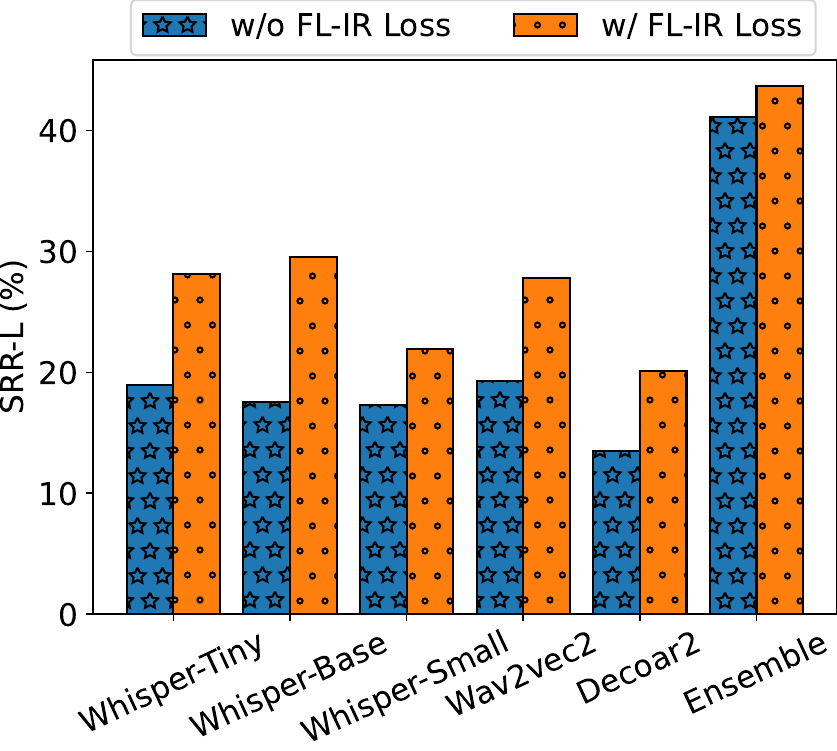}
\label{fig:transfer_lyric_grad_svc}
\caption{Grad-SVC}
\end{subfigure}
\caption{Transferability of \toolname for causing lyric disruption in terms of success reduction rate (SRR). 
}
    \label{fig:transfer_lyric}
\end{figure*}

\begin{figure*}\centering
\begin{subfigure}{0.28\textwidth}
        \centering
       \includegraphics[width=.95\textwidth]{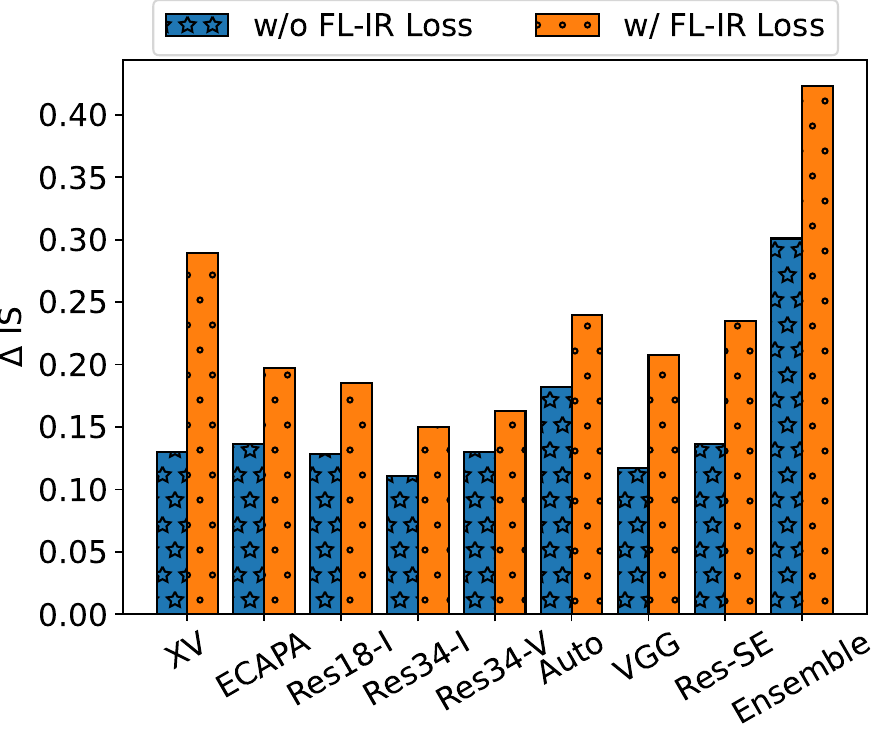}
\label{fig:transfer_identit_lora_svc_is}
\caption{Lora-SVC}
\end{subfigure}\quad
\begin{subfigure}{0.28\textwidth}
        \centering
       \includegraphics[width=.95\textwidth]{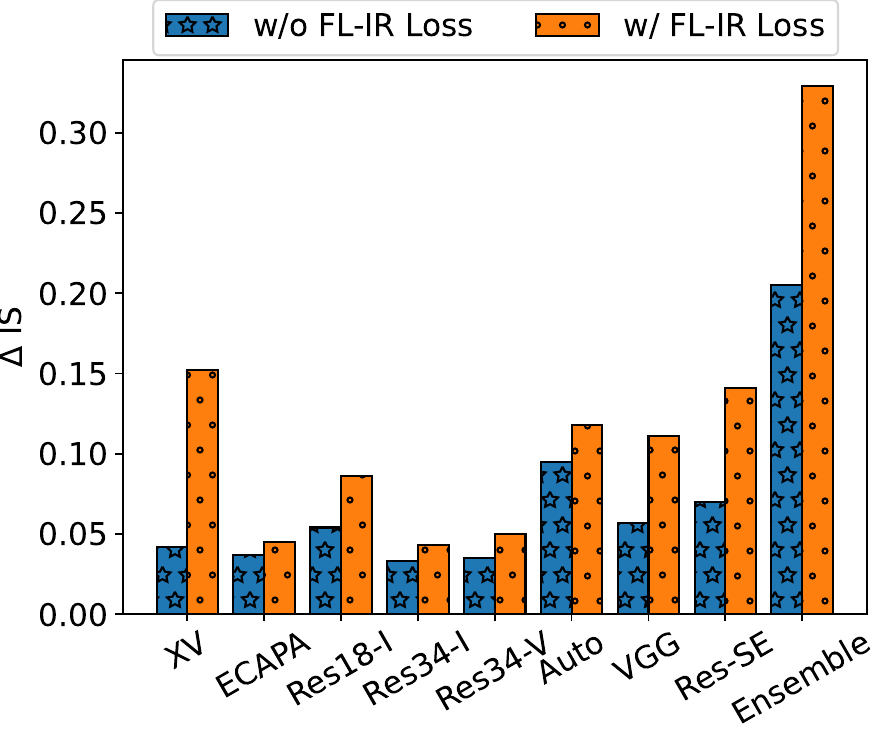}
\label{fig:transfer_identit_vits_svc_is}
\caption{Vits-SVC}
\end{subfigure}\quad
\begin{subfigure}{0.28\textwidth}
        \centering
       \includegraphics[width=.95\textwidth]{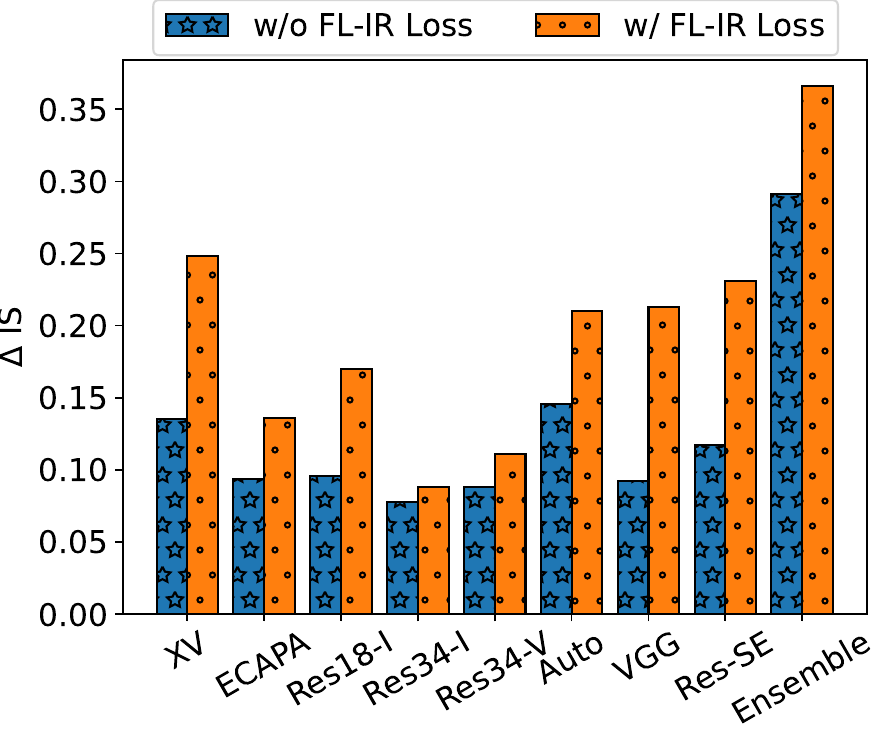}
\label{fig:transfer_identit_grad_svc_is}
\caption{Grad-SVC}
\end{subfigure}
\caption{Transferability of \toolname for causing identity disruption in terms of change of identity similarity ($\Delta$IS).  
}
    \label{fig:transfer_identity_is}
\end{figure*}

\begin{figure*}\centering
\begin{subfigure}{0.28\textwidth}
        \centering
       \includegraphics[width=.95\textwidth]{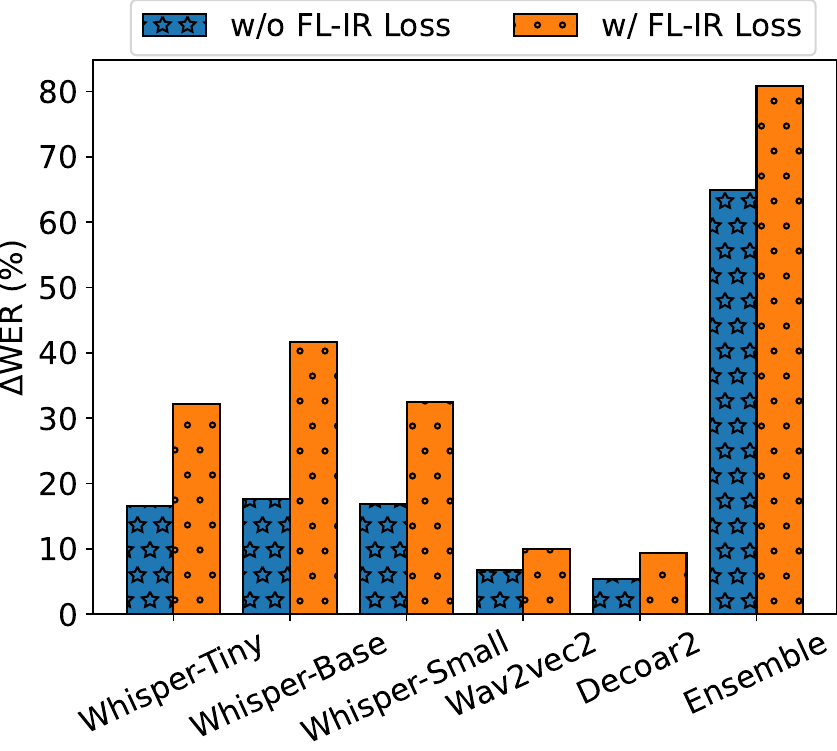} 
\label{fig:transfer_lyric_lora_svc_lwer}
\caption{Lora-SVC}
\end{subfigure}\quad
\begin{subfigure}{0.28\textwidth}
        \centering
       \includegraphics[width=.95\textwidth]{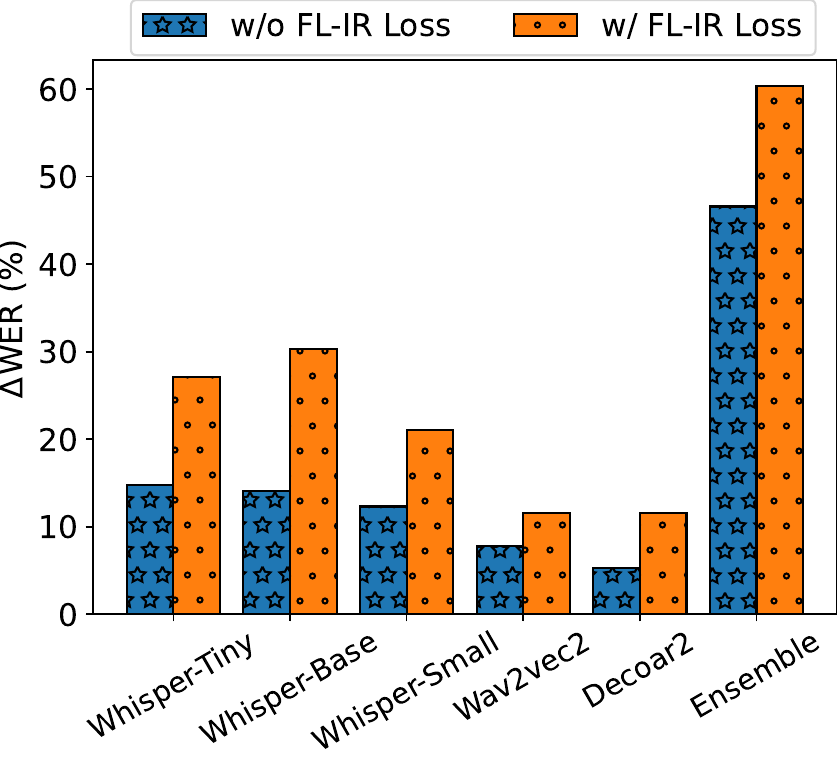}
\label{fig:transfer_lyric_vits_svc_lwer}
\caption{Vits-SVC}
\end{subfigure}\quad
\begin{subfigure}{0.28\textwidth}
        \centering
       \includegraphics[width=.95\textwidth]{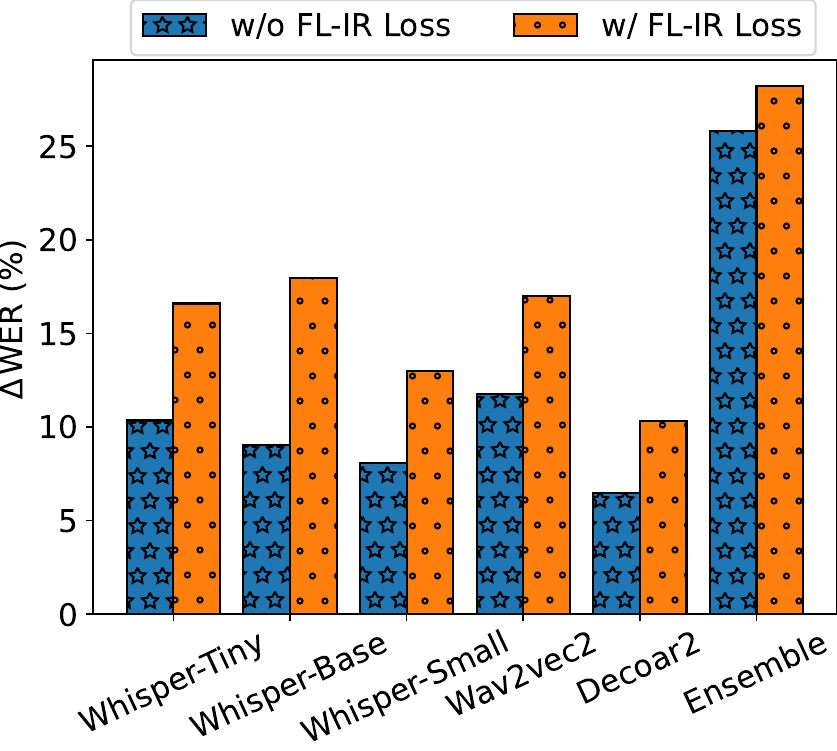}
\label{fig:transfer_lyric_grad_svc_lwer}
\caption{Grad-SVC}
\end{subfigure}
\caption{Transferability of \toolname for causing lyric disruption in terms of change of lyric word error rate ($\Delta$WER).  
}
    \label{fig:transfer_lyric_lwer}
\end{figure*}

\subsection{Impact of the Frame Length/Shift $w_l/w_s$ for Transferability Enhancement}\label{sec:frame_length_exper}
\toolname utilizes a frame-level interaction reduction-based (FL-IR) loss $f_{\Theta}^{te}$ and $f_{\Phi}^{te}$ to enhance the transferability to unknown encoders (cf.~\cref{sec:tr-enloss}). To evaluate the impact of the frame length/size $w_l/w_s$ in $f_{\Theta}^{te}$ and $f_{\Phi}^{te}$, we set $w_l$ and $w_s$ to $\frac{L}{50},\frac{L}{100},\frac{L}{200},\frac{L}{300}$ and $\frac{L}{400}$, where $L$ is the number of sample points of a singing voice. We demonstrate the impact on the transferability of causing identity disruption to the Lora-SVC model without applying our encode ensemble. (Recall that the effectiveness of our encode ensemble has been confirmed in~\cref{sec:exper_transfer}.)
We use XV and Res34-V for crafting protected target singing voices, the most and the least transferable identity encoders according to the results in \figurename~\ref{fig:transfer_identity}. 

The results are shown in \figurename~\ref{fig:frame_length}. 
We can observe that although the identity similarity and SRR vary with the frame length/size, they are significantly lower and higher than that of \toolname without using the 
FL-IR loss 
$f_{\Theta}^{te}$ and $f_{\Phi}^{te}$. In general, $\frac{L}{100}$ and $\frac{L}{200}$ are better
 than others, thus we use $\frac{L}{200}$ in the other experiments.
 
\subsection{More Results on the Transferability}\label{sec:more_results_transfer}
We have demonstrated the transferability of \toolname to the SVC model Lora-SVC in terms of the SVC success reduction rate (SRR) in \cref{sec:exper_transfer}. 
Here we report in 
\figurename~\ref{fig:transfer_identity}, \figurename~\ref{fig:transfer_lyric}, \figurename~\ref{fig:transfer_identity_is}, and \figurename~\ref{fig:transfer_lyric_lwer} 
more transferability results on all three SVC models in terms of identity similarity (IS), lyric word error rate (WER), SRR-I, SRR-L, and SRR-T. 
We can draw the same conclusion as in \cref{sec:exper_transfer}.

\begin{figure*}
\centering
\begin{subfigure}{0.325\textwidth}
        \centering
       \includegraphics[width=.95\textwidth,height=.65\textwidth]{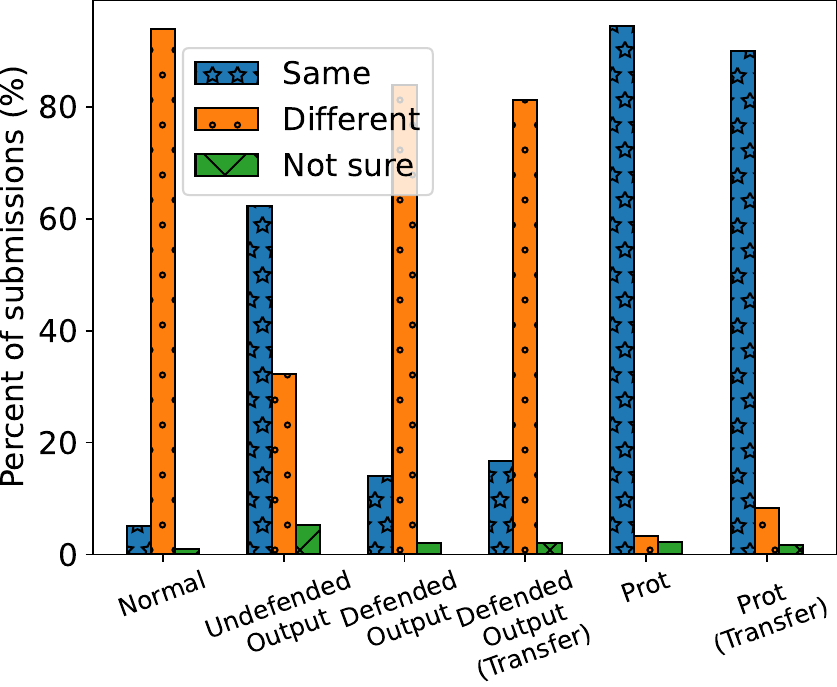}
\caption{Identify Singer}\label{fig:human_study_ID_en}
\end{subfigure}\quad
\begin{subfigure}{0.325\textwidth}
        \centering
\includegraphics[width=.9\textwidth,height=.65\textwidth]{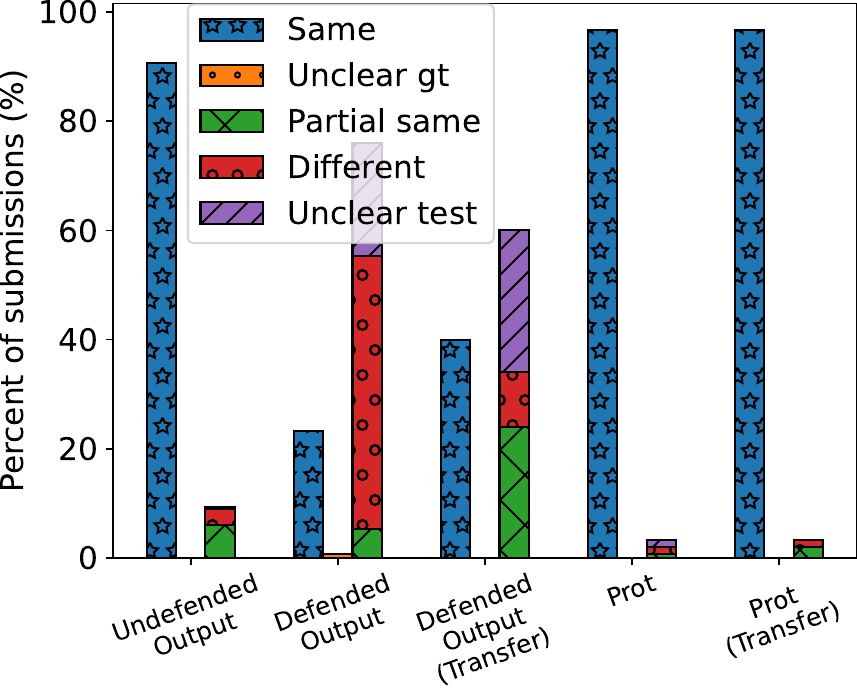}    
\caption{Identify Lyric}\label{fig:human_study_LD_en}
\end{subfigure}\quad
\begin{subfigure}{0.31\textwidth}
        \centering
       \includegraphics[width=.95\textwidth,height=.65\textwidth]{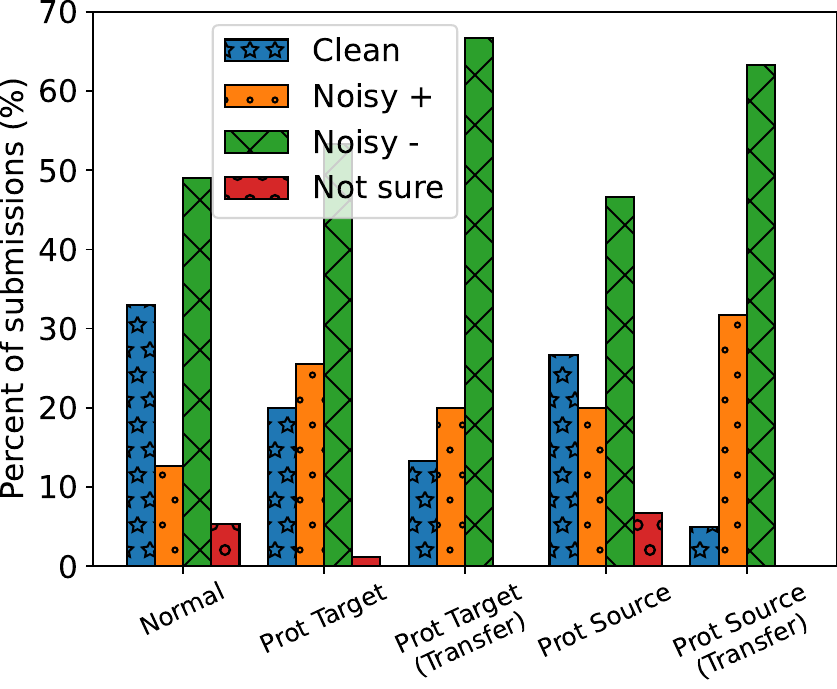}
\caption{Clean or Noisy}\label{fig:human_study_clean_noisy_en}
\end{subfigure}
\caption{{Results of human study on the English dataset NUS-48E. ``Noise +'' and ``Noise -'' denote the answers ``{\it noisy w/ influence}'' and ``{\it noisy w/o influence}''.}}\label{fig:human_study_en}
\end{figure*}

\subsection{{Human Study using the English Dataset NUS-48E}}\label{sec:humn_study_english}
{In \cref{sec:human_study}, we confirmed the prevention effectiveness and harmlessness of \toolname from the perspective of human perception using the Chinese dataset OpenSinger. 
To be diverse, we also conduct a human study on the same Credamo platform using the English dataset NUS-48E. 
}

\noindent 
{{\bf Singing voices/songs.} 
The (pairs of) singing voices/songs of all three tasks including the special questions for low-quality answer filtering are the same as for
the human study using the Chinese dataset OpenSinger (cf. \cref{sec:human_study}) except that they are of English language.}

\noindent 
{{\bf Participants.} 
We recruited 120 USA-located participants (after filtering) for each task (non-overlapping among tasks), resulting in 360 participants in total for the human study using the English dataset NUS-48E. 
While Credamo does not allow the collection of participants' demographic information outside China, we argue that given the high number of participants, they offer a reasonable representation.}

\noindent 
{{\bf Spent time.}
Statistically, they spent $17\pm 8$, $26\pm 10$, and $9\pm 3$ minutes for Task 1, Task 2, and Task 3, respectively. In contrast, filtered participants by special questions spent $9\pm 6$, $13\pm 5$, and $5\pm 4$ minutes for Task 1, Task 2, and Task 3, respectively, indicating a positive correlation between spent time and answer quality.}

\noindent
{{\bf Results.} 
The results are reported in \figurename~\ref{fig:human_study_en}.
They are similar to that of the human study using the Chinese dataset OpenSinger (cf. \cref{sec:human_study}) and we can draw the same conclusion. These results indicate the prevention effectiveness and harmlessness of \toolname across different languages from the perspective of human perception.}